\documentstyle[11pt,amssym,emulateapj_rj,epsfig]{article}

\topline{The Astrophysical Journal Supplement Series, 127:000--000,
2000 {\rm March}\hspace*{6cm}(MS-50061-ApJS)}
\submitted{Received 1999 July 12; accepted 1999 September 28}
\received{July 12, 1999}
\accepted{September 28, 1999}
\journalid{127}{2000 March \#1 issue}
\articleid{000}{000}
\paperid{MS-50061-ApJS}

\lefthead{JANSEN $ET\, AL$.}
\righthead{SPECTROPHOTOMETRY OF NEARBY FIELD GALAXIES}

\newcommand{\ie}{\mbox{i.e.\ }}%
\newcommand{\eg}{\mbox{e.g.\ }}%
\newcommand{\cf}{\mbox{cf.\ }}%
\newcommand{\etal}{\mbox{et al.\ }}%
\newcommand{\muB}{\mbox{$\mu_{{}_B}$}}%
\newcommand{\dmaj}{\mbox{$D_{maj}$}}%
\newcommand{\dmin}{\mbox{$D_{min}$}}%
\newcommand{\MB}{\mbox{M$_B$}}%
\newcommand{\BR}{\mbox{$(B\!-\!R)$}}%
\newcommand{\br}{\mbox{$B\!-\!R$}}%
\newcommand{\BRe}{\mbox{\BR$_e$}}%
\newcommand{\BV}{\mbox{$(B\!-\!V)$}}%
\newcommand{\VR}{\mbox{$(V\!-\!R)$}}%
\newcommand{\persecsq}{\mbox{$\;$arcsec$^{-2}$}}%
\newcommand{\Ho}{\mbox{H$_0$}}%
\newcommand{\kmsMpc}{\mbox{km s$^{-1}$ Mpc$^{-1}$}}%
\newcommand{\tpm}{\mbox{$\pm$}}%
\newcommand{\ttimes}{\mbox{$\times$}}%
\newcommand{\tsim}{\mbox{$\sim$}}%
\newcommand{\tsimeq}{\mbox{$\simeq$}}%
\newcommand{\fluxdens}{\mbox{ergs$\,$s$^{-1}\,$cm$^{-2}\,$\AA$^{-1}$}}%
\newcommand{\Dd}{\mbox{\arcdeg}}%
\newcommand{\Dm}{\mbox{\arcmin}}%
\newcommand{\HII}{\mbox{H{\sc ii}$\;$}}%
\newcommand{\Ha}{\mbox{H$\alpha $}}%
\newcommand{\Hb}{\mbox{H$\beta $}}%
\newcommand{\Hd}{\mbox{H$\delta $}}%
\newcommand{\NII}{\mbox{[N{\sc ii}]}}%
\newcommand{\HaNII}{\mbox{\Ha+\NII}}%
\newcommand{\OI}{\mbox{[O{\sc i}]}}%
\newcommand{\OII}{\mbox{[O{\sc ii}]}}%
\newcommand{\OIII}{\mbox{[O{\sc iii}]}}%
\newcommand{\SII}{\mbox{[S{\sc ii}]}}%
\newcommand{\HeI}{\mbox{He\,{\sc i}}}%
\newcommand{\lam}{\mbox{$\lambda $}}%

\begin{document}

\title {\large\bf Spectrophotometry of nearby field galaxies: the data}

\author{Rolf A. Jansen$^{1,2}$, Daniel Fabricant$^{2}$,
	Marijn Franx$^{1,3}$, \& Nelson Caldwell$^{4,2}$}
\affil {$^1$Kapteyn Astronomical Institute, Postbus~800, NL-9700~AV
	Groningen, The Netherlands}
\affil {$^2$Harvard-Smithsonian Center for Astrophysics, 60 Garden St.,
	Cambridge, MA~02138}
\affil {$^3$Leiden Observatory, Postbus 9513, NL-2300~RA Leiden,
        The Netherlands}
\affil {$^4$F.L.~Whipple Observatory, Amado, AZ~85645}
\affil {jansen@astro.rug.nl, dfabricant or ncaldwell@cfa.harvard.edu,
	franx@strw.leidenuniv.nl}


\begin{abstract}

We have obtained integrated and nuclear spectra, as well as $U, B, R$
surface photometry, for a representative sample of 196 nearby galaxies. 
These galaxies span the entire Hubble sequence in morphological type, as
well as a wide range of luminosities ($\MB=-14$ to $-22$).  Here we
present the spectrophotometry for these galaxies.  The selection of the
sample and the $U, B, R$ surface photometry is described in a companion
paper (Paper~I).  Our goals for the project include measuring the
current star formation rates and metallicities of these galaxies, and
elucidating their star formation histories, as a function of luminosity
and morphology.  We thereby extend the work of Kennicutt (1992a) to
lower luminosity systems.  We anticipate that our study will be useful
as a benchmark for studies of galaxies at high redshift. 

We describe the observing, data reduction and calibration techniques,
and demonstrate that our spectrophotometry agrees well with that of
Kennicutt (1992b).  The spectra span the range 3550---7250\AA\ at a
resolution (FWHM) of \tsim 6\AA, and have an overall relative
spectrophotometric accuracy of \tsim\tpm6\%.  We present a
spectrophotometric atlas of integrated and nuclear rest-frame spectra,
as well as tables of equivalent widths and synthetic colors.  The atlas
and tables of measurements will be made available electronically. 

We study the correlations of galaxy properties determined from the
spectra and images.  Our findings include: (1) galaxies of a given
morphological class display a wide range of continuum shapes and
emission line strengths if a broad range of luminosities are considered,
(2) emission line strengths tend to increase and continua tend to get
bluer as the luminosity decreases, and (3) the scatter on the general
correlation between nuclear and integrated \Ha\ emission line strengths
is large. 

\end{abstract}


  
\section{Introduction}

With the advent of the current generation of very large telescopes,
galaxies are now routinely studied at fainter magnitudes and higher
redshifts than was previously possible.  A major difficulty with the
interpretation of these high-z spectroscopic data is the lack of good
comparison samples from the local universe.  Distant galaxies subtend
small angles on the sky, comparable to spectrograph slit widths, and
their spectra tend to be integrated spectra.  The same slits sample only
the nuclear regions of nearby galaxies.  A direct comparison of distant
and nearby galaxy spectra, therefore, is difficult. 

In a pioneering effort, Kennicutt (1992a) obtained integrated
spectrophotometry for 90 galaxies spanning the entire Hubble sequence. 
This study continues to have broad application for the study of galaxy
spectral properties at both high and low redshift.  Kennicutt's study,
however, is limited to the brightest galaxies of each morphological
type, and no uniform, multiple filter, surface photometry is available
for these galaxies.  Also, only half of Kennicutt's sample was observed
at 5--7\AA\ spectral resolution, the remaining half was observed at
15--20\AA\ resolution.  Kinney \etal 1996, McQuade \etal 1995, and
Storchi-Bergmann \etal 1994, have constructed spectral energy
distributions of samples of star-forming, quiescent and active galaxies
that extend to the UV, and to the X-ray and IR/radio (Schmitt \etal
1997).  Their apertures are smaller than Kennicutt's, however, and their
study is limited to the brightest galaxies. 

The goal of our Nearby Field Galaxy Survey (NFGS) is to significantly
extend Kennicutt's pioneering work.  We have obtained integrated and
nuclear spectroscopy, and $U,B,R$ surface photometry, for a sample of
196 galaxies in the nearby field.  This sample includes galaxies of all
morphological types and spans 8 magnitudes in luminosity.  We include
galaxies from a broad range of local galaxy densities, attempting to
avoid a bias towards any particular cosmic environment.  Our use of the
term ``field'' thus corresponds to that of Koo \& Kron (1992) and Ellis
(1997).  We will use these observations to study the emission and
absorption line strengths, metallicities, star formation rates and star
formation histories, morphologies, structural parameters, and colors of
the sample galaxies.  These data can be used as an aid in understanding
the spectra and imagery of galaxies at larger distances, and in
measuring the changes in their properties over time.  Furthermore, our
sample can be used as a benchmark for galaxy evolution modeling and
comparison with observations of high redshift galaxies, as will result
from future observations with large ground based telescopes and the NGST
(\eg Kennicutt 1998). 

In this second paper we present the integrated and nuclear
spectrophotometry.  The 196 target galaxies in this survey were
objectively selected from the CfA redshift catalog (CfA~I, Huchra \etal
1983) to span a large range of $-14$ to $-22$ in absolute $B$ magnitude,
while sampling fairly the changing mix of morphological types as a
function of luminosity.  For the details of the galaxy selection, a
discussion of the merits and limitations of the sample, and for the
$U,B,R$ photometry, we refer to Jansen \etal (1999a; hereafter Paper~I). 
Detailed analysis of the complete data set will be presented in future
papers.  The structure of this paper is as follows. 
Section~\ref{S-obsredcal} deals with the observing strategy, reduction
and calibration of the data.  In section~\ref{S-Accuracy} we assess the
data quality and spectrophotometric errors.  In section~\ref{S-Results}
we present the primary data products (this section contains the atlas of
integrated and nuclear spectra).  We conclude with a short discussion
(section~\ref{S-Summary}).  Notes on individual objects in the sample
are collected in appendix~A. 

  
\section{Observations and Data Reduction}
\label{S-obsredcal}

\subsection{Observations}
\label{S-observations}

The spectroscopic observations reported here were made with the FAST
spectrograph at the F.L.~Whipple Observatory's 1.5~m Tillinghast
telescope\footnote{The F.L.~Whipple Observatory is operated by the
Smithsonian Astrophysical Observatory, and is located on Mt.~Hopkins in
Arizona.}.  The data were acquired during 41 predominantly moonless
nights between 1995 March and 1997 March.  The

\newlength{\txw}
\setlength{\txw}{\textwidth}

\noindent\vspace*{3mm}\leavevmode
\framebox[0.475\txw]{
   \centerline{\parbox[c]{0.475\txw}{
       \centerline{\rule[-0.20\txw]{0pt}{0.450\txw}\Large\it ``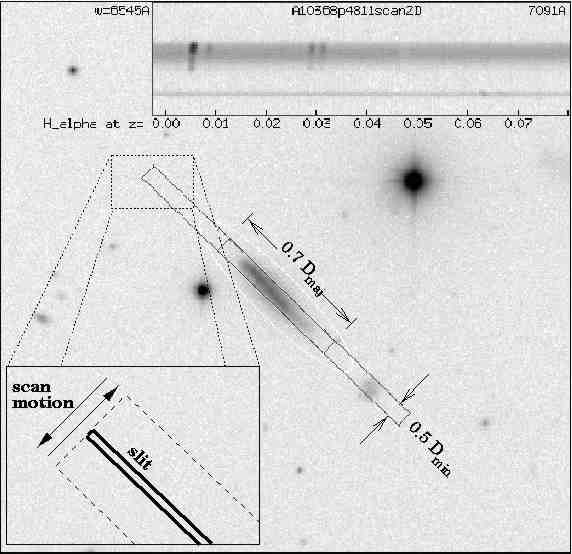''}}
   }
}\par

\noindent\makebox[0.475\txw]{
\centerline{
\parbox[t]{0.475\txw}{\footnotesize {\sc Fig.~1 ---} An example of the
geometry of the periodic scan used to obtain an integrated spectrum for
galaxy A10368$+$4811.  The slit is scanned over a distance equal to half
the blue minor axis optical diameter.  Integrated spectra were extracted
using an objectively defined aperture of size $0.7\, D_{26}$ (the major
axis diameter at $\mu_B=26$ mag\persecsq) as determined from our
B-filter photometry.  The ellipse drawn into the image was fit to the
$B_{26}$ isophote.  In this case, \tsim 80\% of the light within this
isophote enters into the integrated spectrum, or 68\% of the total
galaxian light. } }
}\vspace*{0.5cm}
%
   
\noindent vast majority of the observations were obtained during
transparent or photometric conditions, the remainder through thin
clouds. 

The FAST spectrograph is equipped with a thinned back-side illuminated
Loral CCD with 2720$\times $512, 15\micron\ pixels.  We used a 300 line
mm$^{-1}$ grating blazed at 4750\AA\ in first order and a 3\arcsec\ slit
to obtain a spectral coverage of \tsim 3940\AA\ (centered at \tsim
5450\AA), at a resolution of 6\AA\ FWHM.  The spatial resolution is
0.57\arcsec\ per pixel, although we usually binned the images on
read-out either by 4 or by 8 pixels along the slit to improve the
signal-to-noise (S/N) ratio.  The unvignetted slit length is 3\arcmin. 
The FAST spectrograph is described in Fabricant \etal (1998). 

Our procedure for measuring the nuclear spectra was to align the
spectrograph slit with the major axis of each galaxy, centering the slit
on the nucleus.  We are then able to extract a spectrum from the desired
central portion of the galaxy.  To reduce the overhead due to the (at
that time) manual rotation of the spectrograph we opted to bin the
galaxy position angles in bins of $<$20\arcdeg.  For most galaxies,
therefore, the alignment is better than 10\arcdeg.  Exposure times
ranged from 900 to 1800~s, and we typically binned the CCD by 4 pixels
in the spatial direction. 

To obtain integrated spectra we used the telescope drives to scan the
spectrograph slit across the galaxy.  First, the slit was offset from
the major axis by a distance of 0.25 times the blue minor axis diameter,
\dmin, as listed in the UGC catalogue.  \dmin\ loosely corresponds to
the diameter measured at \muB=26 mag\persecsq\ for most galaxies
(Paper~I).  We then scanned the slit back and forth across the face of
the galaxy over a total distance of $0.5\,\dmin$, as indicated in
figure~1.  The direction of the scan was set at the galaxy's minor axis
position angle even if the slit position angle was offset slightly from
that of the galaxy's major axis.  Backlash in the telescope drives
sometimes caused the scan to assume a slight ``S'' shape, with a
sideways offset of up to 5\arcsec.  By programming the telescope control
system appropriately, we ensured that the slit made a minimum of 20
passes over the galaxy per exposure.  We integrated 1200 to 1800~s per
exposure with total exposure times of 1800 to 7200~s.  Here, we binned
the CCD by 8 in the spatial direction. 

Three of the target galaxies have major axis blue diameters that exceed
the spectrograph slit length (NGC~3279, IC~708, A12446\-$+$5155).  For
these galaxies the slit was rotated to the position angle of the minor
axis, and the scanning was performed over a total distance of 0.5 times
the major axis diameter. 

The scan length was chosen to avoid degrading the S/N by integrating for
excessive amounts of time on the parts of the galaxies with surface
brightnesses below the detection limit.  As will be discussed in
section~\ref{S-extraction} we were able to reach to fainter surface
brightnesses than we had estimated.  We therefore use the data along the
slit to increase our sampling of the faint outer parts of the galaxies. 

During most nights we also observed flux standard stars from Massey
\etal (1988) to flux calibrate our spectra.  At least one of the
well-observed secondary standards BD+28\Dd 4211 and Feige~34 was
monitored throughout these nights.  For each standard star observation
we rotated the spectrograph to align the slit with the parallactic
angle.  In most cases we took several exposures with the star at
different locations along the slit.  The 3\arcsec\ slit is wide compared
to the seeing FWHM.  We observed additional standard stars when their
parallactic angles matched the current position angle of the slit.  The
importance of aligning the slit with the parallactic angle for these
exposures is discussed by Filippenko (1982). 

Before and after each object exposure we observed a He-Ne-Ar lamp to
provide wavelength reference lines.  At the beginning and end of each
night we obtained a series of twilight spectra and halogen lamp spectra 
for flatfielding, as well as a series of bias frames.  At the end of 
each night and whenever bad weather prevented observations we obtained
dark exposures.

In table~1 we give an overview of the sample and observations.  Columns
(1) through (3) contain: the galaxy numbers that we use internally,
their common names (NGC, IC, or IAU anonymous notation), and UGC catalog
number (Nilson 1973), respectively.  Columns (4) and (5) list the
numerical morphological types and their translation onto the Hubble
sequence, as reclassified using our CCD imagery (Paper~I).  Column (6)
gives the absolute $B$ filter magnitude calculated directly from total
apparent $B$ filter magnitude and the galaxy redshifts, assuming a
simple Hubble flow and \Ho=100 \kmsMpc.  The blue photographic major and
minor axis diameters in columns (7) and (8) are from the UGC; the
diameter at the \muB=26 mag\persecsq\ isophote in column (9) is from
Paper~I.  In column (10) we give the apertures for the integrated
spectra, first along the slit direction and then the total scan length
perpendicular to the slit.  Column (11) indicates what fraction of the
total $B$-filter light is sampled in the integrated spectra.  Columns
(12) and (13) list the position angles of galaxy and spectrograph slit,
respectively, measured from North through East.  The galaxy position
angles are from the UGC, with the exception of A12446$+$5155 (75 instead
of 5\arcdeg) and A11336$+$5829 (171 instead of 9\arcdeg).  Column (14)
lists the available data, where abbreviations ``i'', ``n'', ``$U$'',
``$B$'', and ``$R$'' are used for integrated and nuclear spectra, and
$U,B$ and $R$ photometry, respectively.  The final column contains
references to the notes for the table entries.  We include data for two
additional galaxies, A01047\-$+$1625 and NGC~784, that are not part of
the statistical sample.  In table~1 these galaxies are placed within
parentheses. 

In order to compare our spectrophotometry with Kennicutt's (1992b), we
reobserved nine of the galaxies in his sample with our observational
strategy: scans over $0.5\,\dmin$($0.5\,\dmaj$) with the spectrograph
slit aligned to the position angle of the galaxy major(minor) axis. 
Data for these galaxies is given in table~2 in the same format as
table~1.

\subsection{Data reduction}
\label{S-reduction}

The data were reduced within the IRAF environment, following standard
techniques.  The sequence of steps is: (1) interpolation over bad
columns, dead and hot pixels, (2) bias subtraction, (3) dark current
subtraction\footnote{This was important if the CCD had been UV-flooded
recently; otherwise the dark current in a half hour exposure is less
than 2 e$^{-}$/pix.} using a median of the dark exposures, \vfill


\newcommand{\inubr}{\mbox{\makebox[0.60em][c]{i}\makebox[0.69em][c]{n}\makebox[0.69em][c]{$U$}\makebox[0.69em][c]{$B$}\makebox[0.69em][c]{$R$}}}%
\newcommand{\xnubr}{\mbox{\makebox[0.60em][c]{-}\makebox[0.69em][c]{n}\makebox[0.69em][c]{$U$}\makebox[0.69em][c]{$B$}\makebox[0.69em][c]{$R$}}}%
\newcommand{\xxubr}{\mbox{\makebox[0.60em][c]{-}\makebox[0.69em][c]{-}\makebox[0.69em][c]{$U$}\makebox[0.69em][c]{$B$}\makebox[0.69em][c]{$R$}}}%
\newcommand{\xxxbr}{\mbox{\makebox[0.60em][c]{-}\makebox[0.69em][c]{-}\makebox[0.69em][c]{-}\makebox[0.69em][c]{$B$}\makebox[0.69em][c]{$R$}}}%
\newcommand{\inxbr}{\mbox{\makebox[0.60em][c]{i}\makebox[0.69em][c]{n}\makebox[0.69em][c]{-}\makebox[0.69em][c]{$B$}\makebox[0.69em][c]{$R$}}}%
\newcommand{\xnxbr}{\mbox{\makebox[0.60em][c]{-}\makebox[0.69em][c]{n}\makebox[0.69em][c]{-}\makebox[0.69em][c]{$B$}\makebox[0.69em][c]{$R$}}}%
\newcommand{\inxxx}{\mbox{\makebox[0.60em][c]{i}\makebox[0.69em][c]{n}\makebox[0.69em][c]{-}\makebox[0.69em][c]{-}\makebox[0.69em][c]{-}}}%
\newcommand{\ixxxx}{\mbox{\makebox[0.60em][c]{i}\makebox[0.69em][c]{-}\makebox[0.69em][c]{-}\makebox[0.69em][c]{-}\makebox[0.69em][c]{-}}}%
\newcommand{\xnxxx}{\mbox{\makebox[0.60em][c]{-}\makebox[0.69em][c]{n}\makebox[0.69em][c]{-}\makebox[0.69em][c]{-}\makebox[0.69em][c]{-}}}%
\newcommand{\xxxxx}{\mbox{\makebox[0.60em][c]{-}\makebox[0.69em][c]{-}\makebox[0.69em][c]{-}\makebox[0.69em][c]{-}\makebox[0.69em][c]{-}}}%

\newcommand{\paslit}{\mbox{PA$_{\hbox{slit}}$}}

\setcounter{table}{0}

\setlength{\tabcolsep}{3pt}

\normalsize

\noindent (4) flat fielding, correcting for the pixel-to-pixel
variations with halogen lamp flats and for illumination variations with
twilight flats, and (5) wavelength calibration and subsequent resampling
of the data on a linear wavelength grid.  For one run, due to a problem
with the CCD preamplifier, the bias level was corrected as a function of
the signal level in the previously read-out pixels, using an empirical
model of the preamplifier behavior. 

We removed cosmic ray hits as follows.  Because we binned the CCD on a
fairly coarse grid in the spatial direction, the profile of a cosmic ray
hit tends to become more symmetric and the signal is diluted by summing
with unaffected pixels.  Prior to the wavelength calibration step, we
used a (non-IRAF) routine to automatically flag and remove \tsim 95\% of
the cosmic ray hits by fitting a data model and using a rejection
criterion based on the relative signal levels in the data and model, and
on the CCD parameters.  The remaining cosmic ray hits were flagged
manually and were subsequently removed by interpolation. 

\subsubsection*{\footnotesize Extraction of one-dimensional spectra}
\label{S-extraction}

We extracted nuclear and integrated spectra from the processed pointed
and scanned exposures using objectively defined apertures.  In general,
the peak of the galaxian light distribution was used to trace the
extraction aperture, but occasionally, a star in the (scanned) aperture
was used.  For the nuclear spectra, we used a fixed aperture,
6.84\arcsec\ along the slit centered on the 3 brightest pixels in the
portion of the spectrum between 5100--5600\AA.  The second aperture
dimension, the slit width, was always 3\arcsec.  We assigned fractional
pixel values if the aperture did not include an entire pixel. 

Because the S/N at low surface brightness was better than anticipated,
for our integrated spectra we use an aperture 0.7 times the major axis
diameter at \muB = 26.0 mag\persecsq\ (as determined from our $B$-filter
photometry, Paper~I) rather than $0.5\, D_{26}$ (\cf
section~\ref{S-observations}).  We thereby better sample the total
galaxian light distribution.  For asymmetric profiles the center of the
aperture is offset from the peak of the galaxy surface brightness.  We
computed errors for the nuclear and integrated galaxy spectra by summing
the contributions of read noise and shot noise in quadrature. 

For the flux standard star exposures we defined apertures interactively
and conservatively to contain \emph{all} the stellar light.  Again we
traced the peak of the light as a function of wavelength.  If multiple
exposures along the slit were available, the extracted spectra were
summed. 

A few of the two-dimensional galaxy exposures were contaminated by the
spectra of foreground stars.  If the contamination occurred within the
extraction aperture, the affected part was excluded from summation into
the final one-dimensional spectra. 

\subsubsection*{\footnotesize Sampled fraction of the galaxian light}
\label{S-scanfrac}

Using our imagery, we measured the fraction of the galaxian light we
include in the integrated spectra, and the dependence of this fraction
on morphological type and luminosity.  Our integrated spectra include
52--97\% of the light enclosed within the $B_{26}$ isophote.  The
average is 82\% with an RMS scatter of \tpm6.4\%.  They include 76\% of
the total $B$ light, with an RMS scatter of \tpm8\%. 

Because later type galaxies tend to have somewhat lower surface
brightnesses than early type galaxies, one might expect a trend with
morphological type.  Although 18 out of 21 galaxies for which we sample
less than 75\% of the light within $B_{26}$ are of type Sc or later, no
clear trend is visible as a function of type for the bulk of the sample. 
The sampled fraction also does not depend strongly on galaxy color or
absolute magnitude. 

The nuclear spectra include on average 10\% of the light enclosed within
the $B_{26}$ isophote, ranging from 0.4--72\% and with an RMS scatter of
\tpm11\%.  For these small nuclear extraction apertures the sampled
fraction depends on the radial light profile.

\subsection{Flux calibration, de-redshifting and normalization}
\label{S-calibration}

The extracted spectra were flux calibrated on a relative flux scale
using flux standard stars from the sample of Massey \etal (1988).  Cubic
spline sensitivity functions of 21st order were fit interactively for
each of the standard star observations.  The sensitivity function (SF)
relates the measured intensity to the (calibrated) flux density (in
\fluxdens) as a function of wavelength, after removing atmospheric
extinction.  SFs for individual observations were offset to match the
observations obtained during the most transparent photometric
conditions. 

Individual SF zeropoints for photometric observations match to \tsim
0.05 mag.  As the combined throughput of telescope, spectrograph and
atmosphere drops sharply below 3700\AA, the derived SF is not reliable
below \tsim 3600\AA.  The spectrophotometric errors also increase
longwards of 7100\AA.  Most flux standards are much bluer than a typical
galaxy so we also observed red standards, \eg Cyg~OB2\#9.  SFs derived
for blue and red standards differ systematically due to second order
contamination (see section~\ref{S-secorder}). 

After fitting an initial solution for the SF using a KPNO standard
atmospheric extinction curve, we attempted to measure the atmospheric
extinction curve, and to subsequently refit the combined SF.  The
accuracy of this extinction measurement is limited by the number of
available standard star observations, but we determined that the actual
extinction curve does not deviate from the KPNO standard one by more
than 0.07 mag airmass$^{-1}$ at any wavelength between 3700 and 7100\AA. 
We adopt the SF fitted using the KPNO standard extinction curve.

Redshifts were measured for each galaxy by either averaging emission
line redshifts (\OII\lam3727, \OIII\lam\lam4959,5007, \Ha, \NII\lam6584,
and \SII\lam\lam6718,6731), or by cross-correlating the galaxy spectra
with a composite stellar template spectrum.  The stellar template was
constructed from several K and G star spectra of various metallicities,
observed for this purpose.  Using these redshifts the calibrated spectra
were shifted to rest frame wavelengths. 

The rest frame spectra were normalized to the average flux in a 50\AA\
interval centered at 5500\AA\ and subsequently resampled on a uniform
wavelength grid spanning the range 3250--7500\AA\ at \tsim
1.36\AA/pixel.  Pixel values outside the actual data range are set to
zero.  The pixel scale matches the scale in the rest-frame spectrum of
the highest redshift galaxy in our sample, and the wavelength range
accommodates the extremes in our data.  Since the S/N ratio below
3600\AA\ tends to be low and the number of galaxies with data redward of
7200\AA\ is small, we will restrict the spectra plotted in
section~\ref{S-Results} to this smaller range. 

\subsection{Measurement of spectroscopic indices}

Equivalent widths (EWs) and fluxes of \OII\lam3727, \Hd, \Hb,
\OIII\lam4959, \OIII\lam5007, \HeI\ \lam5876, \OI\lam6300, \Ha,
\NII\lam6548, \NII\lam6584, \SII\lam6718, and \SII\lam6731 were measured
in both the nuclear and integrated spectra of the emission line
galaxies.  We used {\tt SPLOT} within IRAF to interactively define the
bandpasses and measure the EWs and fluxes.  Our spectral resolution is
sufficient to separate the nitrogen lines from \Ha, except where \Ha\ is
broadened (Sy~{\sc i}).  Measurement uncertainties are dominated by the
errors in the sky subtraction and cosmic ray residuals (see below) but
shot noise and in some cases the uncertainty in the continuum level
become important in the fainter lines as well.  Typical errors are
\tpm3\%--10\% for strong emission lines rising to \tpm30\% for EWs
smaller than 2\AA.  A much larger number of emission and absorption
lines, as well as several continuum indices, and synthetic broadband
magnitudes and colors, were measured using an automated routine and
fixed bandpass/filter definitions.  Except for the few galaxies with
broadened Balmer lines, the manual and automated measurements of \Ha\
agree to \tsim\tpm10\%.  We present here only the manually measured
emission line EWs and fluxes which we believe are more accurate as well
as the synthetic broadband filter colors.\vfill

  
\section{Spectrophotometric accuracy}
\label{S-Accuracy}

\subsection{Evaluation of the individual sources of error}
\label{S-erroreval}

The spectrophotometric accuracy can be estimated by adding individual
contributions to the total error in quadrature.  The main sources of
error can be divided into those that affect the spectral shape and those
that affect only small ranges in wavelength.  The first category
contains: (1) the fitting of the SF, (2) the published standard star
fluxes, (3) the adopted atmospheric extinction curve, (4) contam-\newpage

\noindent\leavevmode
\makebox[\txw]{
   \centerline{
      \epsfig{file=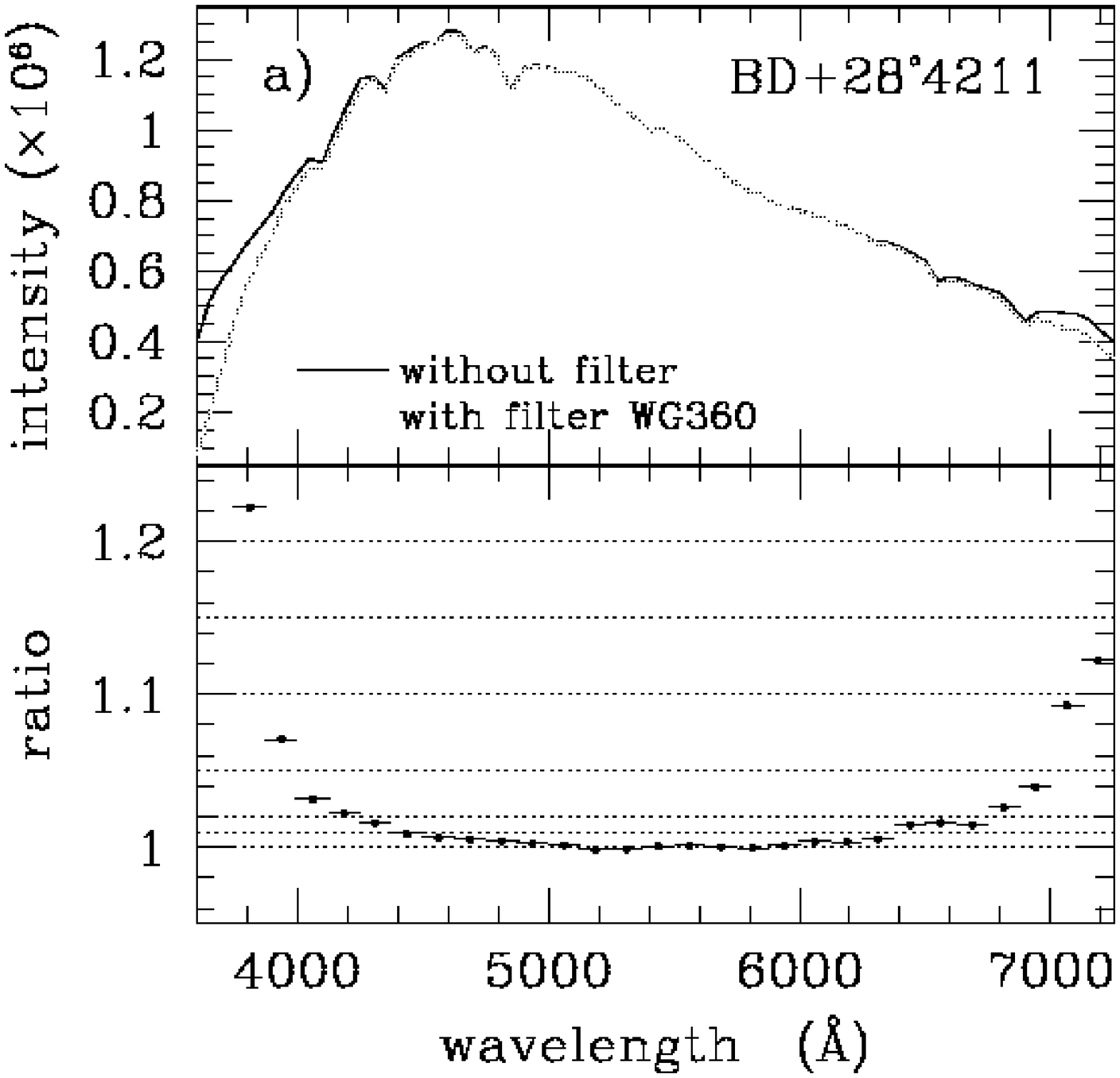,width=0.465\txw,clip=}\hspace*{0.05\txw}
      \epsfig{file=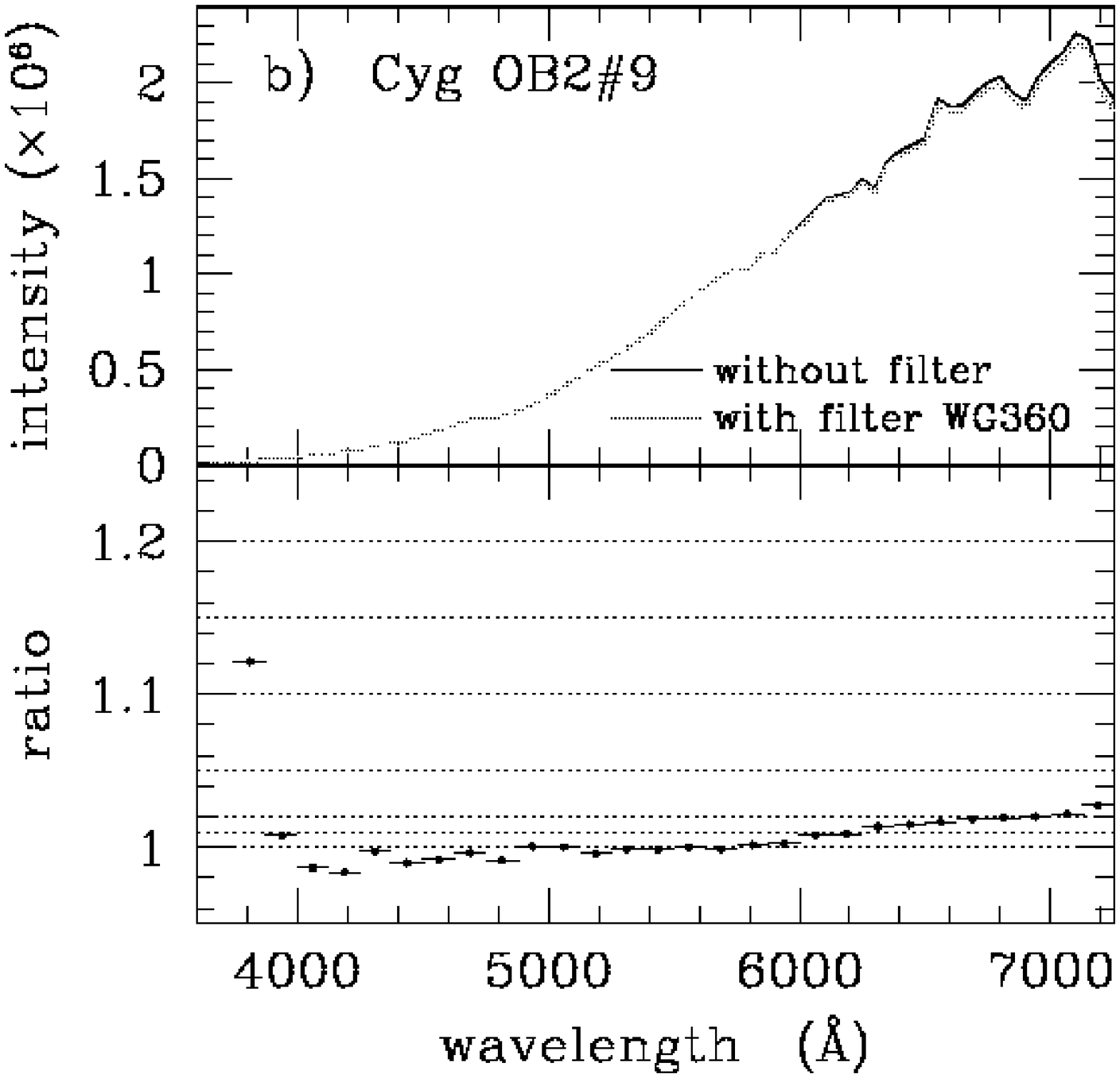,width=0.465\txw,clip=}
   }
}\par\vspace*{0.03\txw}\noindent\makebox[\txw]{
\centerline{
\parbox[t]{\txw}{\footnotesize {\sc Fig.~2 ---} A measurement of the
amount of second order UV light falling onto the red part of the spectra
on the CCD, using observations with and without a WG~360 order-blocking
filter, for a) the extremely blue stellar spectrum of BD+28\Dd 4211, and
b) for the highly reddened spectrum of Cyg~OB2\#9.  In the latter case,
the slow increase of the ratio with wavelength reflects the decreasing
efficiency of the anti-reflection coating on the filter.  Second-order
light becomes noticeable only redward of 6800\AA.  The
spectrophotometric error introduced into the galaxy spectra will not
exceed 5\%, 3.5\% and 3\% at 7250\AA\ for galaxies of type Im, Sab and
E, respectively. } }
}\vspace*{0.5cm}
%

\noindent ination of the red part of the spectrum by second order blue
light, and (5) for the nuclear spectra, the slight misalignment of the
CCD with the dispersion axis in combination with the coarse spatial
sampling.  The second category contains: (1) the flatfielding, (2) the
wavelength calibration, (3) the subtraction of the sky background and
(4) residuals due to cosmic ray hits.  We discuss each of these sources
of error in turn. 

The {\bf error in the fit of the SF} is estimated using the residuals of
individual standard stars from the mean calibration.  These residuals
are dominated by systematic differences in the SF fitted to the
different stars, and range from \tpm2 to 5\%.  {\bf Standard star fluxes},
summed in 50\AA\ bandpasses, are tabulated in Massey \etal (1988) and
are accurate to better than \tsim\tpm3\%, based on a comparison between
measurements for stars observed by both Stone (1977) and Oke (1974). 
Considering this 3\% uncertainty, we find that errors in our SF fits are
likely to be less than \tpm2\%. 

Application of the KPNO {\bf atmospheric extinction curve} to correct
our data introduces a wavelength dependent error into the fitted SF and,
hence, an error in the continuum slope of the spectra.  The best fit
solutions for the actual extinction curves scatter around the KPNO curve
with \tpm10\% slope differences.  Assuming that this scatter is a valid
measure of the error in the adopted extinction curve and given that most
galaxies were observed at low airmasses\vfill

\null\vspace*{0.6\txw}\noindent (the median airmass $A.M.=1.1\pm0.1$;
95.5\% of the spectra were observed at $A.M.<1.42$), we expect the error
in the continuum slope to be $\lesssim$1.2\% in the interval
4500--6200\AA\ and $<4.2$\% at any wavelength. 

\label{S-secorder}Given the large spectral coverage, we expect the red
part of the spectra to be contaminated {\bf by second order blue light}. 
To assess this source of error, we observed a very red (Cyg~OB2\#9,
\bv=1.93) and very blue (BD+28\Dd 4211, \bv=$-$0.34) star both with and
without a WG~360 order-blocking filter.  The result is shown in
figure~2.  The intensities were integrated over 125\AA\ bandpasses and
the spectra taken with the WG~360 filter inserted have been scaled to
the average level of the reference spectra (taken without the filter) in
a 1000\AA\ interval centered on 5500\AA\ (where no second order light
will be detected and filter throughput differences are minimal). 

The second order contamination turns out to be small.  We use the ratio
of the spectra of Cyg~OB2\#9 with and without the order-blocking filter
to measure the decreasing efficiency of the antireflection coating on
the order-blocking filter towards the red, since Cyg~OB2\#9 has no blue
light to produce second order contamination. Making this small
correction (3\% at 7250\AA), we find that second-order contamination in
the BD+28\Dd 4211 spectrum is detectable redward of 6800\AA, rising to
\tsim 10\% at 7250\AA.

We chose not to use the WG~360 filter for the galaxy observations
because it renders spectroscopy blueward of 4000\AA\ impossible.  The
only strong line in the near-UV, \OII\lam3727, will appear at 7454\AA,
\ie outside the wavelength range covered for all but 5 galaxies.  Of the
features in the red part of the spectrum, only the EWs of the redshifted
\SII\lam\lam6717,6731 lines may be underestimated by up to 3\% due to
the superposition of second-order UV light.  On larger wavelength
scales, given the mismatch of the galaxy colors and the colors of the
flux standard stars used in the fitting of the SF, the contribution of
second-order light to the red part of the spectra is expected to be
\tsim 0.47, 0.34 and 0.31 times that measured for BD+28\Dd 4211 for
galaxies of type Im, Sab and E, respectively; the corresponding errors
in the spectrophotometry will not exceed 5\%, 3.5\% and 3\% at 7250\AA. 

In the case of the nuclear spectra, the {\bf misalignment of the CCD
with the dispersion axis} introduces an error in the spectrophotometry
on large scales due to interpolation errors.  Addition of fractional
pixels derived from the aperture intersection with a pixel will not
exactly represent the light within the aperture.  The magnitude of this
error depends on the light profile.  In the scanned spectra errors of
this type are negligible, as the apertures are much larger than the
pixel size and the overall intensity level at the edges of the apertures
is very low compared to the integrated intensity.  Our tests indicate
that typical errors in the nuclear spectrophotometry are \tpm1--2\% over
wavelength ranges of about 1000\AA.  In the worst cases (highly compact
nuclei where the center falls between two pixels or galaxies with small
scale structure in their central regions) errors of up to \tpm5\% are
possible. 

{\bf Flatfielding errors} vary with wavelength and from observing run to
run depending on the quality of the CCD's UV flood.  The flatfield
variations are largest in the far blue (shortward of \tsim 4000\AA\ for
the March 1995 run, \tsim 3750\AA\ for all other runs).  Redward of
\tsim 6900\AA\ the errors increase due to the onset of fringing. 
Differences between flatfields taken on different nights within a run
are small ($\lesssim$1.5\% at any wavelength, typically \tpm0.5\% in the
3750--6900\AA\ interval).  The contribution of the read noise to the
error in the flatfielding is negligible. 

{\bf Errors in the wavelength calibration} introduce small errors in the
inferred flux densities on scales comparable to the distance between
individual calibration lines.  Errors in the dispersion solution range
from 0.25\AA\ in the blue (where the calibration lines are weaker and
blended) to 0.10\AA\ in the red. These dispersion errors produce
spectrophotometric errors of at most \tpm1.7\%.

{\bf Sky subtraction errors} dominate the total error at small scales,
especially for galaxies with $D_{maj}\gtrsim 2.5$\arcmin.  Following sky
subtraction, the noise increases by \tsim 3\% in the nuclear spectra and
10\% in the integrated spectra.  The larger contribution for the
integrated spectra arises from their coarser spatial binning, with fewer
pixels available to establish the sky level.  Large galaxies also leave
fewer pixels to measure the sky level. 

{\bf Cosmic ray residuals} introduce large errors in the extracted
spectra only near emission lines, where the steepness of the local
background renders a clean fit difficult.  Residuals in continuum or sky
portions in the spectra are smaller than \tpm1\% of the local
background.  Where we have multiple exposures and large extraction
apertures (integrated spectra, flux standards) errors are smaller than
\tpm0.4\%.  For the nuclear spectra, errors may be as large as \tpm10\%
per extracted pixel, worst case. \vspace*{\baselineskip}


Combining the above contributions to the errors in quadrature, we can
estimate the spectrophotometric accuracy at small and at large scales. 
Over scales of a few to several tens of \AA\ the expected error
$\varepsilon \approx\tpm8\% $.  Note, that this error should be
interpreted as a non-constant noise source that may affect single or a
small number of adjacent pixels, rather than a systematic offset in the
overall spectrophotometry.  On scales of hundreds of \AA\ or more the
expected spectrophotometric accuracy ${\cal E}\approx\tpm6\% $.  This
error gives us the expected spectrophotometric accuracy if our spectra
are integrated over larger bandpasses and compared with the
spectrophotometry of other workers. 

\subsection{Internal and external checks of the spectrophotometry}

\subsubsection*{\footnotesize Internal checks}
\label{S-compusus}

Five galaxies were observed twice during different observing runs.  In
figure~3 we present the data for the five sets of duplicate spectra. 
The spectra were normalized to the average level in interval
5200---6400\AA\ prior to plotting.  In the upper panels we plot the
spectra, where the solid lined spectra have been offset by 0.5 for
clarity.  The dates of observation, exposure times, effective airmasses
during the exposures, and dimensions of the extraction apertures are
indicated.  The effective apertures for each pair of spectra were very
similar.  In the lower panels we plot the ratio of each pair, measured
in 250\AA\ bins.  The binning ensures that photon noise does not
dominated the ratios.  To guide the eye, lines of \tpm2\% and \tpm5\%
deviation are indicated (dotted lines).  The deviations are rarely
larger than 5\%, and over large spectral regions the match is better
than \tpm2\%.  No systematic differences are seen.  \vfill

\noindent\leavevmode
\makebox[\txw]{
   \centerline{
      \epsfig{file=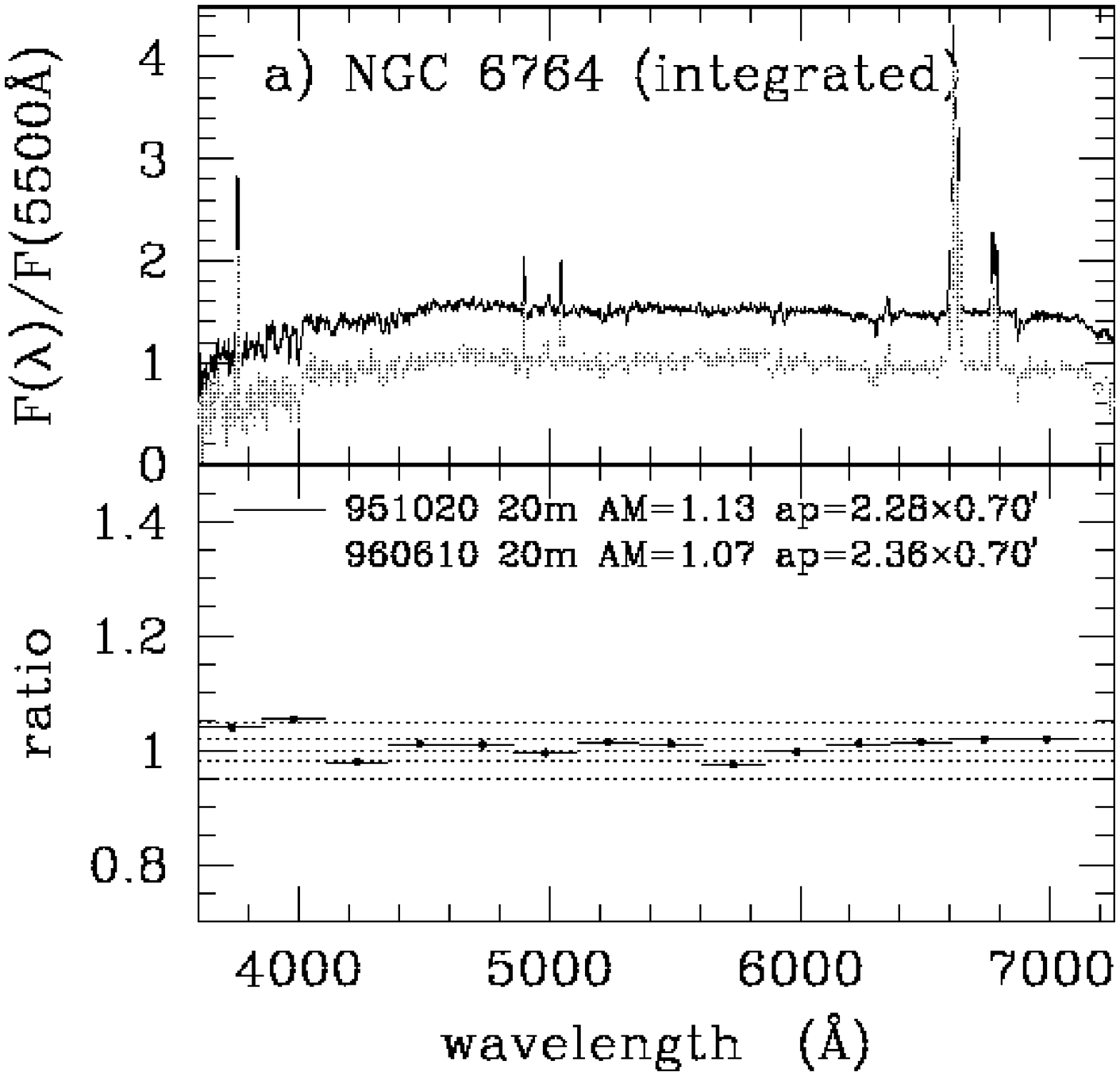,width=0.42\txw,clip=}\hspace*{0.05\txw}
      \epsfig{file=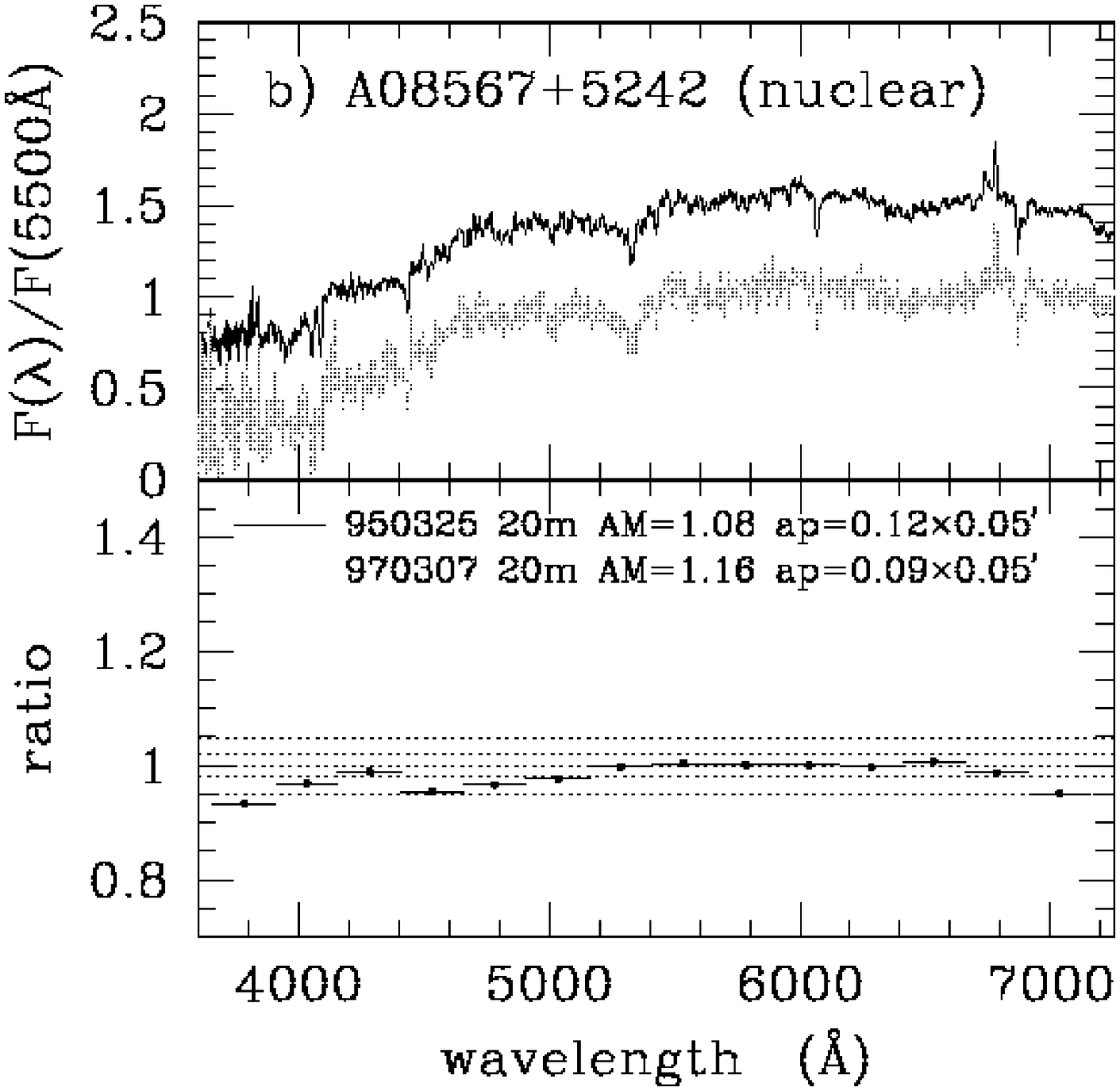,width=0.42\txw,clip=}
   }
}\par\vspace*{0.03\txw}\noindent\leavevmode
\makebox[\txw]{
   \centerline{
      \epsfig{file=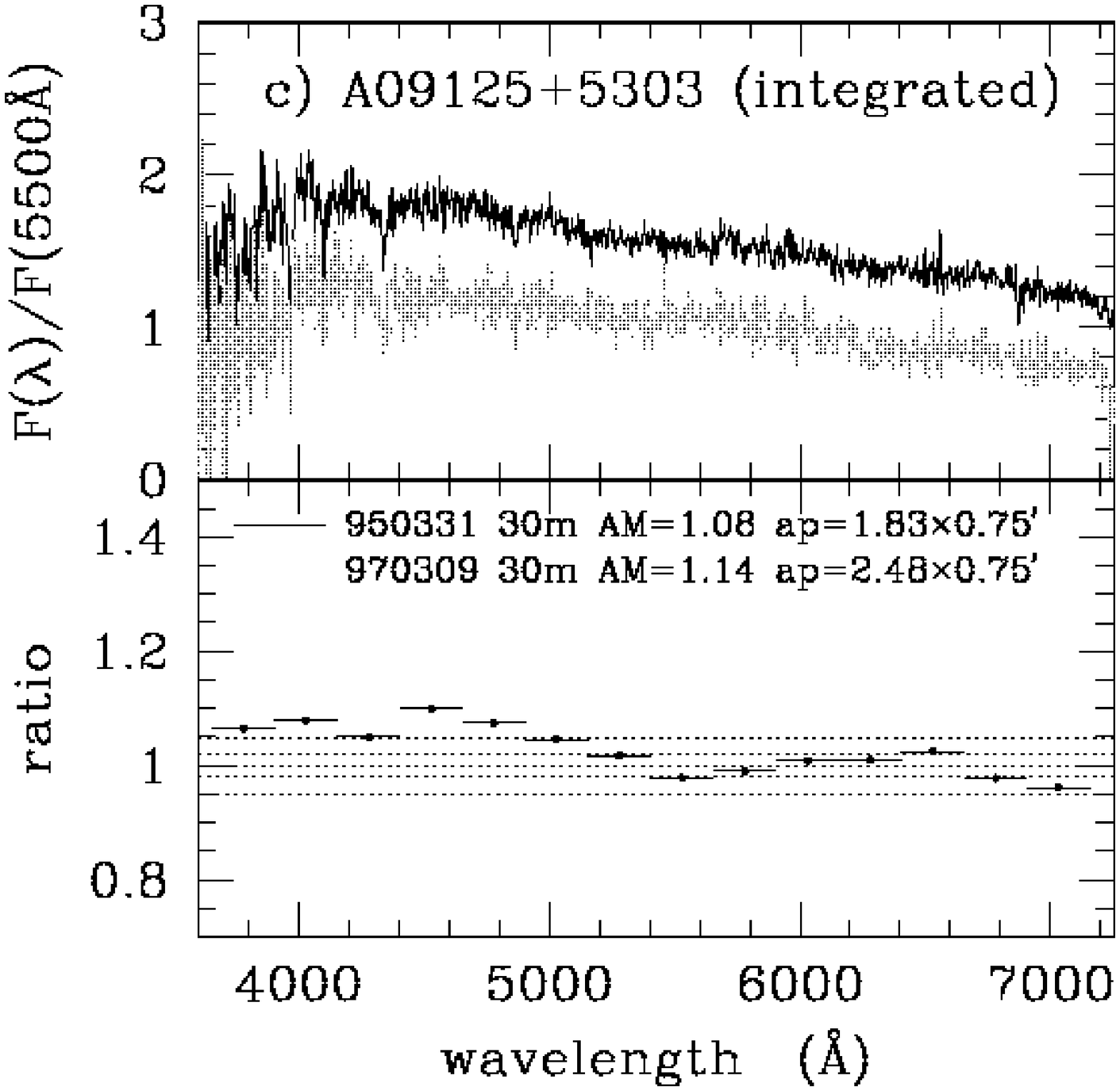,width=0.42\txw,clip=}\hspace*{0.05\txw}
      \epsfig{file=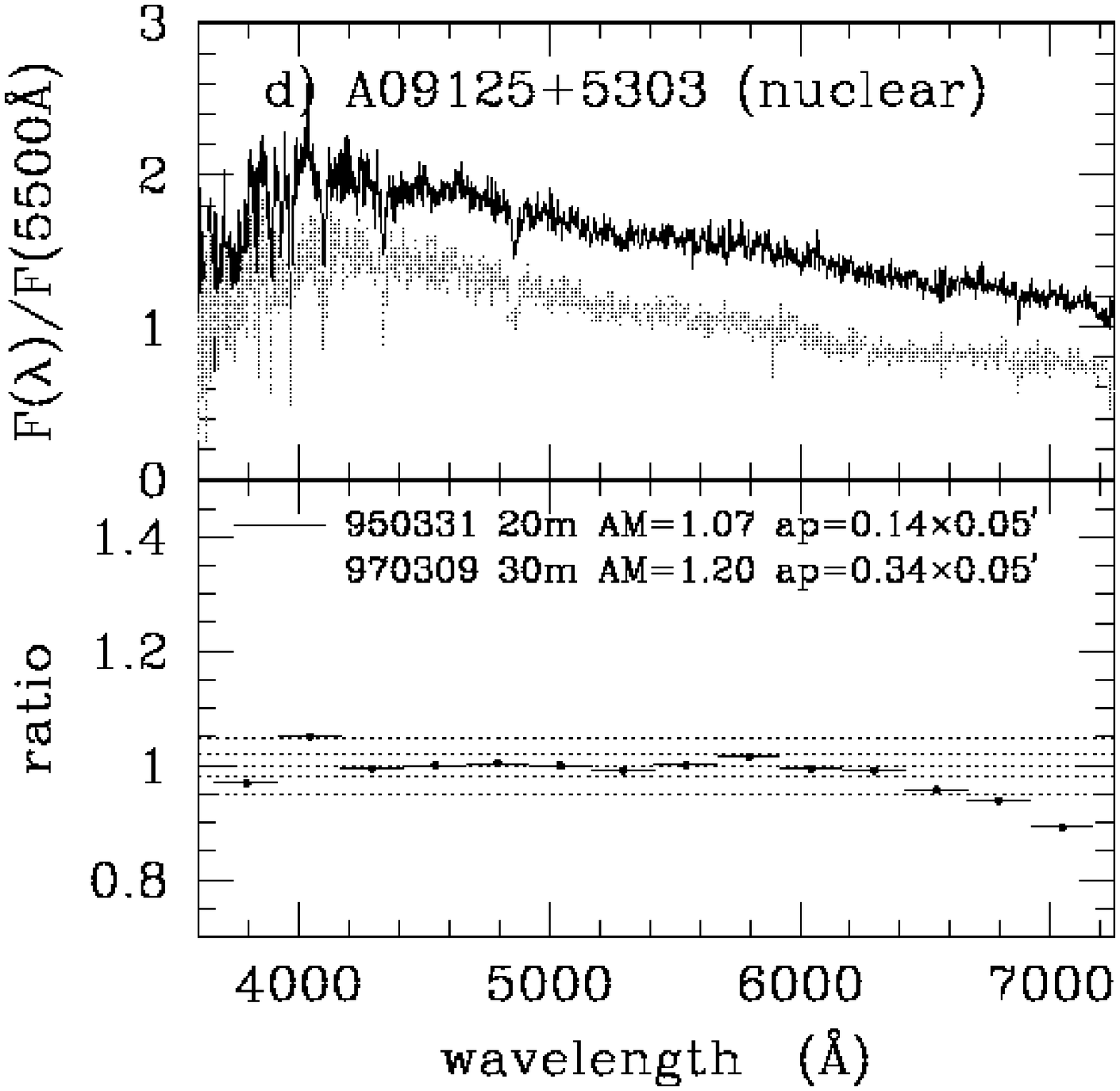,width=0.42\txw,clip=}
   }
}\par\vspace*{0.03\txw}\noindent\leavevmode
\makebox[\txw]{
   \centerline{
      \epsfig{file=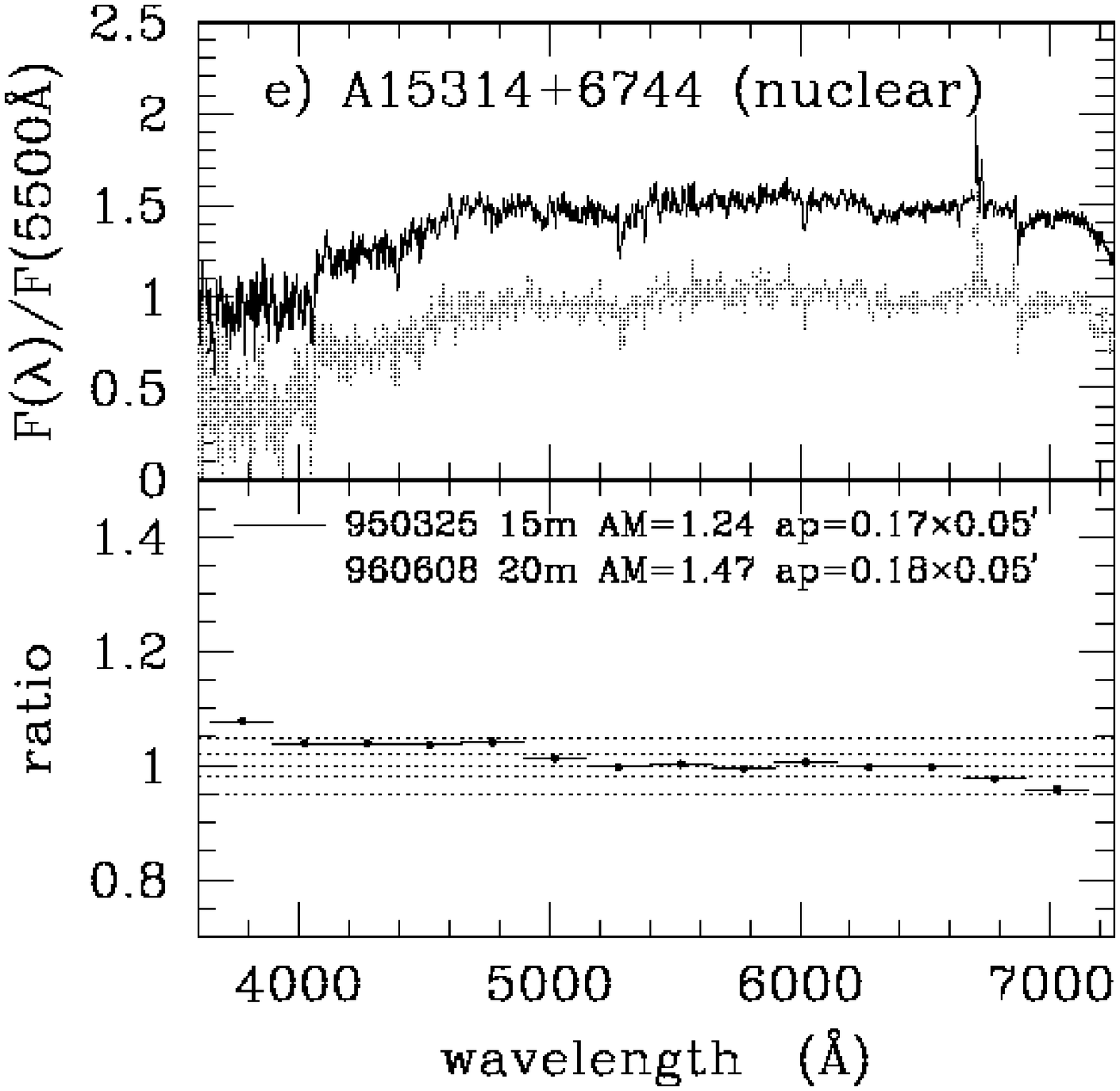,width=0.42\txw,clip=}\hspace*{0.05\txw}
      \makebox[0.43\txw]{\hspace*{0.07\txw}
\parbox[b]{0.35\txw}{\footnotesize {\sc Fig.~3 ---} A check of the
internal consistency of our spectrophotometry using the spectra of the
four galaxies that we observed during two different runs.  a) the
integrated spectra of NGC~6764, b) the nuclear spectra of A08567+5242,
c) the integrated and d) the nuclear spectra of A09125+5303, and e) the
nuclear spectra of A15314+6744.  All spectra were normalized to the
average level in interval 5200--6400\AA.  The upper spectrum in each
pair has been offset by 0.5 for clarity (upper panels).  The ratio of
each pair of spectra is calculated in 250\AA\ bins to ensure that photon
noise does not dominate.  Deviations are rarely larger than \tpm5\%, and
over large spectral regions the match is better than \tpm2\% (lower
panels).\\ \rule{0pt}{0.5cm}\null }
                         }
   }
}\par\vfill

%

\subsubsection*{\footnotesize Comparison with our photometry}

A more potent test of our external spectrophotometric accuracy is the
comparison of synthetic broadband colors measured in our spectra with
the broadband colors measured in the photometric part of this survey
(Paper~I).  The flux calibrated and normalized galaxy spectra were
convolved with the standard $B$ and $R_c$ filter bandpasses as tabulated
in Bessell (1990), and \BR\ colors on the AB system were determined. 
When measuring colors, the absolute flux calibration drops out. 

The $R_c$ bandpass has a red tail out to 9000\AA.  We used linear
interpolation to estimate the galaxy flux in the interval between the
last data point and 7500\AA, and model spectral energy distributions
(Bruzual \& Charlot 1993) to estimate the galaxy flux lost beyond
7500\AA.  The latter correction depends on galaxy color and ranges from
0.5\% (Irr) to 4.0\% (E).  The effective wavelength of the $R_c$ filter
is shifted towards shorter wavelengths with respect to the nominal
Cousins filter (6460\AA) when convolved with the spectrum of a typical
galaxy (see also the remark in Bessell (1979) regarding the large shift
in effective wavelength with stellar spectral type in $R_c$). 
Conversely the effective wavelength of the $B$ filter is shifted a
little towards the red, because of the presence of the 4000\AA\ break in
this filter. 

Before they can be compared with our photometric colors, the synthetic
colors on the AB system have to be placed onto the standard Johnson
$BV$--Cousins $R_c$ photometric system by correcting for the color zero
point differences between these systems.  Using the data in Bessell
(1990) (his table~3), we find: $\BR_{synth} =
\BR^{AB}-(B\!-\!B^{AB})_0+(R_c\!-\!R^{AB})_0 = \BR^{AB}-(-0.102)+0.193$,
\ie a star having a \BR\ color of 0.00~mag in the standard photometric
system would be observed to have a synthetic $(B-R)^{AB}$ color of
$(-0.102-0.193)=-0.295$.  As for the photometric colors, we correct the
synthetic colors for Galactic foreground extinction. 

In figure~4 we plot the measured synthetic \BR\ colors as a function of
photometric effective \BRe\ color.  Apart from a small offset in
zeropoint of $-0.023$ mag of the synthetic colors with respect to the
photometric ones, the synthetic colors reproduce the photometric colors
to within 0.065 mag (RMS) over the full range in galaxy colors.  This
implies that our relative spectrophotometry is accurate to \tsim\tpm6\%
over the range 3700--7200\AA\ as referenced to our photometry.  We note
that the observed offset and part of the remaining scatter of the
individual points arises from the inexact match of the spectroscopic and
photometric apertures and the presence of color gradients in galaxies.\\

\null\noindent\vspace*{5mm}\leavevmode
\makebox[0.475\txw]{
\centerline{\epsfig{file=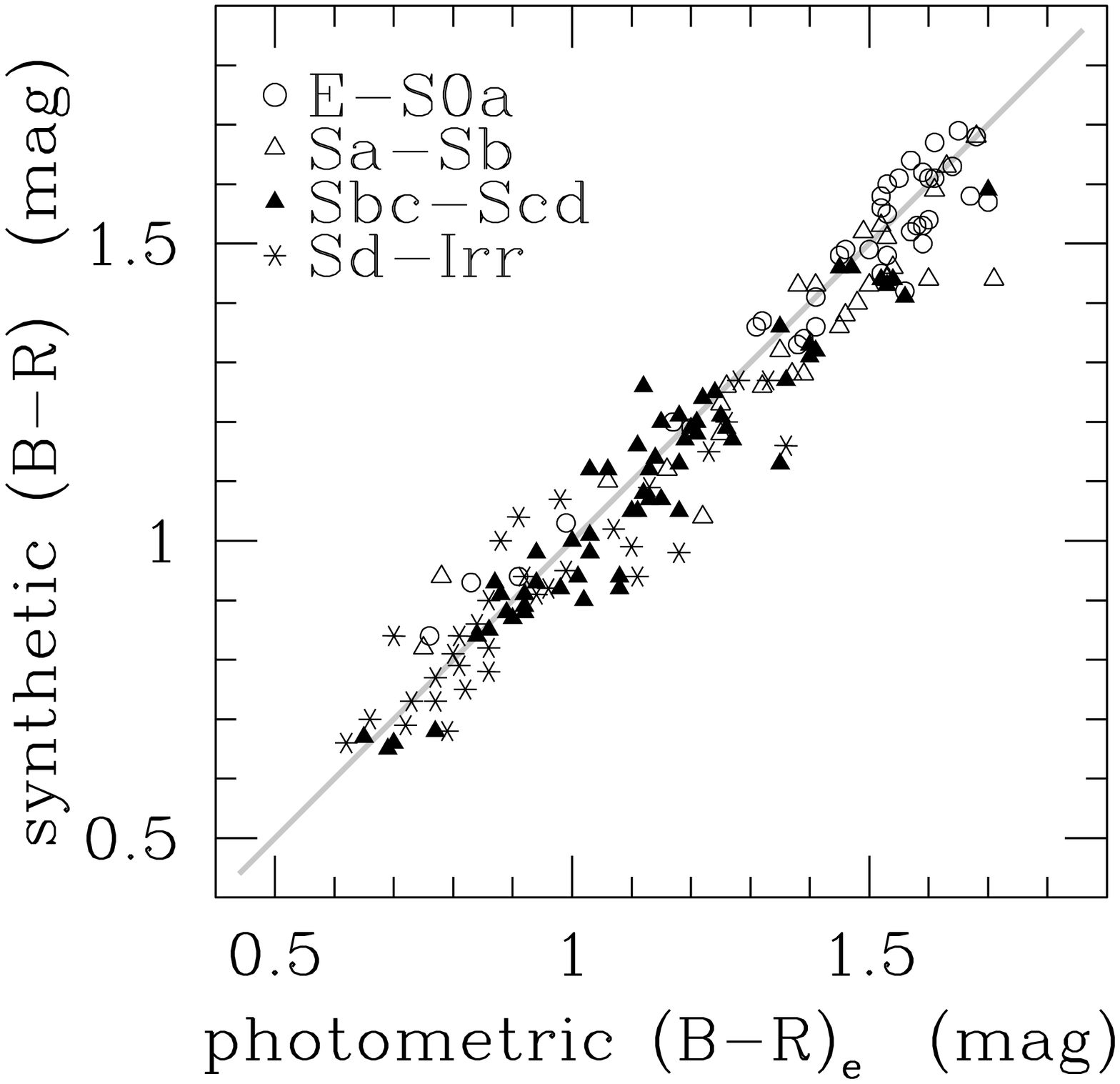,width=0.475\txw,clip=}}
}\par

\noindent\makebox[0.475\txw]{
\centerline{
\parbox[t]{0.475\txw}{\footnotesize {\sc Fig.~4 ---} A comparison of
photometric and synthetic spectrophotometric broadband \BR\ colors. 
Apart from a small residual color zeropoint offset, the synthetic colors
match the photometric ones to 0.065 mag (RMS) over the full range in
galaxy color.  The scatter is consistent with our error analysis (see
text). } }
}\vspace*{0.5cm}
%

\subsubsection*{\footnotesize Comparison with Kennicutt's spectra}

We compare the integrated spectra we obtained for 9 galaxies in
Kennicutt's sample (1992b) with his spectra.  In figures~5{\em a}
through {\em i} we plot both sets of spectra.  The spectra were
normalized to the average level in the 5200--6400\AA\ interval prior to
plotting.  We apply an arbitrary offset of 0.5 to our spectra for
clarity.  In the bottom panels we plot the ratio of the two spectra,
averaged in 250\AA\ bins.  As in figure~3, the binning ensures that
photon noise does not dominated the ratios. 

In the 5100--6700\AA\ region our spectra match Kennicutt's to better
than \tpm2\% over small ranges and to better than \tpm5\% overall (with
the exception of the region around \Ha\ in the case of NGC~6052 and
NGC~6764).  Bluewards of 5000\AA\ differences tend to become larger, up
to about \tpm30\%.  In six of the nine galaxies (five of the eight, if
we exclude NGC~5548) we find evidence that the slopes of the two sets of
spectra differ systematically in the blue with respect to the red half
of the spectra.  This difference does not correlate with either the date
of observation, airmass, aperture size or orientation on the sky. 
Kennicutt's spectra were obtained in two pieces, a blue (3650--5150\AA)
segment and a red (4950--7150\AA) segment.  \vfill

\noindent\leavevmode
\makebox[\txw]{
   \centerline{
      \epsfig{file=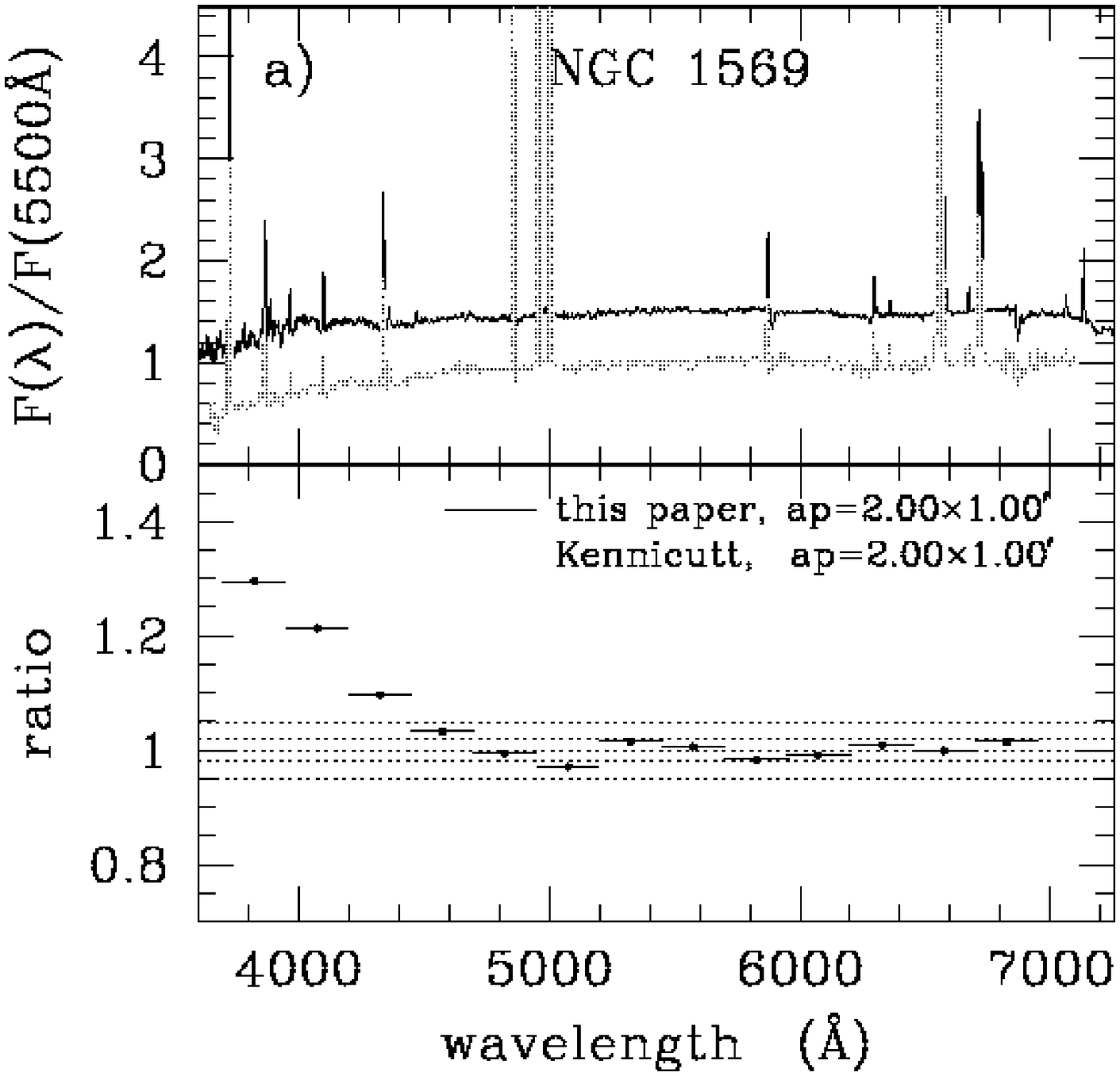,width=0.39\txw,clip=}\hspace*{0.05\txw}
      \epsfig{file=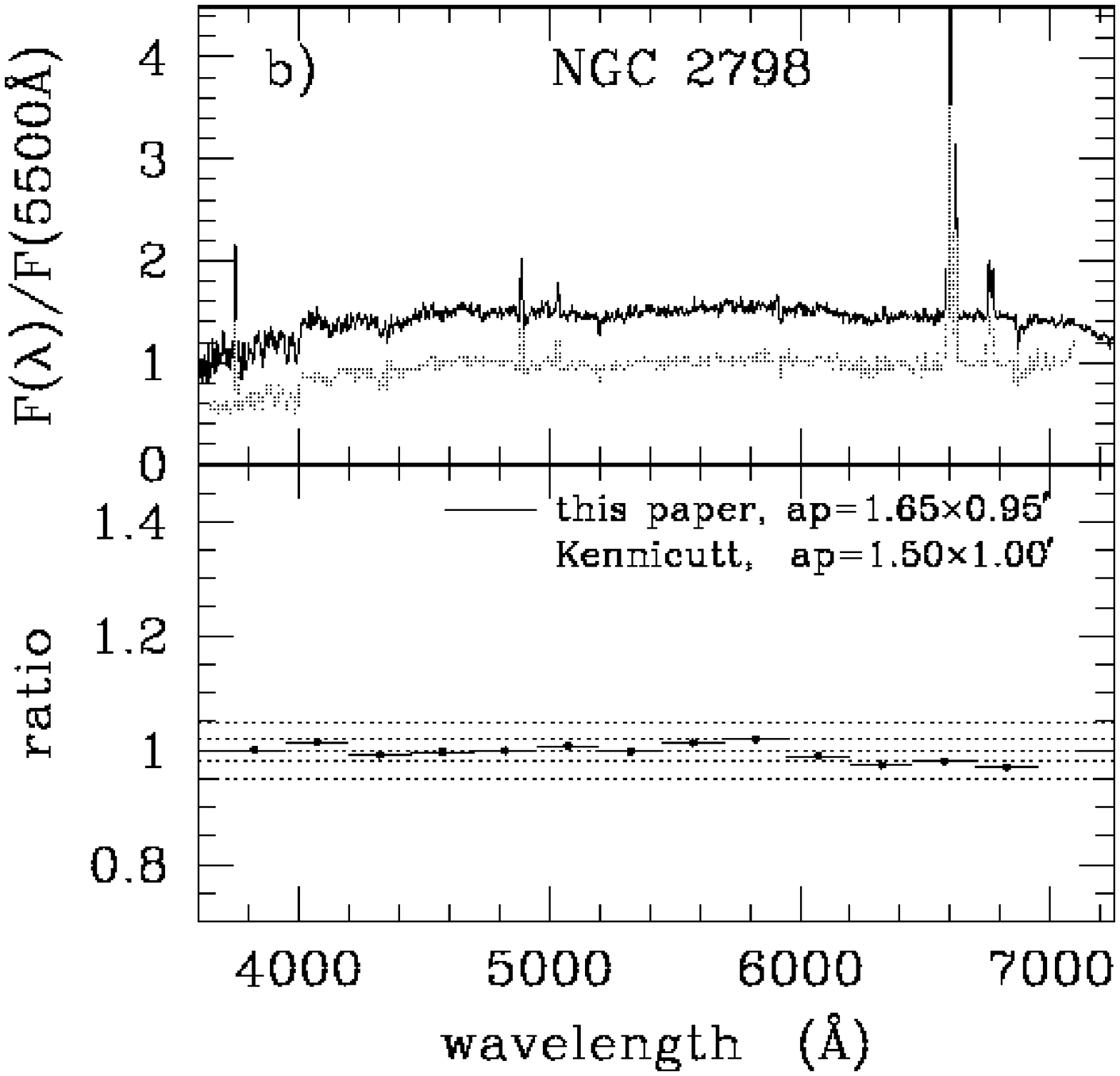,width=0.39\txw,clip=}
   }
}\par\vspace*{0.03\txw}\noindent\leavevmode
\makebox[\txw]{
   \centerline{
      \epsfig{file=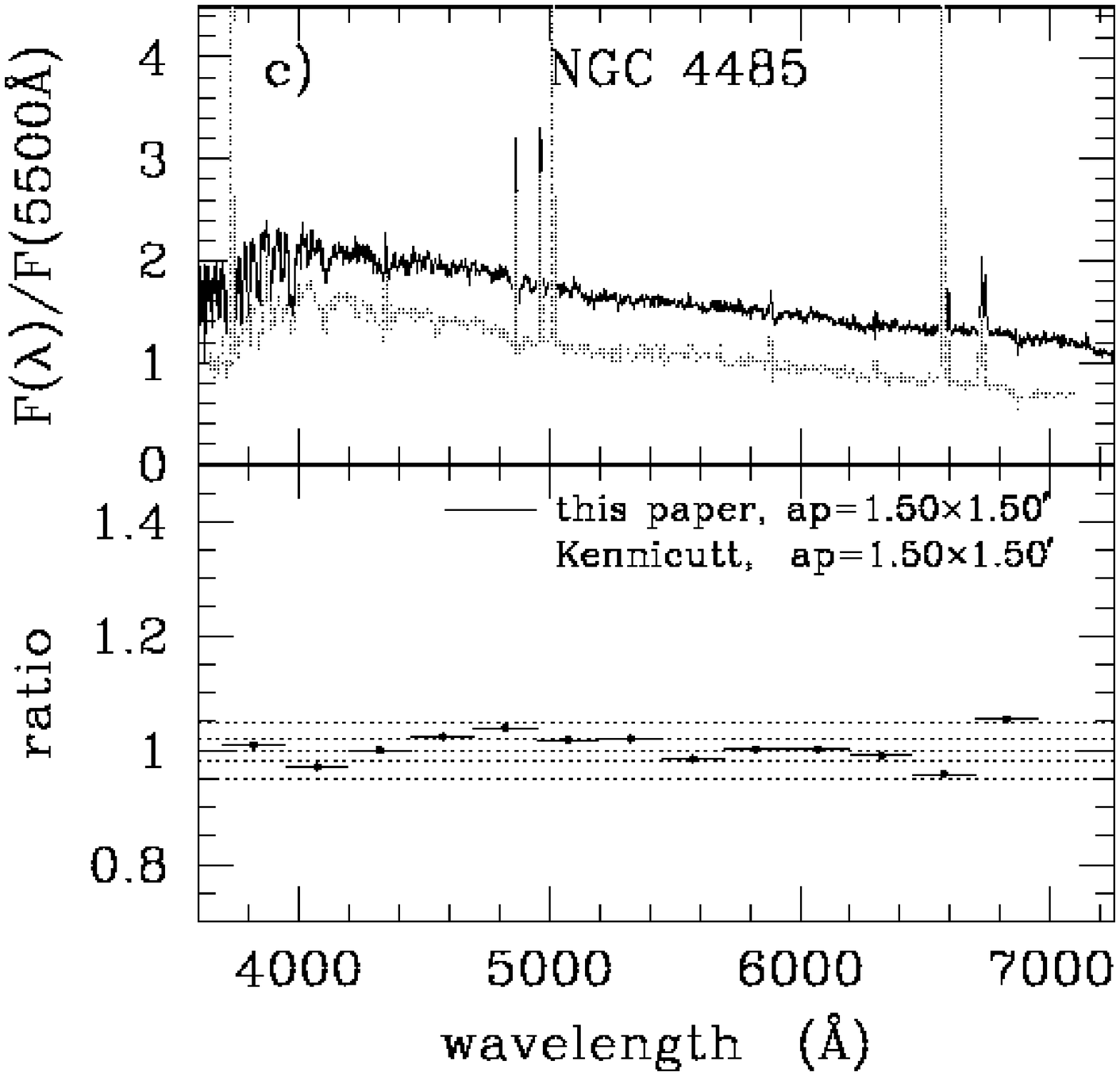,width=0.39\txw,clip=}\hspace*{0.05\txw}
      \epsfig{file=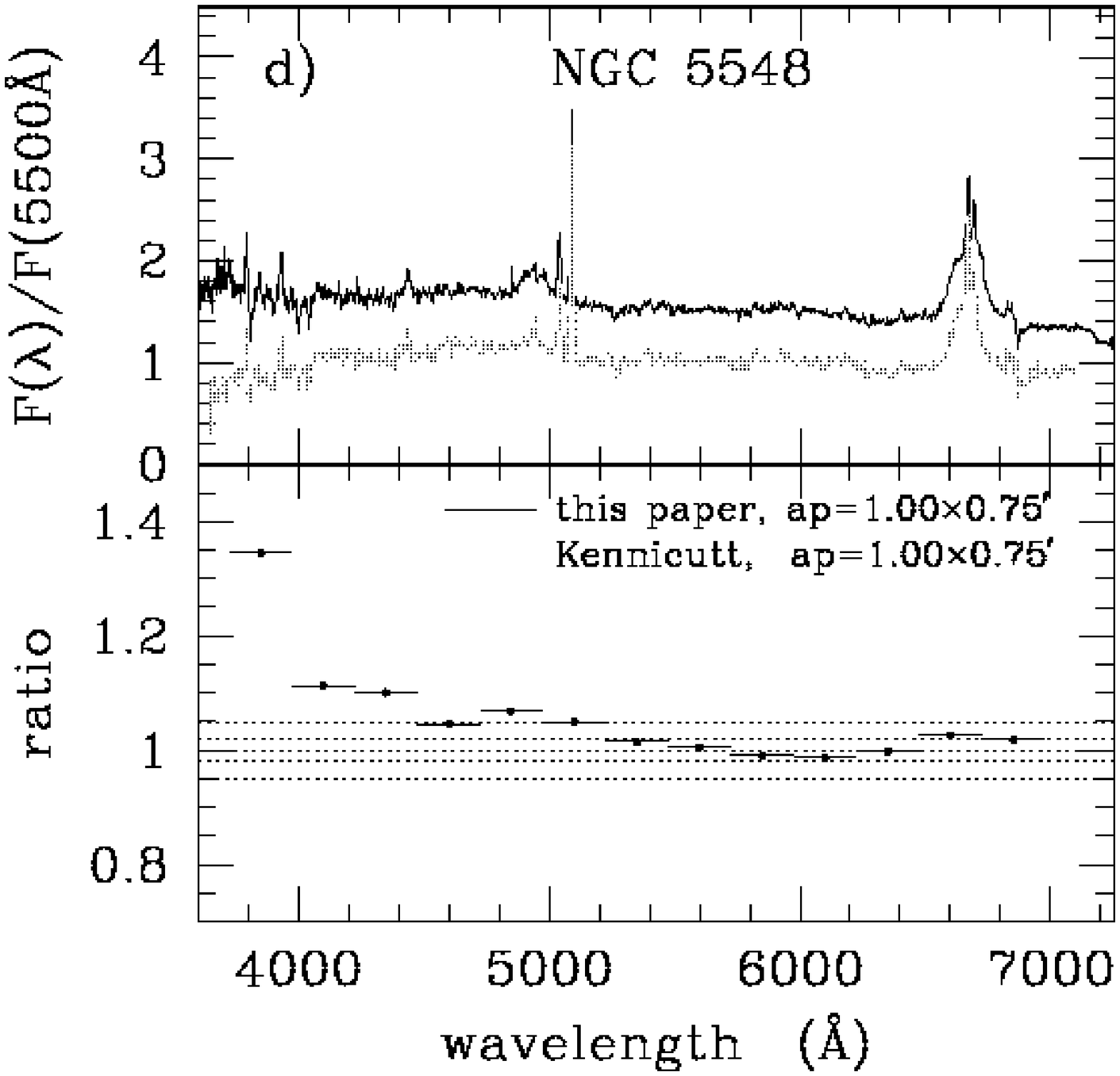,width=0.39\txw,clip=}
   }
}\par\vspace*{0.03\txw}\noindent\leavevmode
\makebox[\txw]{
   \centerline{
      \epsfig{file=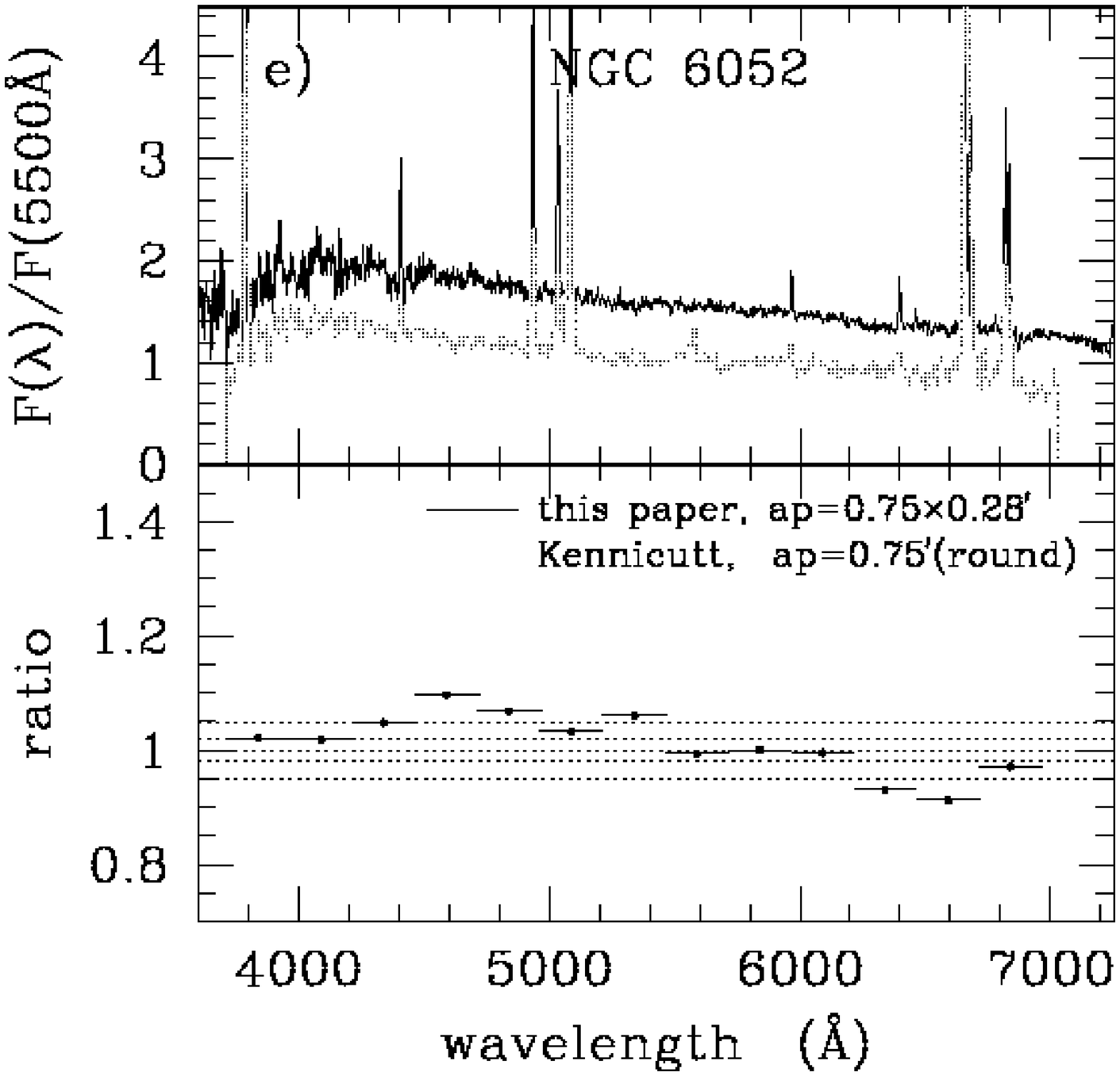,width=0.39\txw,clip=}\hspace*{0.05\txw}
      \epsfig{file=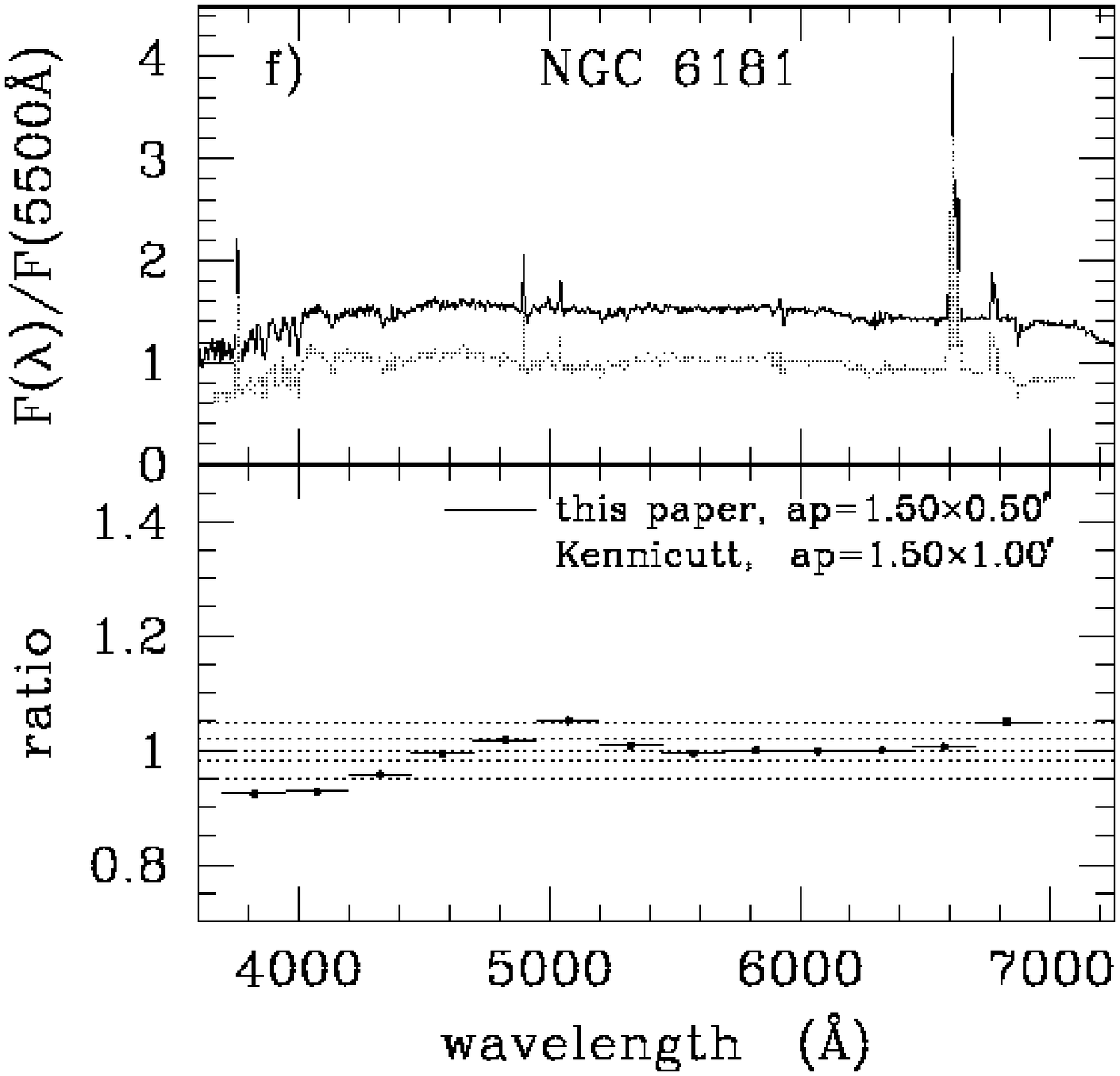,width=0.39\txw,clip=}
   }
}\par\vspace*{0.03\txw}\noindent\leavevmode
\makebox[\txw]{
\centerline{   
\parbox[t]{\txw}{\footnotesize {\sc Fig.~5 ---} A comparison of our
spectrophotometry with Kennicutt's (1992b) for the 9 galaxies from his
sample that we re-observed.  a) NGC~1569, b) NGC~2798, c) NGC~4485, d)
NGC~5548 (due to bright sky background, the errors in our
spectrophotometry in the blue part are larger), e) NGC~6052, and f)
NGC~6181.  Within the quoted spectrophotometric errors, \tsim\tpm6\% for
our study and \tpm10-15\% for Kennicutt's, the spectra agree, excepting
below 4500\AA\ for three galaxies. } }
}\vspace*{0.5cm}

\newpage

\noindent\leavevmode
\makebox[0.45\txw]{
   \centerline{
      \epsfig{file=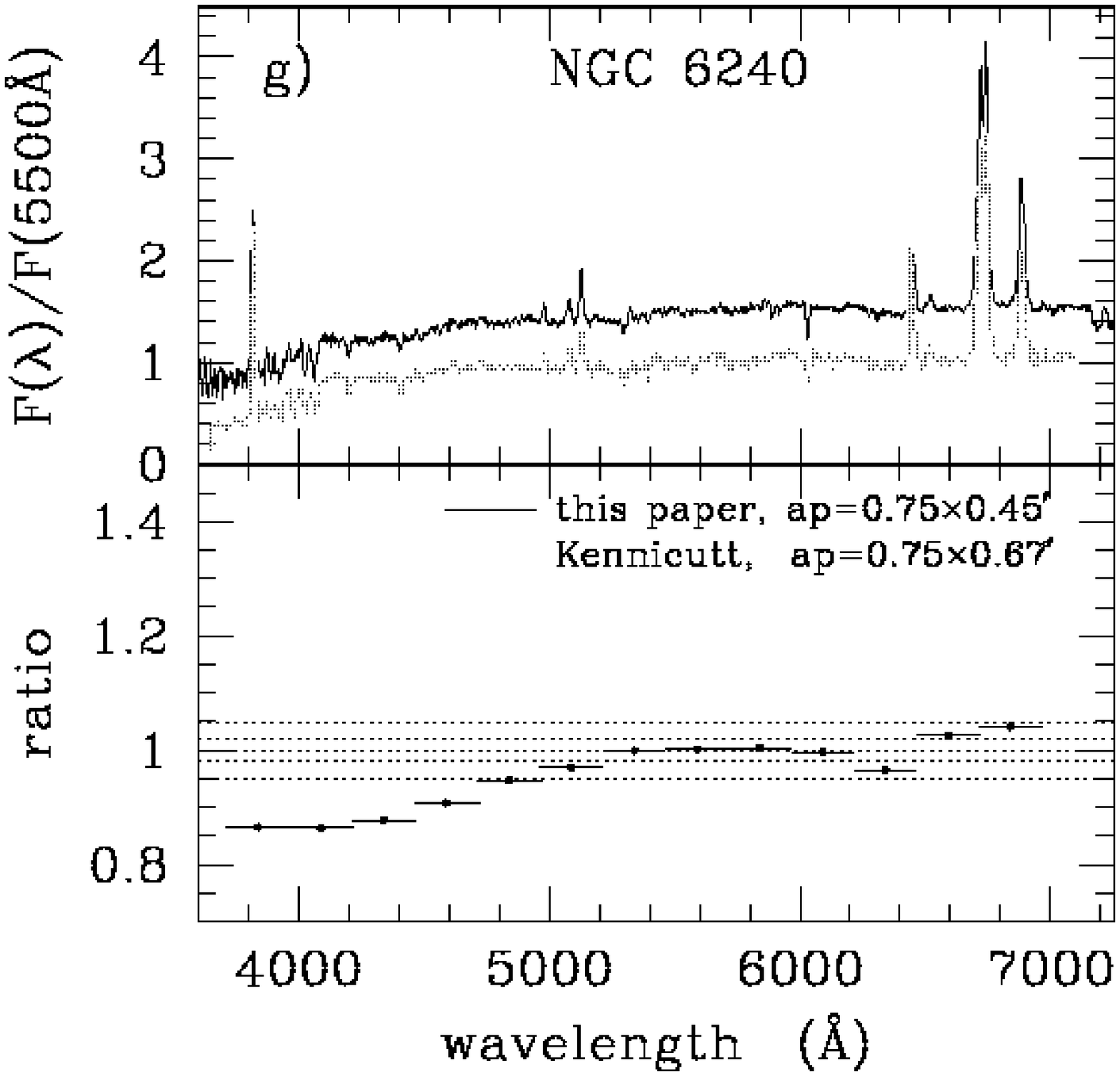,width=0.39\txw,clip=}
   }
}\par\vspace*{0.02\txw}\noindent\leavevmode
\makebox[0.45\txw]{
   \centerline{
      \epsfig{file=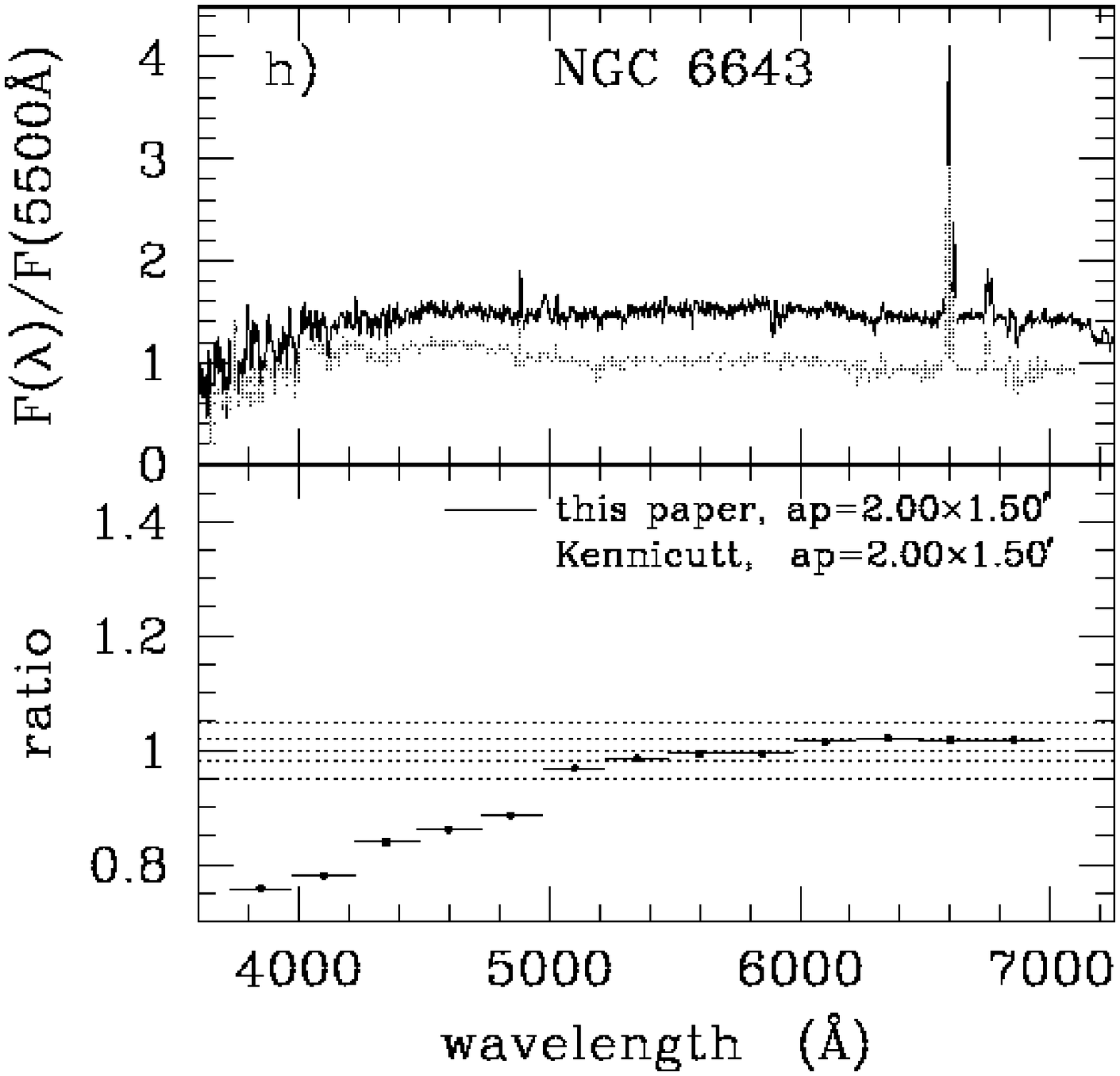,width=0.39\txw,clip=}
   }
}\par\vspace*{0.02\txw}\noindent\leavevmode
\makebox[0.45\txw]{
   \centerline{
      \epsfig{file=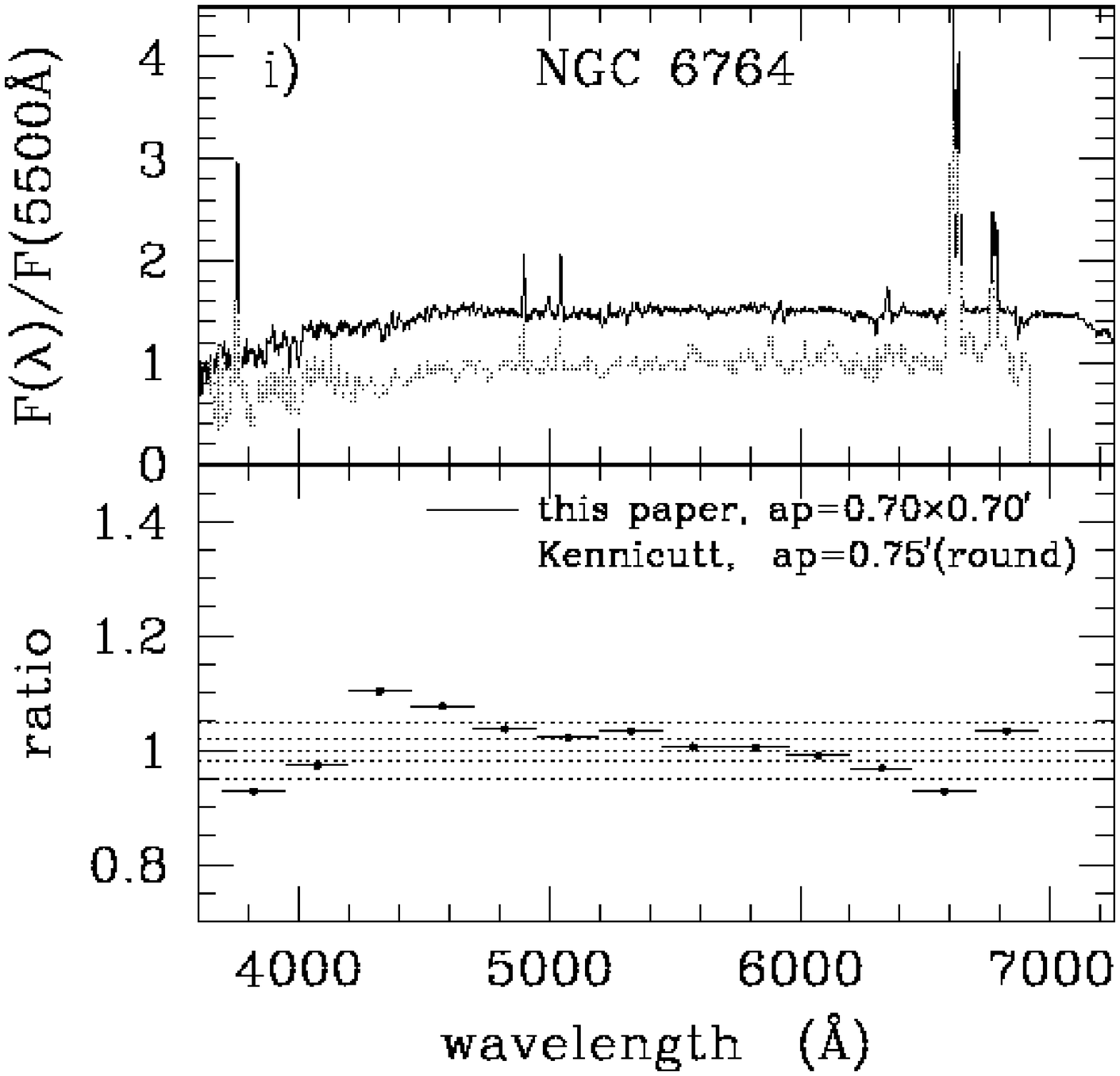,width=0.39\txw,clip=}
   }
}\par\vspace*{0.01\txw}\noindent\leavevmode
\makebox[0.475\txw]{
\centerline{ \parbox[t]{0.475\txw}{\footnotesize {\sc Fig.~5 (cont'd)
---} g) NGC~6240, h) NGC~6643, and i) NGC~6764.  } }
}\vspace*{0.5cm}
%

\noindent The two segments were grafted together using the mean
continuum intensities in the overlap region or

\vbox{
\begin{center}
\setlength{\tabcolsep}{2pt}
\centerline{\sc Table 3}
\centerline{\sc Comparison of emission line EWs.}
\vspace*{0.2cm}
\scriptsize
\begin{tabular}{lrrrrrr}
\hline
\hline\noalign{\vskip 0.1cm}
     & \multicolumn{3}{c}{EW(Kennicutt)} & \multicolumn{3}{c}{EW(this paper)} \\
galaxy & \OII & OIII & \HaNII & \OII & \OIII & \HaNII \\\noalign{\vskip 0.1cm}
\hline\noalign{\vskip 0.1cm}
NGC~1569 & 49 & 182 & 202~~~ & 62.6\tpm 3.3 & 186.8\tpm 7.3 & 206.0\tpm 6.2 \\
NGC~2798 & 17 &   3 &  53~~~ & 18.8\tpm 0.4 &   2.3\tpm 0.1 &  53.0\tpm 0.7 \\
NGC~4485 & 48 &  33 &  65~~~ & 32.7\tpm 4.0 &  27.7\tpm 0.5 &  50.6\tpm 4.2 \\
NGC~5548 &  6 &  26 & 113~~~ &  3.9\tpm 0.9 &  18.9\tpm 0.4 & 101.4\tpm 2.6 \\
NGC~6052 & 54 &  53 & 181~~~ & 60.5\tpm 4.8 &  42.7\tpm 0.6 & 161.3\tpm 5.2 \\
NGC~6181 & 15 & 3.5 &  47~~~ & 12.4\tpm 2.0 &   2.3\tpm 0.5 &  45.6\tpm 0.9 \\
NGC~6240 & 67 &  11 & 104~~~ & 44.0\tpm 16. &   7.1\tpm 0.6 &  85.4\tpm 2.4 \\
NGC~6643 & 11 & 1.5 &  35~~~ & 11.4\tpm 2.4 &   1.0\tpm 0.6 &  34.0\tpm 1.9 \\
NGC~6764 & 45 &   7 &  86~~~ & 50.3\tpm 4.3 &   5.1\tpm 1.2 &  88.4\tpm 2.3 \\
\noalign{\vskip 0.1cm}
\hline
\end{tabular}
\end{center}
}\vspace*{0.1cm}

\noindent in the \OIII\lam5007 emission line.  He notes that errors of
\tsim\tpm5\% in the flux scale between regions longward and shortward of
5100\AA\ are possible, which may account for some of the differences
between our spectra. 

We interactively measured the rest-frame EWs of \OII\lam3727,
\OIII\lam5007, and \HaNII\lam\lam6548,6584 to see whether the
differences in slope also turn up in the emission line indices.  The
resulting measurements, as well as those published in Kennicutt (1992a),
are collected in table~3.  The EWs agree well with one another,
suggesting that differing apertures are unlikely to account for the
spectral differences, because these might also be expected to affect the
relative contributions of line and continuum emission. 

Given the stringent limits that the good agreement of our photometric
and synthetic broadband colors place on errors in our spectrophotometry,
we suspect that at least part of the observed difference must be
attributed to Kennicutt's data.  We note, however, that the two sets of
spectra do generally agree within the spectrophotometric errors of the
two samples, \tsim\tpm6\% in our case, and \tpm10--15\% for Kennicutt's. 
Larger discrepancies are only seen in three of the nine spectra below
4500\AA.


\section{Spectrophotometric results}
\label{S-Results}

\subsection{Presentation of the data}
\label{S-Atlas}

In this section we briefly comment on our primary data products, the
nuclear and the integrated spectra, and present our spectrophotometric
measurements.  Notes on individual galaxies and interesting
spectroscopic features are collected in appendix~A. 

The atlas (figure~6) consists of pairs of rest-frame spectra for the 198
galaxies in our sample.  The spectra have been ordered according to
galaxy type, and within each morphological type according to their
$B$-filter luminosity.  The morphological types and \MB\ are indicated
in the atlas, as are their radial velocities.  The spectra have all been
normalized to the flux in a 50\AA\ wavelength interval centered on
5500\AA.  In~ each~ panel,~ we~ plot~ the~ spectrum~ twice:~ at~ the 
proper vertical scaling, and scaled up by a factor of

\renewcommand{\textfraction}{0.000}
\noindent\leavevmode\framebox[\txw]{
   \centerline{\parbox[c]{\txw}{
       \centerline{\rule[-0.37\txw]{0pt}{0.80\txw}
       \Large\it Spectrophotometric atlas --- ``fig6\_01.gif''}}
   }
}\par\noindent\leavevmode
\makebox[\txw]{
\centerline{   
\parbox[t]{\txw}{\footnotesize {\sc Fig.~6 ---}
The atlas of rest-frame nuclear (top panel) and integrated (bottom
panel) spectra.  The galaxies have been ordered according to their
morphological type and within each type according to their absolute $B$
magnitude.  The spectra span the range 3600--7200\AA\ and the resolution
(FWHM) is \tsim 6\AA.  The fluxes are normalized to a 50\AA\ region
centered on 5500\AA.  Each spectrum is plotted twice: at the proper
vertical scaling, and scaled up by a factor of 2--8 which was chosen to
sufficiently separate the two spectra (indicated in the upper right
corner of each panel).  The scaled up spectra allow closer examination
of the continuum and absorption lines, as well as the fainter emission
lines.  The top labels list the galaxy ID and common name, Hubble type,
total absolute $B$ magnitude and recessional velocity in the observers
frame (these velocities were used to de-redshift the spectra).  The name
of galaxy A01047$+$1625 is placed in parentheses, as it is not part of
the statistical sample.\\
Compact ellipticals (cE).  } }
}\vspace*{0.5cm}

%

\noindent 2--8 (chosen to sufficiently separate the two spectra) to
allow a better examination of the continua and fainter spectral
features.  The spectra span the range 3600--7200\AA\ and have a
resolution (FWHM) of \tsim 6\AA. 

The spectrophotometric measurements presented in this paper are
collected in tables~4 through 6.  In table~4 we present the emission
line equivalent widths measured in both integrated and nuclear spectra. 
To conserve space and improve readability, we use positive numbers to
indicate emission.  In column (1) we give the galaxy identification
number in this survey.  For the common names we refer to table~1.  The
emission line EWs of \OII\lam3727, \Hd, \Hb, \OIII\lam4959, \vfill

\null\vspace*{0.98\txw} \noindent
\OIII\lam5007, \NII\lam6548, \Ha, \NII\lam6584, \SII\lam6718 and
\SII\lam6731 are listed in columns (2) through (11) for the integrated
spectra, in columns (12) through (21) for the nuclear ones. 

Line flux ratios can be measured accurately even in (distant) galaxies
where the continuum is too faint to measure EWs reliably.  To facilitate
comparison with the present sample, in table~5 we present fluxes for the
emission lines listed in table~4.  We give the emission line fluxes
relative to the \Hb\ flux ($f(\Hb) \equiv 1$).  If \Hb\ is not detected
or too faint to measure reliably (EW$(\Hb) \gtrsim -0.5$\AA), the
emission line fluxes are given relative to $f(\Ha) \equiv 1$. 


\noindent\leavevmode\framebox[\txw]{
   \centerline{\parbox[c]{\txw}{
       \centerline{\rule[-0.60\txw]{0pt}{1.25\txw}
       \parbox[c]{0.7\txw}{\Large\it 
           Spectrophotometric atlas (cont'd) --- pages 16 through 49\\
           (``fig6\_02.gif'' through ``fig6\_35.gif'')\\$\quad$\\
           A 600 dpi postscript version of this preprint, including all
           figures and the spectrophotometric atlas of galaxies, can be
           retrieved from URL\  http://www.astro.rug.nl/\tsim nfgs/}}}
   }
}\par\noindent\leavevmode\makebox[\txw]{\centerline{
\parbox[t]{\txw}{\footnotesize {\sc Fig.~6 (Cont'd) ---} Ellipticals (page 16) through Irregulars (page 49).
} } }\newpage

\setcounter{page}{50}

\setlength{\tabcolsep}{2pt}

\normalsize
\clearpage

In table~6 we list the synthetic colors from the integrated spectra. 
After the identifications, column (2) lists the ``41-50'' continuum
color index defined in Kennicutt (1992a).  In columns (3) through (5) we
present the integrated \BV, \VR\ and \BR\ colors.  Column (6) contains
the effective \BRe\ colors as measured in our photometry (Paper~I). 

\subsection{Spectrophotometric properties of the sample}

In figure~7 we plot the logarithm of the integrated emission line
strengths of \Ha\ and \OII\lam3727 versus morphological type, absolute
$B$ magnitude, and effective $(B-R)$ color.  Upper limits are shown for
galaxies without significant emission (EW(\Ha)$>$$-1$\AA).  The
distribution of \Ha\ EW as a function of galaxy morphological type
(figure~7{\em a}) shows the expected trend of larger EWs towards later
morphological types.  However, a number of early type galaxies show
significant emission as well.  These galaxies are a magnitude or more
fainter than the characteristic absolute $B$ magnitude of the local
galaxy luminosity function ($M_*\sim -19.2$, de Lapparent \etal 1989;
$M_*\sim -18.8$, Marzke \etal 1994).  A similar behavior is seen for
\OII\ EW in figure~7{\em b}.  The scatter around the mean trend is
virtually identical for \OII\lam3727 and for \Ha\ (MAD\footnote{MAD =
Median Absolute Deviation = $\frac{1}{N} \sum_1^N |x_i-\langle
x\rangle|$.  This statistic is similar to the RMS, but is less sensitive
to outlying data values, $x_i$.  It also uses the median, $\langle
x\rangle$, rather than the mean.}=0.188 and 0.189, respectively). 

In figure~7{\em c} we plot \Ha\ EW versus absolute $B$ magnitude.  Lower
luminosity systems tend to have larger EWs, but the range in EW at a
given luminosity is large.  The galaxies responsible for this tendency
are intermediate type spirals and earlier type systems.  For \OII\ we
observe a more pronounced trend towards larger EWs at lower luminosities
(figure~7{\em d}).  Although galaxies with faint emission lines have a
large spread in luminosity, galaxies with the strongest \OII\ lines are
of low luminosity. 

Given the correlations of both \Ha\ EW (figure~7{\em a}) and color with
morphological type (Paper~I), one might expect a fairly good relation
between the color of a galaxy and the \Ha\ emission line strength. 
Indeed, in figure~7{\em e} most galaxies appear to follow a relation of
bluing \BRe\ colors with increasing \Ha\ EW.  Again a similar
correlation exists for \OII\ with \BRe\ color, but with a somewhat
larger scatter (figure~7{\em f}).  The trend found in figure~7, of bluer
colors and stronger emission lines at lower luminosities, can also be
observed directly in the spectrophotometric atlas, if one compares
spectra of galaxies of the same morphological type but different
absolute magnitudes.  It is particularly apparent for the early type
galaxies and early and intermediate type spiral galaxies. 

Kennicutt (1992a) found that galaxies with emission spectra dominated by
photoionization (normal \HII-like star formation) showed a relation
between the EW of \OII\ and \Ha+\NII\ with a slope of 0.4, albeit with a
rather large dispersion.  For an average \NII/\Ha\ \tsimeq\ 0.5 in his
sample this implies EW(\OII) \tsimeq\ 0.6 \ttimes\ EW(\Ha).  In
figure~8{\em a} and {\em b} we plot for our sample \OII\ versus \Ha\ and
\HaNII\ EW, respectively.  There is a significant scatter of \tsim 0.20
dex (\tsim 0.22 dex in figure~8{\em b}) in logarithmic units, that is
almost constant over the range in EW sampled here.  More galaxies
scatter towards higher than towards lower \OII\ EWs with respect to
Kennicutt's relation.  Although most galaxies brighter than \MB \tsim
$-$19 follow his relation reasonably well, we find that fainter galaxies
follow a much steeper relation.  In the lowest luminosity galaxies the
\OII\ EW approximates that of \Ha.  \NII, on the other hand, tends to
scatter towards lower equivalent widths, as shown in figure~8{\em c}. 
Again, there is a dependence on luminosity: the lowest luminosity
systems have the smallest \NII/\Ha\ ratios.  In figure~8{\em d} we plot
\NII\ versus \OII\ EW.  The lowest \NII/\OII\ ratios correspond to the
lowest luminosities.  The opposite sign of the dependences of \NII\ and
\OII\ EW on luminosity causes the slight increase in scatter in
figure~8{\em b} with respect to figure~8{\em a}.  The stronger \OII\ and
weaker \NII\ (with respect to \Ha) in the lower luminosity galaxies
reflects the relation between galaxy luminosity and metallicity (\cf
Zaritsky \etal 1994; Kobulnicky \& Zaritsky 1999). 

Figures~9 support the conclusion that the galaxies that deviate most
from Kennicutt's linear \OII--\Ha\ relation are the lowest luminosity
systems.  In figure~9{\em a} we plot the logarithm of the ratio of \OII\
and \Ha\ EW versus absolute magnitude for the galaxies with sufficiently
strong emission (EW(\Ha)$<$$-10$\AA) that their \OII/\Ha\ ratios could
be measured reliably.  \OII/\Ha\ ratios increase at low luminosities
although the scatter on this relation is still sizable (0.17 dex).  In
figure~9{\em b} we see that the \NII/\Ha\ ratio for high luminosity
galaxies (\MB$<$$-19$) agrees with Kennicutt's value of 0.5, with small
scatter.  The \NII/\Ha\ ratios tend to decrease in low luminosity
systems, although the scatter is very large, ranging from 0.03 to nearly
0.5. 

In figures~9{\em c} and {\em d} we plot the logarithms of the \OII/\Ha\
and \NII/\Ha\ ratios versus effective \BR\ color.  Blue galaxies tend to
show somewhat larger \OII/\Ha\ ratios than red galaxies with a scatter
of \tsim 0.18 dex about the mean relation.  Galaxies bluer than \BRe\ =
1 show a very steep dependence of \NII/\Ha\ on color that is almost
absent for redder galaxies. \newpage

\noindent\vspace*{-3mm}\leavevmode
\makebox[\txw]{
   \centerline{
      \epsfig{file=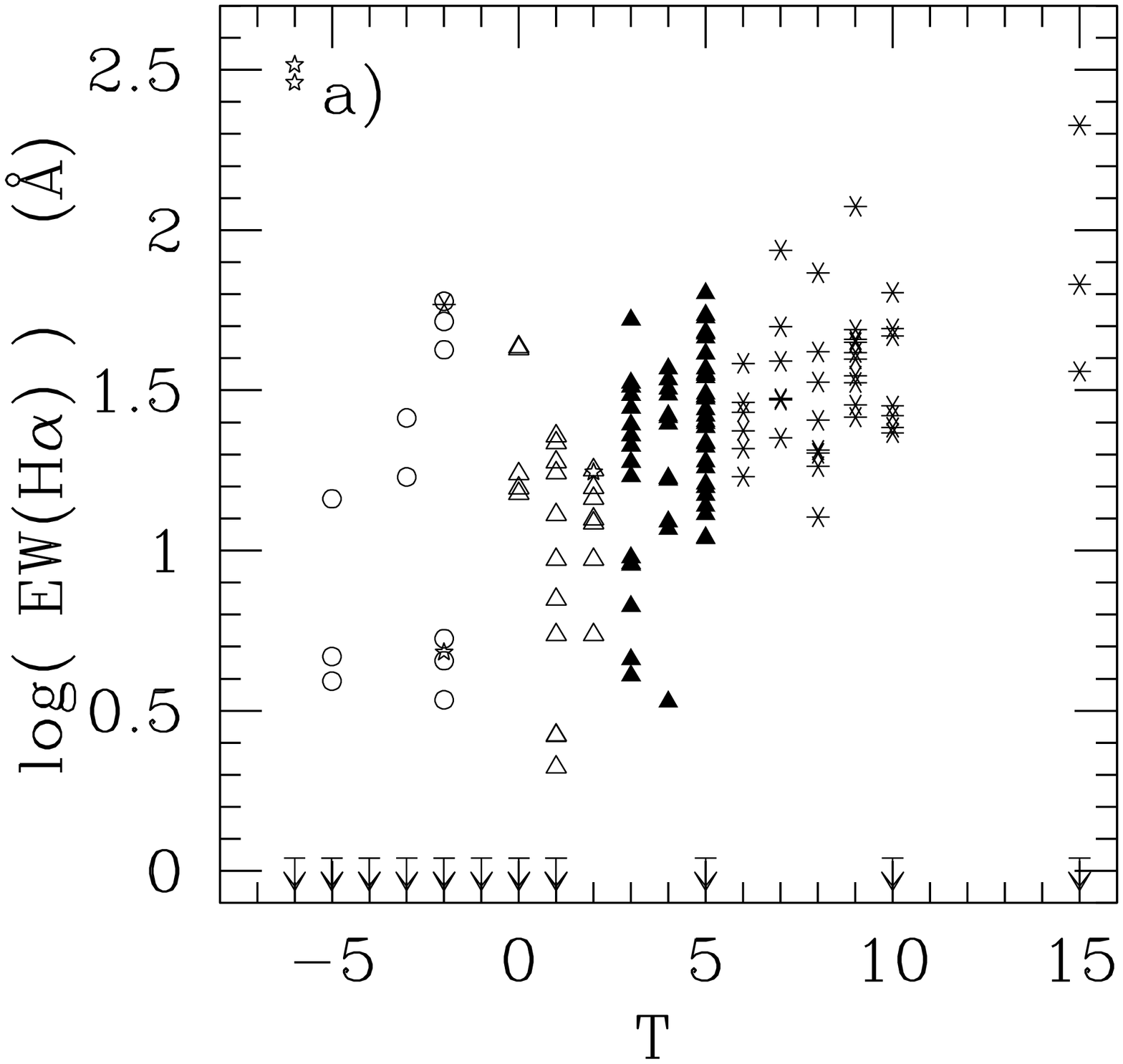,width=0.39\txw,clip=}\hfill
      \epsfig{file=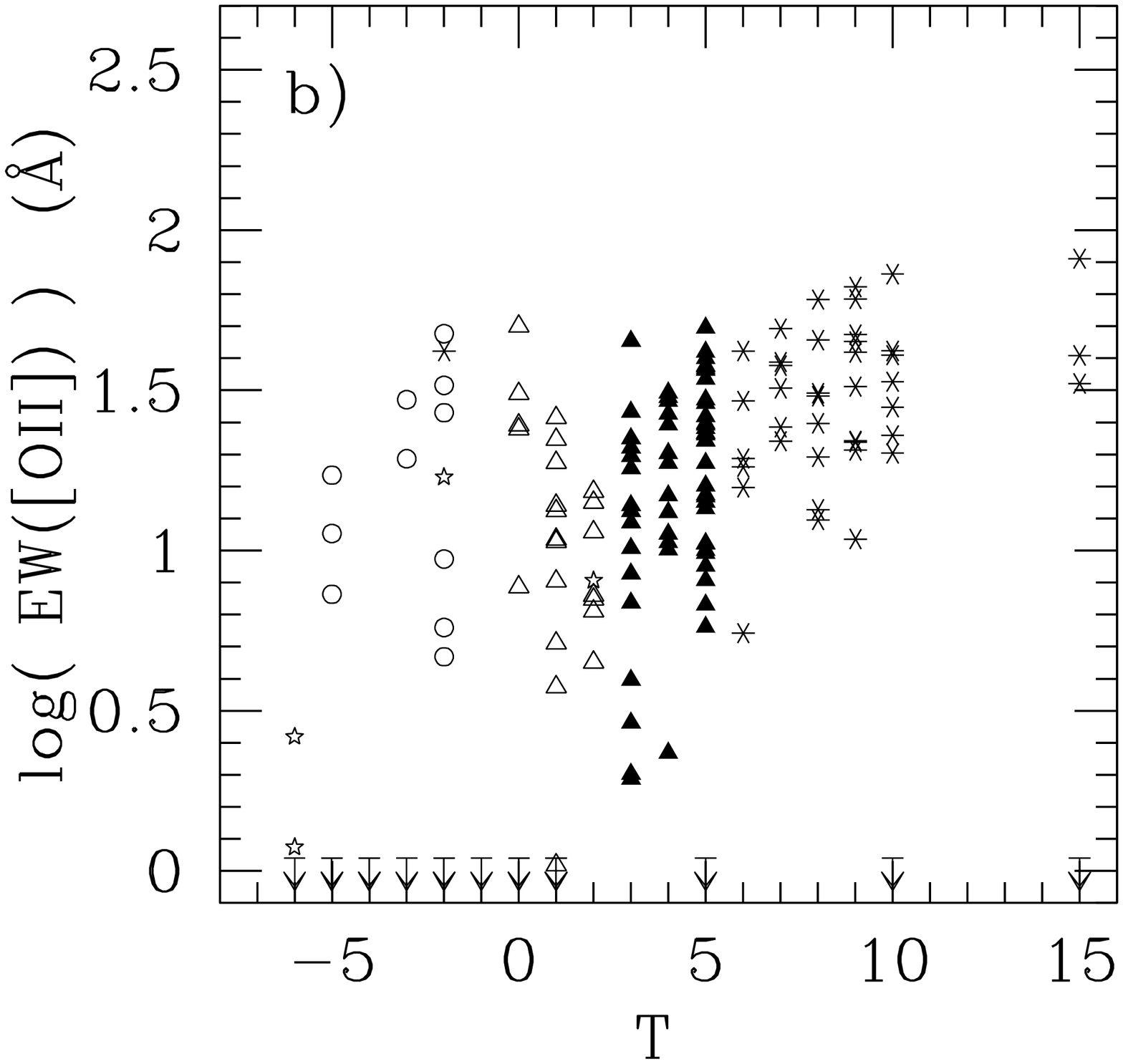,width=0.39\txw,clip=}
   }
}\par\noindent\vspace*{-3mm}\leavevmode
\makebox[\txw]{
   \centerline{
      \epsfig{file=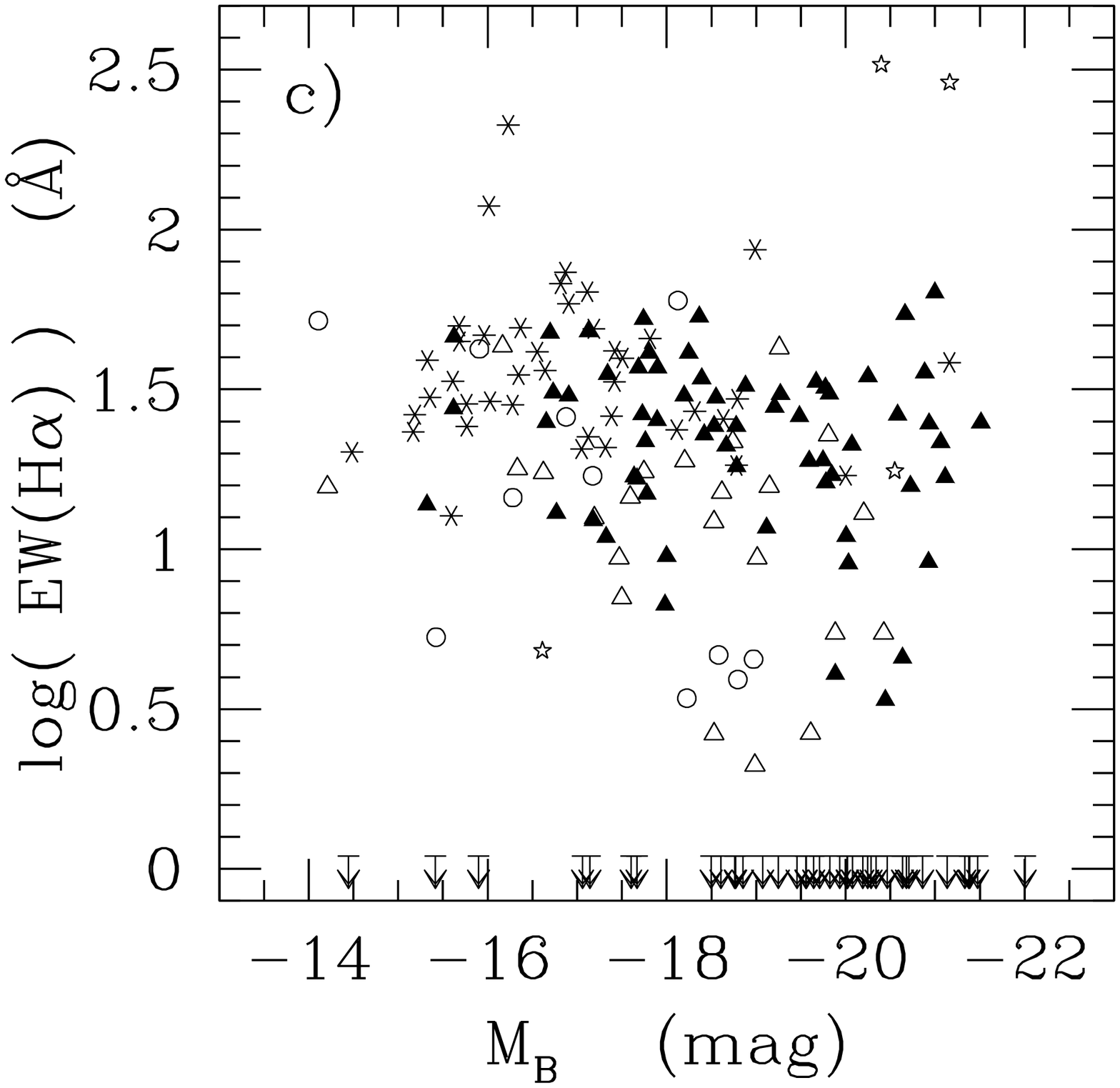,width=0.39\txw,clip=}\hfill
      \epsfig{file=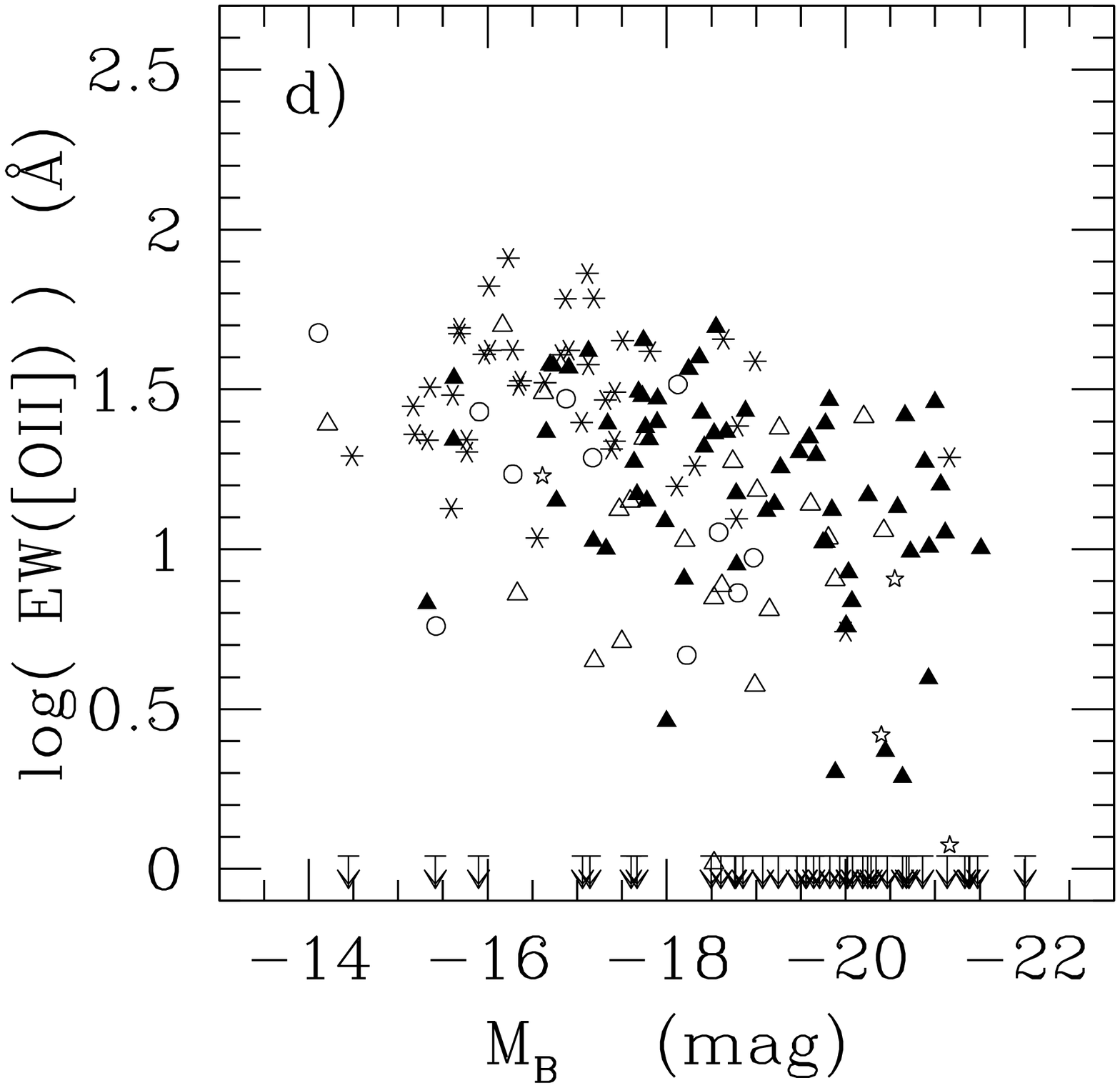,width=0.39\txw,clip=}
   }
}\par\noindent\vspace*{-3mm}\leavevmode
\makebox[\txw]{
   \centerline{
      \epsfig{file=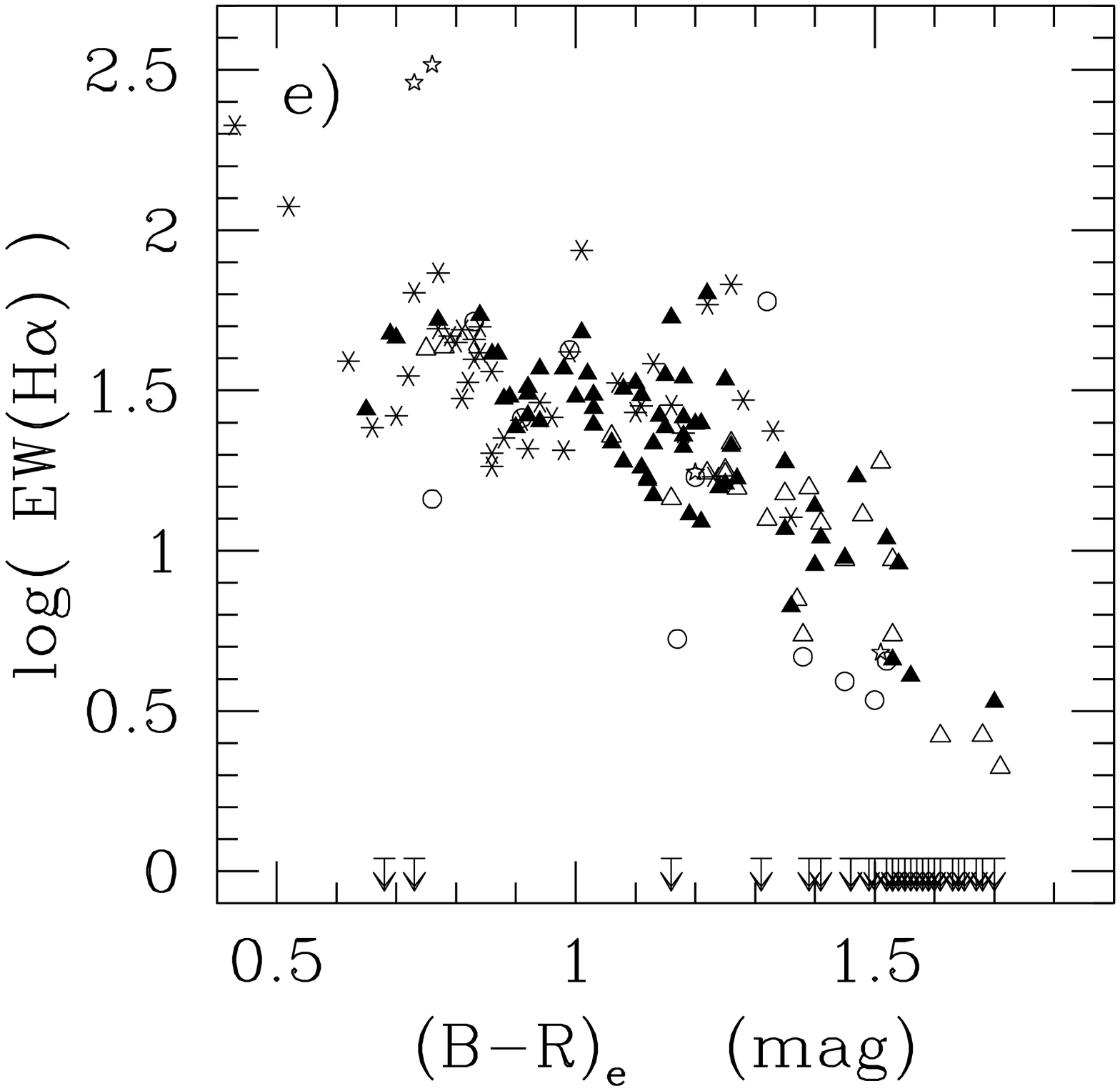,width=0.39\txw,clip=}\hfill
      \epsfig{file=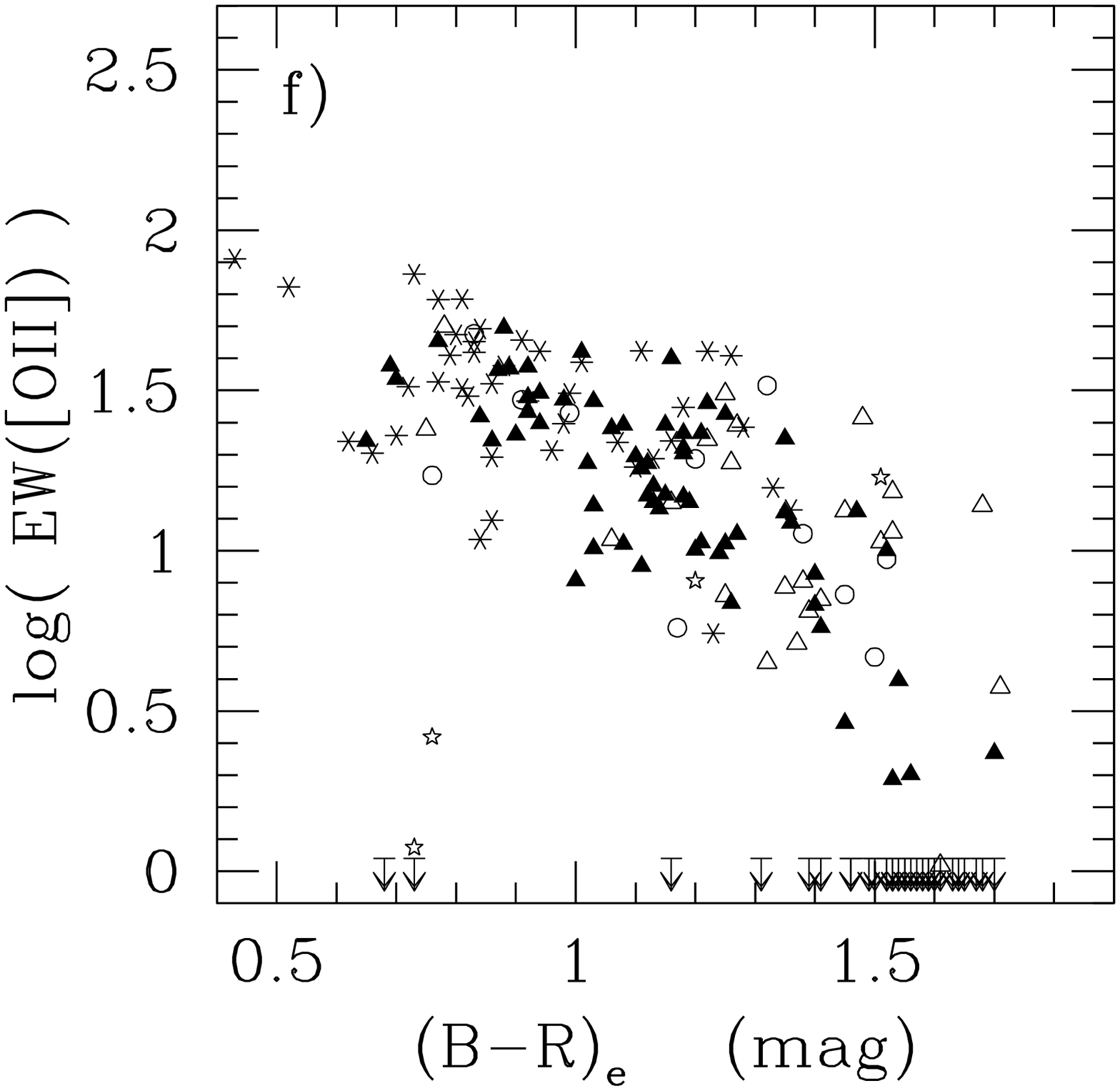,width=0.39\txw,clip=}
   }
}\par\noindent\vspace*{5mm}\leavevmode
\makebox[\txw]{
\centerline{   
\parbox[t]{\txw}{\footnotesize {\sc Fig.~7 ---} The logarithm of
integrated \Ha\ and \OII\ emission line EWs plotted versus morphological
type, absolute $B$ filter magnitude, and effective \BR\ color.  Galaxies
without significant emission (EW(\Ha)$>$$-1$\AA) are indicated by upper
limits (arrows).  The plotting symbols are coded according to
morphological type as in figure~4, with the exception that here galaxies
with AGN are included (open star symbols).  a) \Ha\ EW versus numeric
morphological type.  Emission line EWs increase towards later
morphological types.  b) As a) for \OII.  c) \Ha\ EW versus absolute $B$
magnitude.  A tendency towards larger EWs in lower luminosity systems is
seen.  d) As c) for \OII.  The systems with strong \OII\ lines are
generally at lower luminosity.  e) \Ha\ EW versus effective \BR\ color. 
As expected, bluer galaxies tend to have larger \Ha\ EWs.  f) As e) for
\OII.} } }\newpage


\noindent\leavevmode
\makebox[\txw]{
   \centerline{
      \epsfig{file=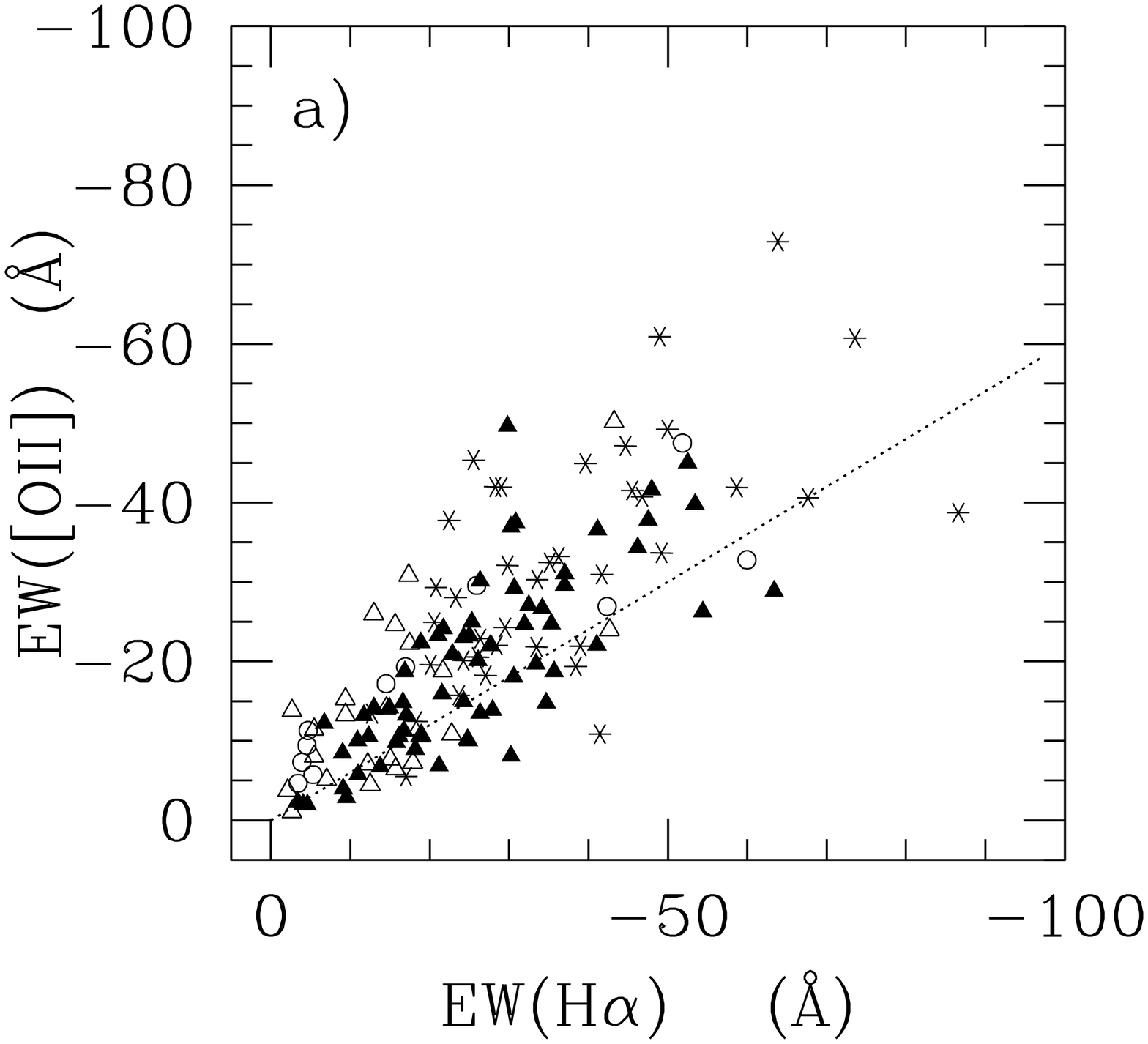,width=0.40\txw,clip=}\hfill
      \epsfig{file=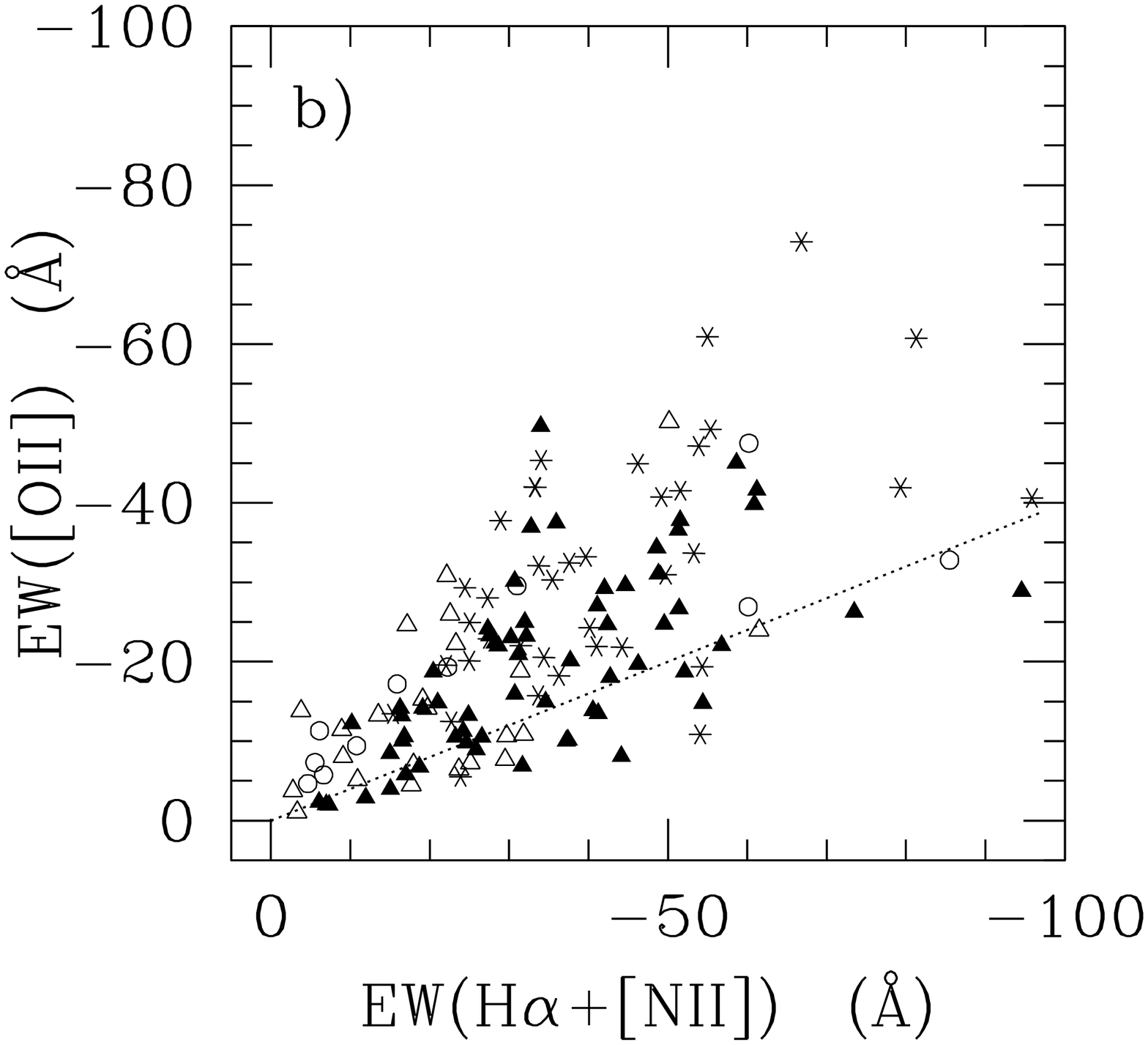,width=0.40\txw,clip=}
   }
}\par\noindent\leavevmode
\makebox[\txw]{
   \centerline{
      \epsfig{file=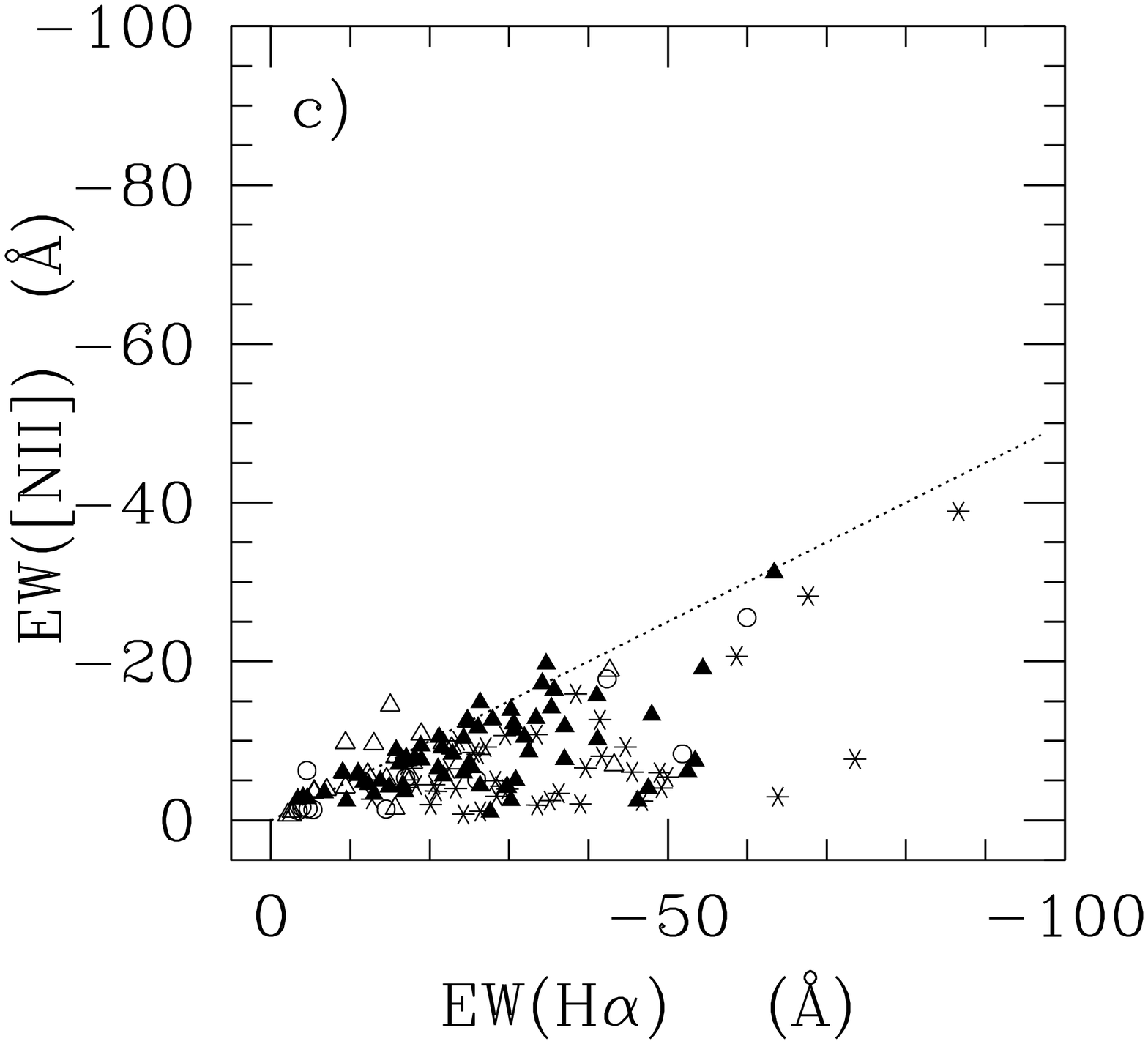,width=0.40\txw,clip=}\hfill
      \epsfig{file=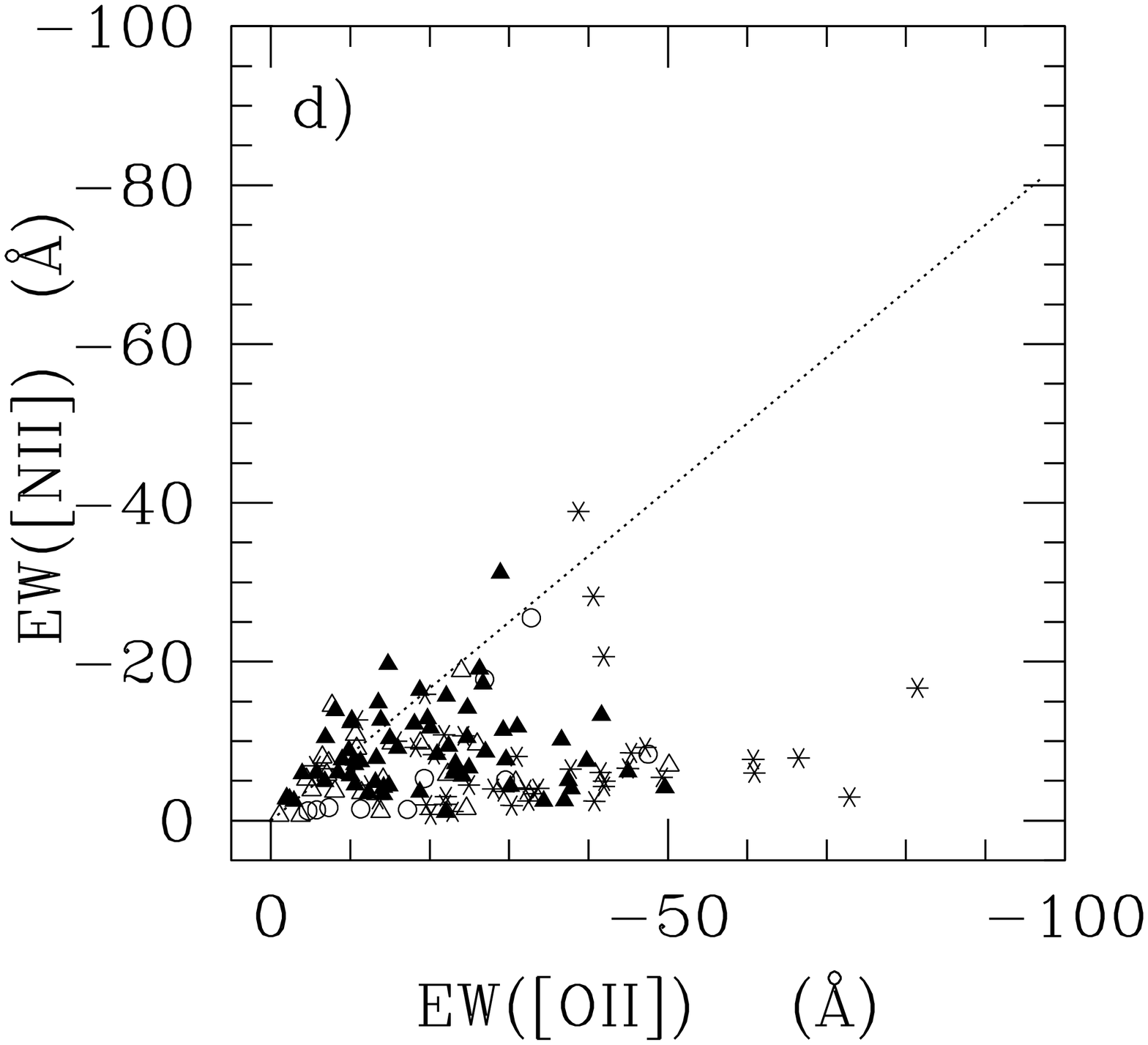,width=0.40\txw,clip=}
   }
}\par\noindent\leavevmode
\makebox[\txw]{
\centerline{
\parbox[t]{\txw}{\footnotesize {\sc Fig.~8 ---} A study of the
correlations between \OII, \Ha\, and \NII6548$+$6584 equivalent widths. 
The relations derived from Kennicutt (1992a), are overlaid (dotted
lines) for reference.  Plotting symbols are coded according to
morphological type as in figure~4.  a) \OII\ versus \Ha.  Most high
luminosity galaxies follow Kennicutts relation reasonably well, but most
low luminosity galaxies are located at higher \OII\ EWs.  The magnitude
of the scatter, \tsim 0.2 dex, is almost constant in logarithmic units
as a function of \Ha\ EW.  b) \OII\ versus \HaNII.  The scatter is
larger than in panel a).  c) \NII\ versus \Ha.  Most of the galaxies
that lie below line \NII=0.5\Ha\ have absolute magnitudes fainter than
\MB=$-19$.  d) \NII\ versus \OII.  As in panel c), the galaxies below
line \NII=0.833\OII\ tend to be intrinsically faint.  } }
}\vspace*{0.5cm}

%

In figure~10 we plot the distributions of \Ha\ and \OII\ EW as a
function of \MB\ for both the integrated and the nuclear spectra.  The
EWs are averaged in two magnitude wide bins, and the average and
quartile range (25\% to 75\%) are shown.  We distinguish between
galaxies with significant emission (EW(\Ha)$<$$-1$\AA) and the general
sample, excluding AGN in both cases.  In the case of the integrated
spectra, we see that the \OII\ EWs depend more strongly on \MB\ than do
the \Ha\ EWs.  For the nuclear spectra, the trend for \Ha\ EWs in
figure~10{\em c} is due to only part of the galaxy population, as
evidenced by the increase in the width of the distribution towards lower
luminosities.  At the lowest luminosities, the average

\null\vspace*{0.95\txw}\noindent \OII\ EW equals the average \Ha\ EW in
the integrated spectra; in the nuclear spectra the \OII\ EW remains
smaller than that of \Ha. 

From figures~9 and 10 it follows that if \OII\ is to be used as an
estimator of the current star formation rate, the accuracy may be
improved by including an estimate of the luminosity into the analysis. 
We will return to this issue, and complications in its interpretation
and use due to interstellar reddening in a future paper (Jansen \etal
1999b; in prep.). 

Figure~11{\em a} compares the \Ha\ equivalent widths in the integrated
and nuclear spectra of 138 galaxies meeting the conditions: (1) both
nuclear and integrated spectra are available, (2) emission lines are

\noindent\leavevmode
\makebox[\txw]{
   \centerline{
      \epsfig{file=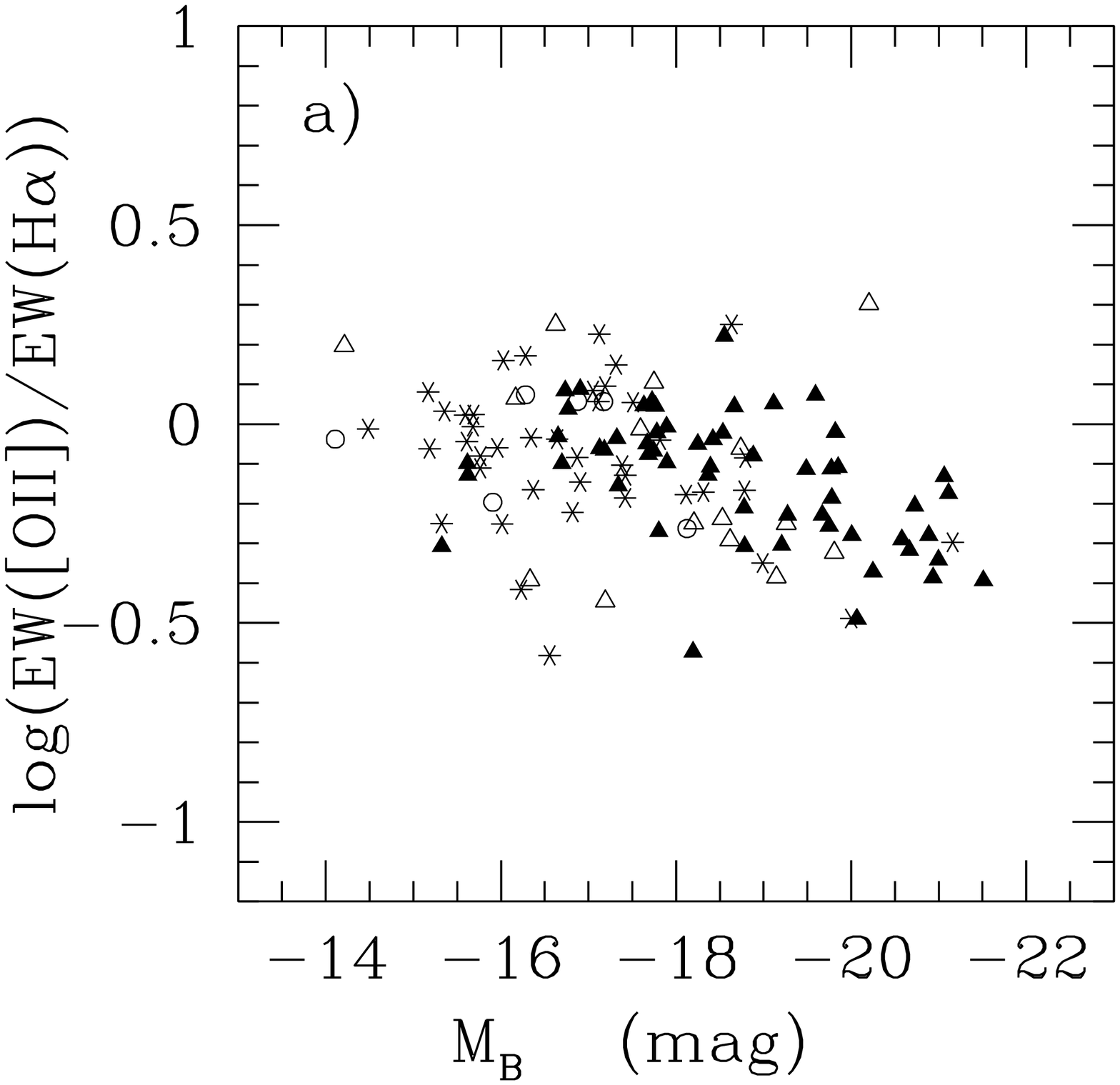,width=0.40\txw,clip=}\hfill
      \epsfig{file=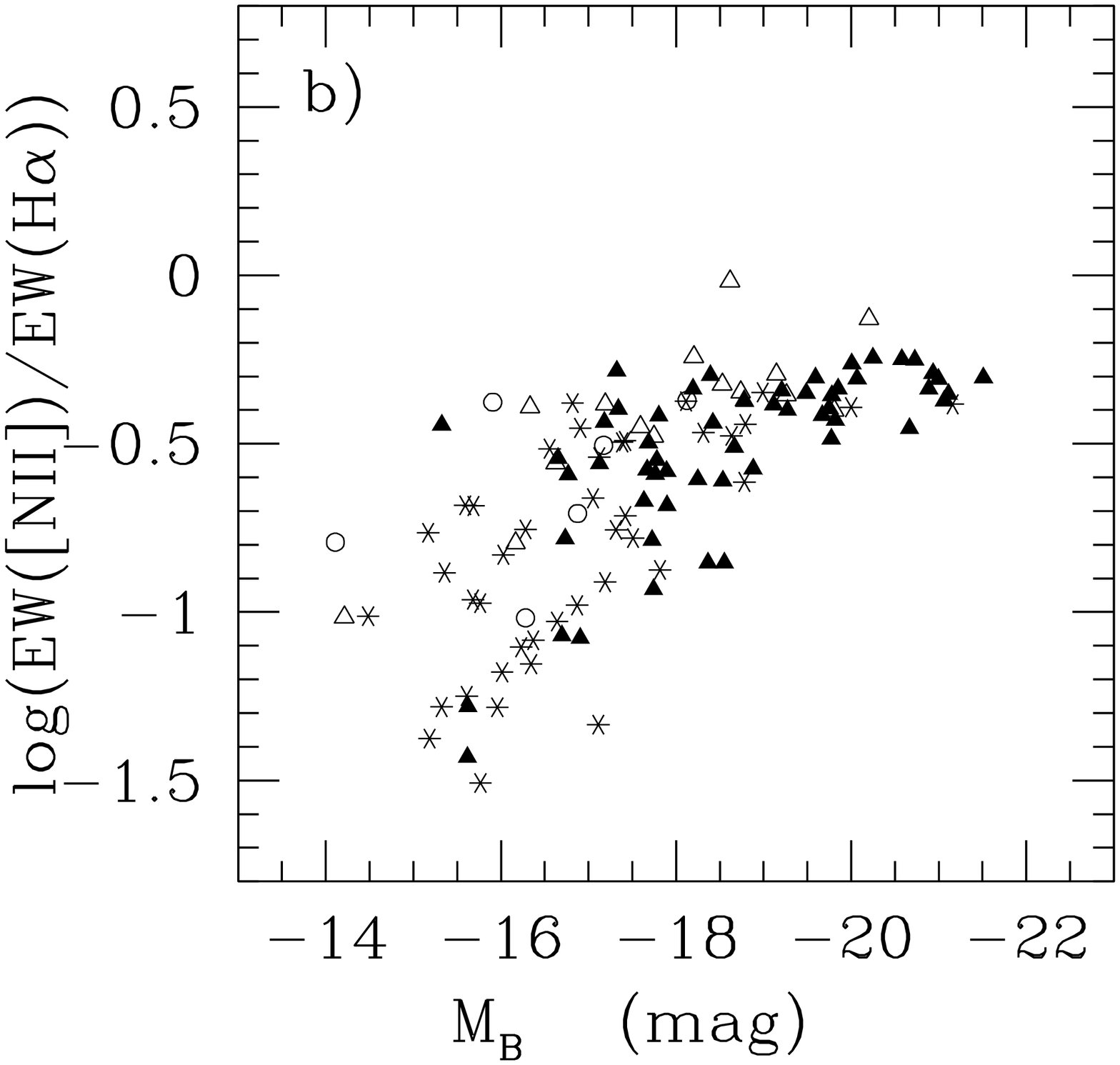,width=0.40\txw,clip=}
   }
}\par\noindent\leavevmode
\makebox[\txw]{
   \centerline{
      \epsfig{file=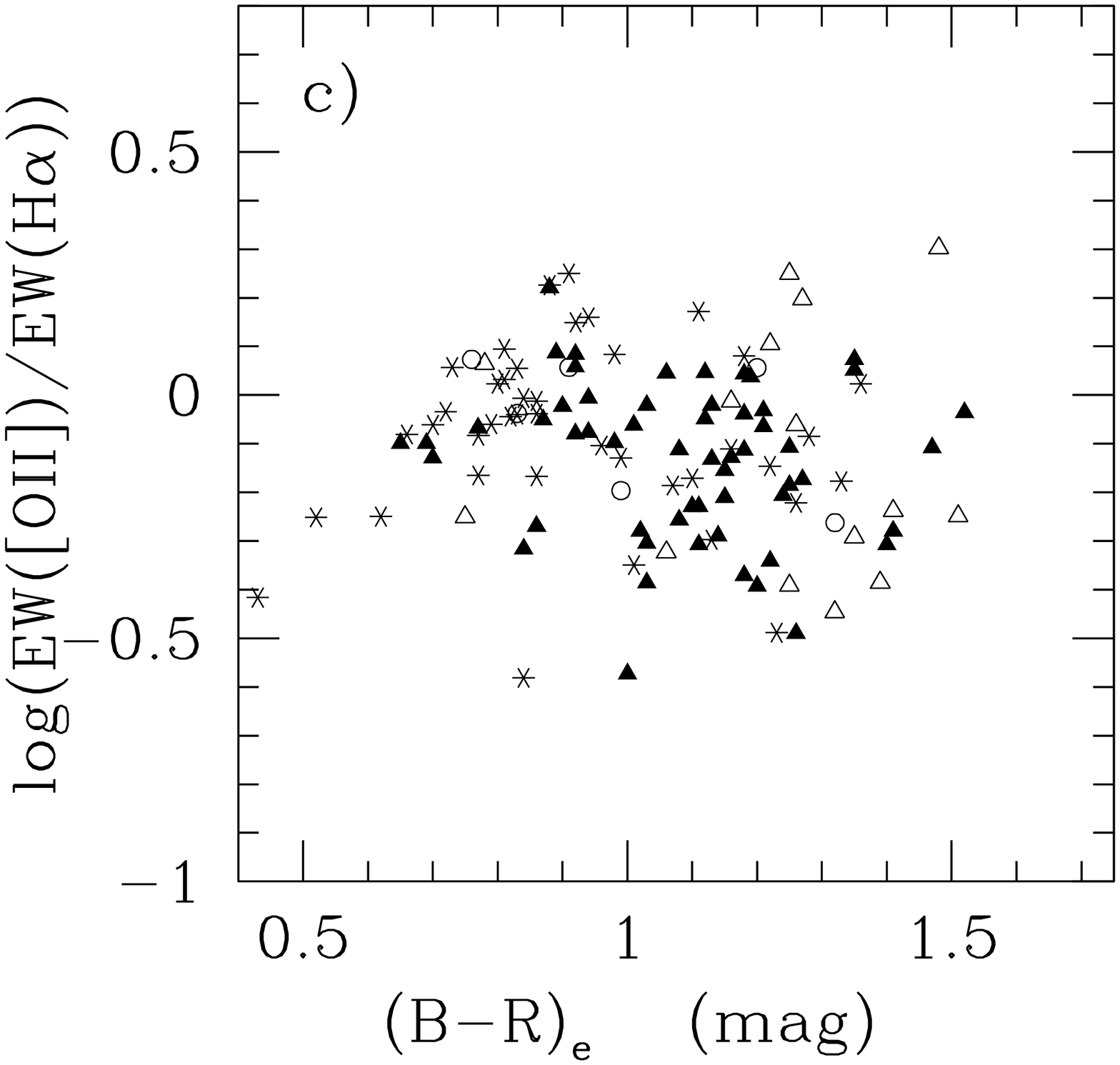,width=0.40\txw,clip=}\hfill
      \epsfig{file=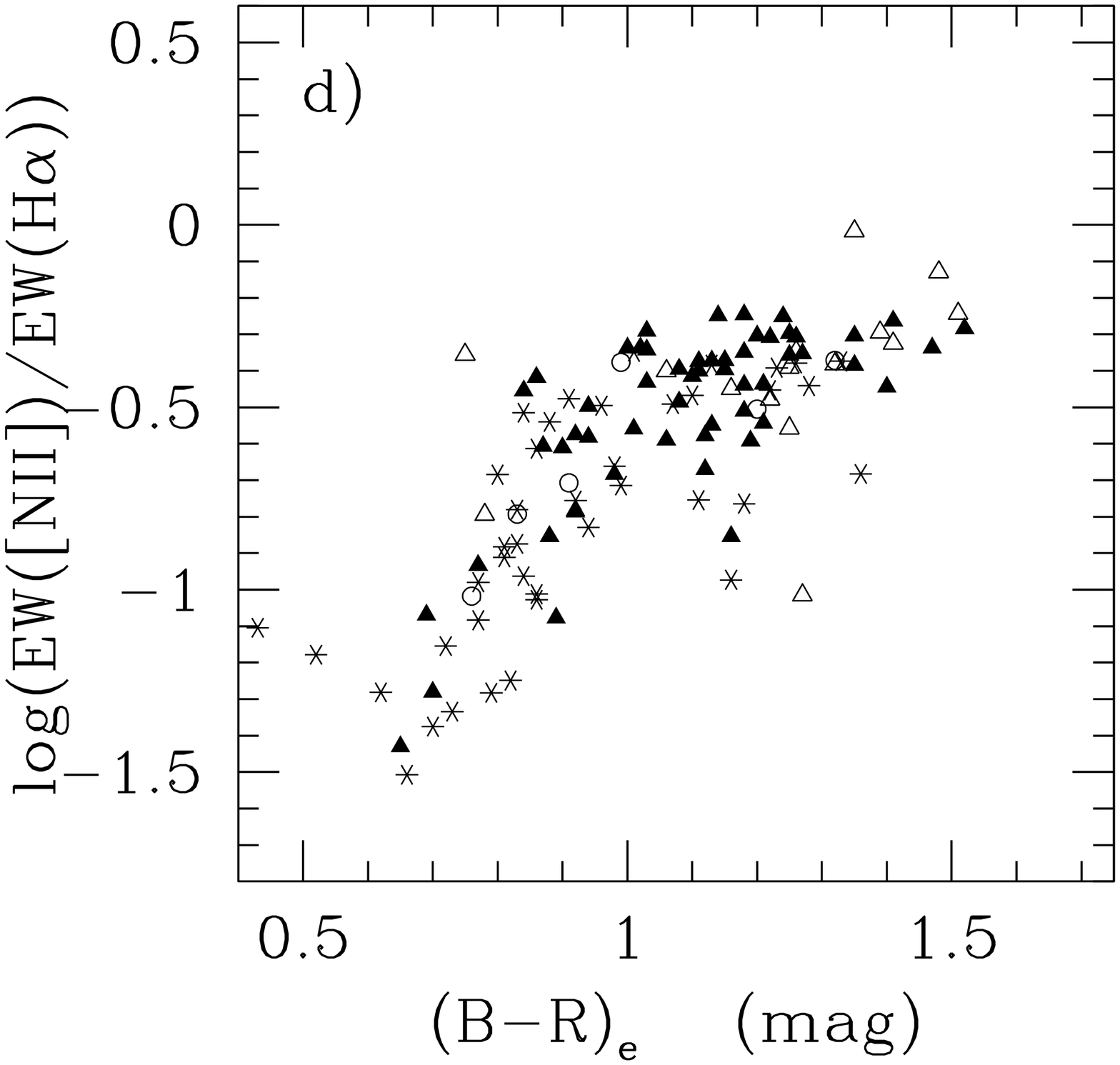,width=0.40\txw,clip=}
   }
}\par\noindent\leavevmode
\makebox[\txw]{
\centerline{
\parbox[t]{\txw}{\footnotesize {\sc Fig.~9 ---} A study of the
distribution of \OII/\Ha\ and \NII/\Ha\ ratios as a function of absolute
$B$ filter magnitude and effective \BR\ color.  Only data for the
galaxies with fairly strong emission (EW(\Ha)$<$$-10$\AA) are shown to
ensure reliable emission line ratios.  Plotting symbols are coded
according to morphological type as in figure~4.  a) \OII/\Ha\ ratios
tend to increase towards lower luminosities, b) \NII/\Ha\ ratios are
smaller in low luminosity galaxies, c) blue systems show somewhat larger
\OII/\Ha\ ratios than red ones, but the scatter is large, d) blue
galaxies tend to have smaller \NII/\Ha\ ratios than red ones, but the
trend is mild for systems redder than \BRe$\;\gtrsim 1$.  } }
}\vspace*{0.7cm}

%

\noindent visible in either nuclear or integrated spectra, and (3) an
active nucleus is not present.  Although galaxies that show nuclear star
formation will {\it ipso facto} show emission in their integrated
spectra as well, the converse need not hold.  Indeed, most galaxies with
only moderately strong emission (EW(\Ha)$\gtrsim$$-30$\AA) lie below the
line of equality.  At larger EWs, the galaxies scatter around this line. 
The scatter is surprisingly large, a result one must bear in mind when
using nuclear emission line strengths to infer the star formation
activity in a galaxy. 

In figure~11{\em b} we check whether this scatter correlates with the
difference in color between the inner and outer parts of the galaxy. 
For this purpose we \vfill

\null\vspace*{0.94\txw}\noindent use the color difference $\Delta
(B-R)_{25-75}$, defined in Paper~I as the difference of the \br\ color
measured in the central region of a galaxy containing 25\% of the light
and that containing the next 50\% of the light.  Indeed a trend is
found: galaxies with stronger \Ha\ in the integrated spectra than in the
nuclear spectra are redder in their inner than in their outer parts
(negative $\Delta (B-R)_{25-75}$).  The galaxies that are bluer in their
central regions tend to have strong central emission.  We note that
trends of this kind can also reflect differences in stellar population
and star formation history, or differences in extinction.  Weaker trends
are found with \MB\ and \BRe\ color (figures~11{\em c} and {\em d}). 
Similar conclusions regarding the nature of \newpage

\noindent\leavevmode
\makebox[\txw]{
   \centerline{
      \epsfig{file=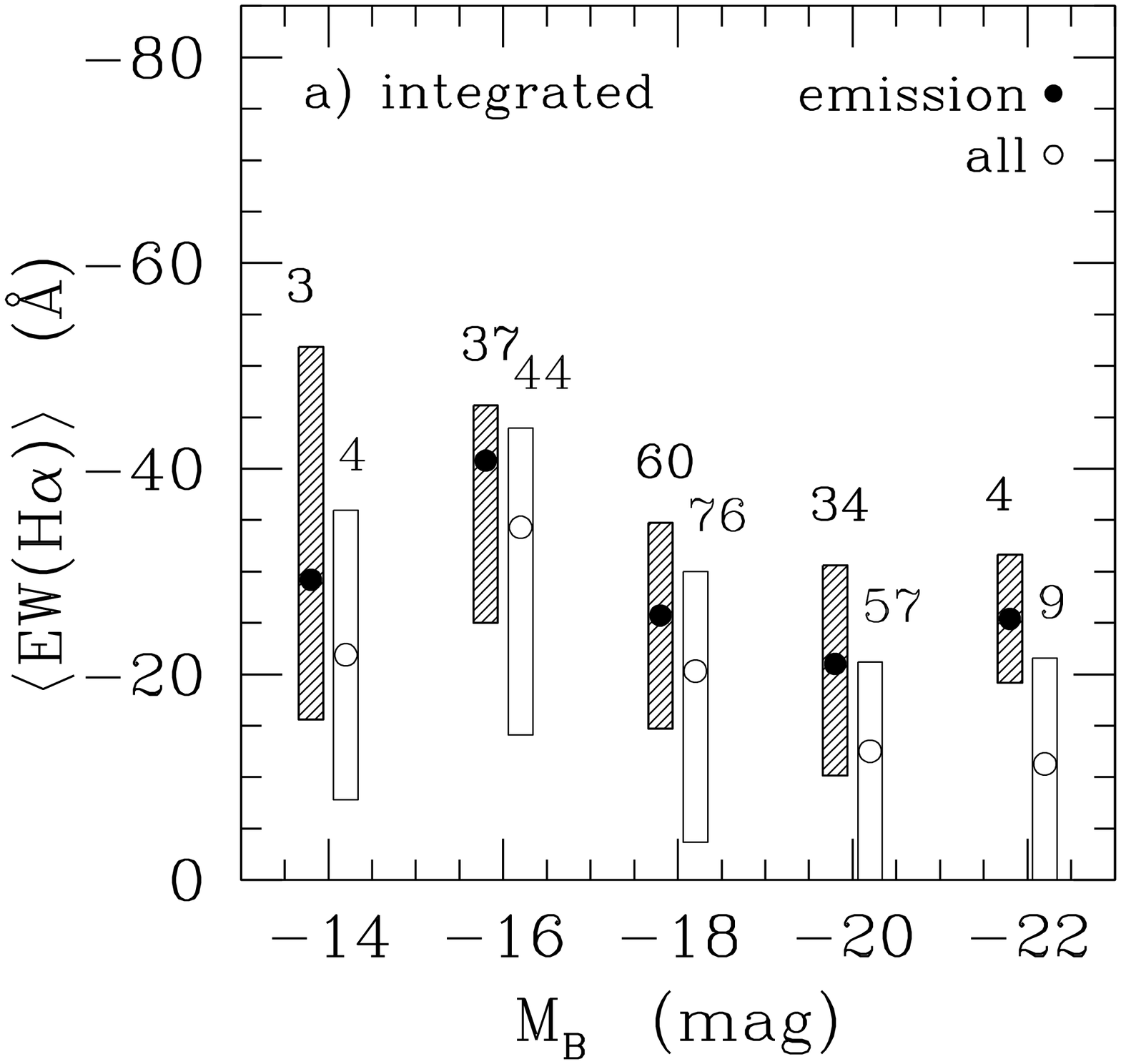,width=0.40\txw,clip=}\hfill
      \epsfig{file=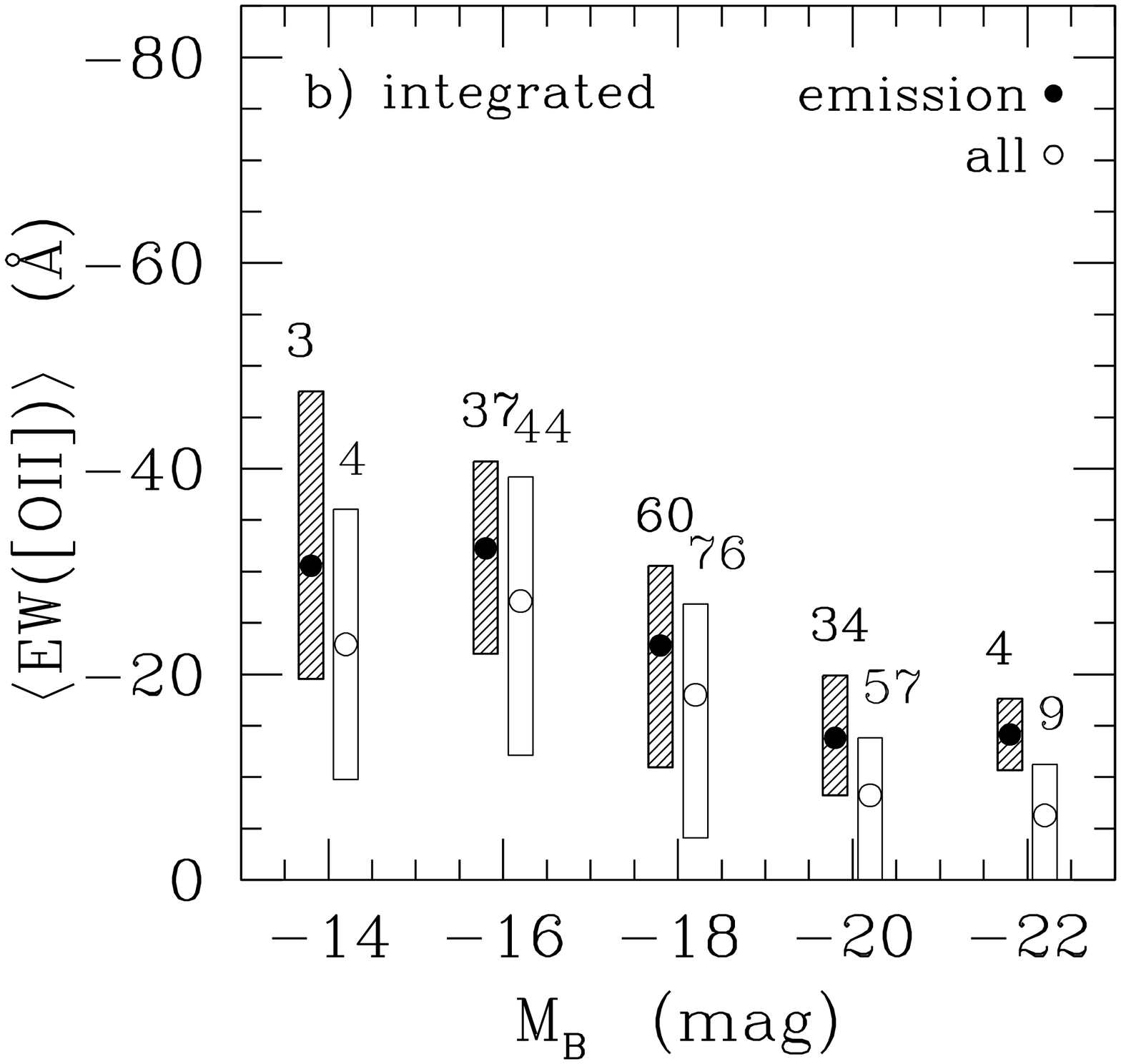,width=0.40\txw,clip=}
   }
}\par\noindent\leavevmode
\makebox[\txw]{
   \centerline{
      \epsfig{file=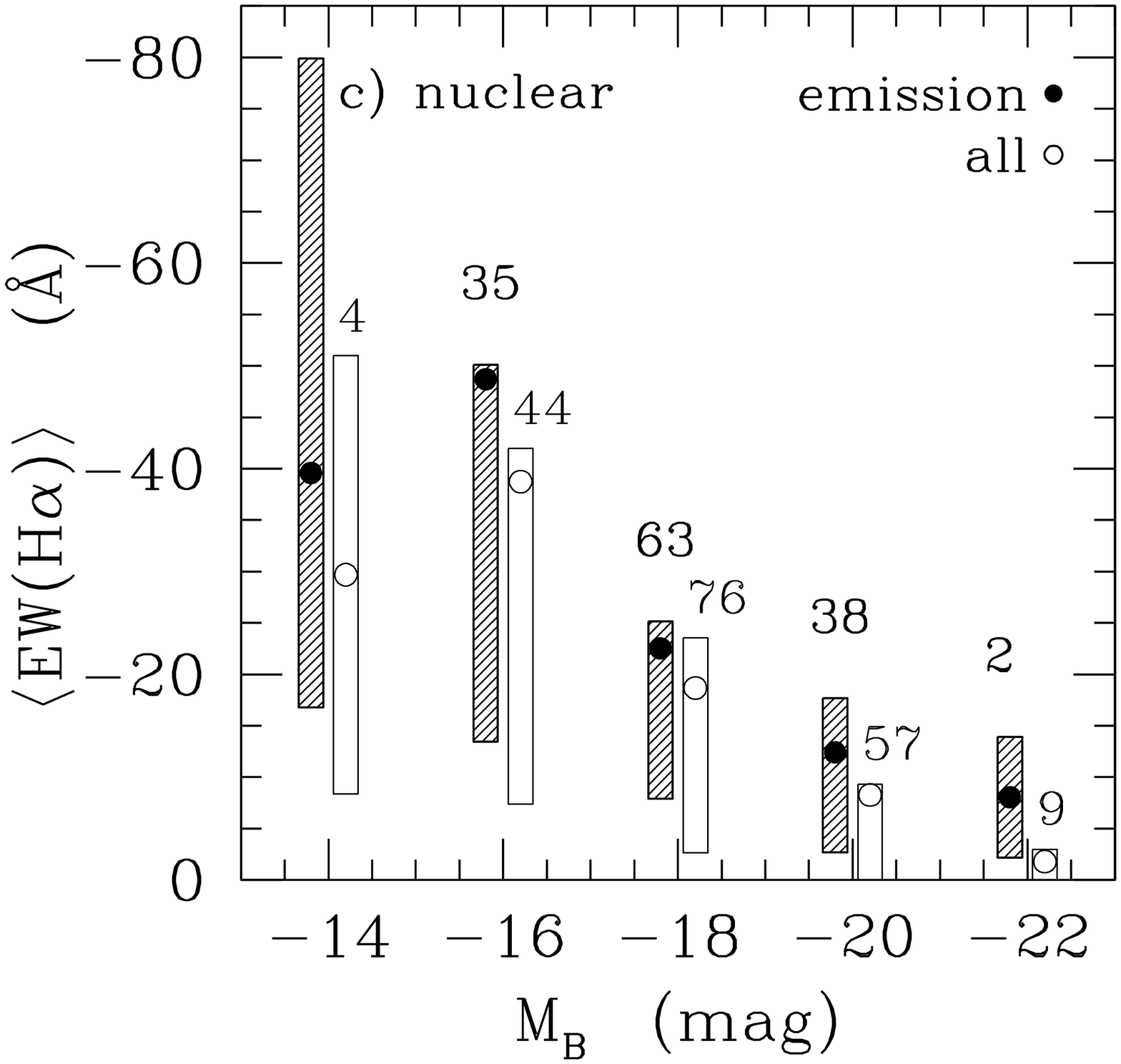,width=0.40\txw,clip=}\hfill
      \epsfig{file=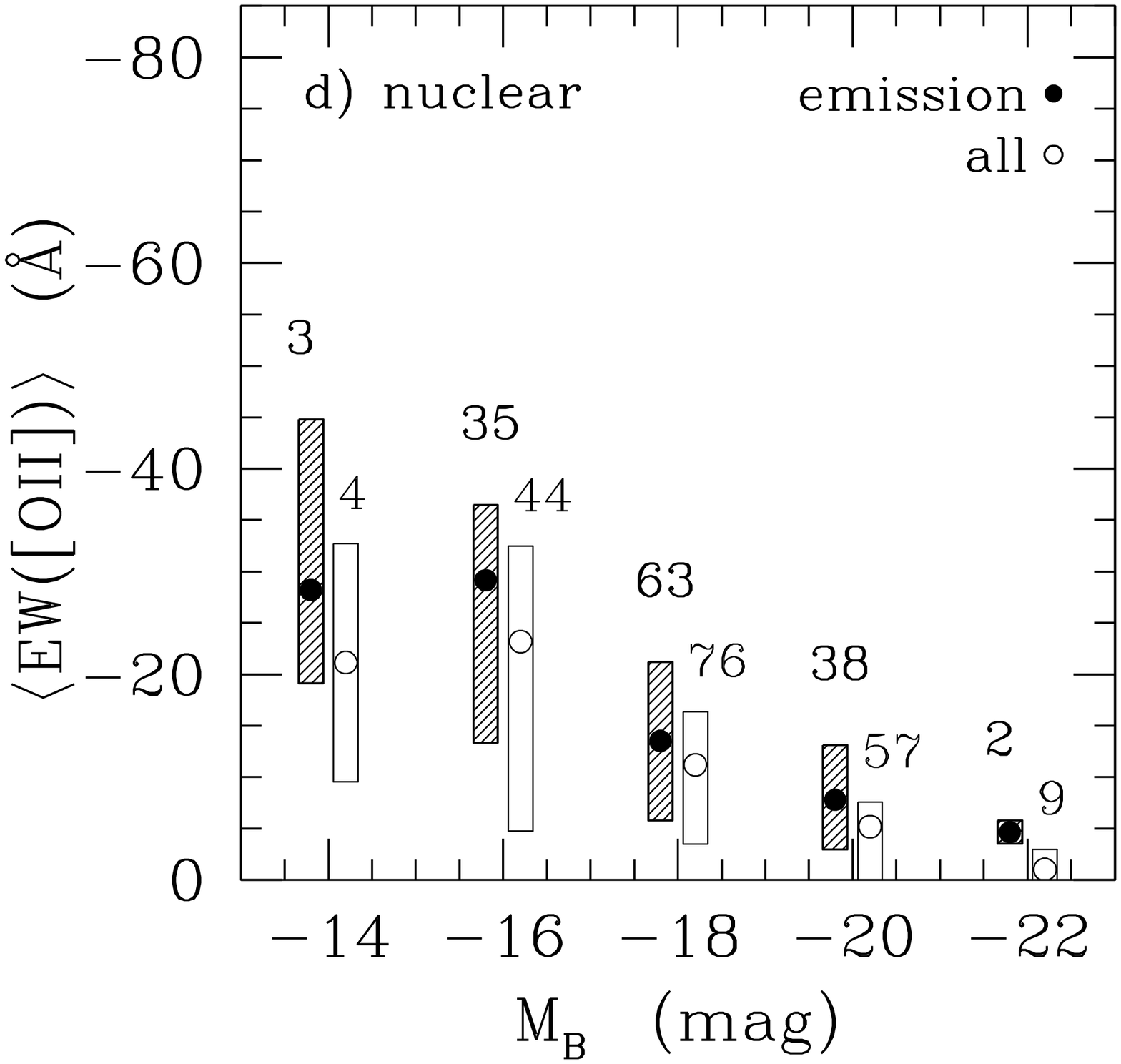,width=0.40\txw,clip=}
   }
}\par\noindent\leavevmode
\makebox[\txw]{
\centerline{
\parbox[t]{\txw}{\footnotesize {\sc Fig.~10 ---} Distributions of the
average EWs of both integrated and nuclear \Ha\ and \OII\ emission as a
function of \MB, averaged in two magnitude wide bins, for the emission
line galaxies and for the general sample.  AGN are excluded in both
cases.  The bars indicate the quartile range (between the 25\% and 75\%
points), a measure of the width of the distribution in each magnitude
bin and indication of the location of the bulk of the galaxies.  The
number of galaxies entering a bin is indicated as well.  a) The average
integrated \Ha\ EW does not depend strongly on absolute magnitude.  b)
The average integrated \OII\ EW does show a trend towards larger EWs in
lower luminosity systems.  c) The nuclear \Ha\ EW does show a moderate
trend towards larger EWs at lower luminosities.  d) As in b) the nuclear
\OII\ EW depends on luminosity.  } }
}\vspace*{0.7cm}

%

\noindent galaxies with bluer inner than outer colors were reached in
Paper~I based on the photometry and morphology alone. 


\section{Summary}
\label{S-Summary}

We have completed a photometric and spectrophotometric survey of 198
nearby galaxies spanning a range of 8 magnitudes in \MB\ and a wide
range of morphologies.  Our main goal is to study the variation in star
formation rates, star formation histories, and metallicities as a
function of galaxy type and luminosity, for a well defined sample of
galaxies in the local field. 

Here we have described the techniques used to ob-\vfill

\null\vspace*{0.955\txw}\noindent tain nuclear and integrated
spectrophotometry, the data reduction, and the spectrophotometric
quality of the data.  We have presented an atlas of nuclear and
integrated spectra, calibrated on a relative flux scale, and tables of
emission line measurements and colors.  We have made extensive use of
the photometric data obtained as part of this survey and described in
Paper~I. 

Our spectra cover wavelengths from 3550--7250\AA, at a resolution of
\tsim 6.0\AA\ (FWHM).  The spectrophotometry is expected to be accurate
to \tsim\tpm6\% on scales of several hundreds of \AA\ or more, and to
\tsim\tpm8\% on small scales of a few \AA.  We have demonstrated the
consistency of our spectrophotometry with our $B$ and \newpage

\noindent\leavevmode
\makebox[\txw]{
   \centerline{
      \epsfig{file=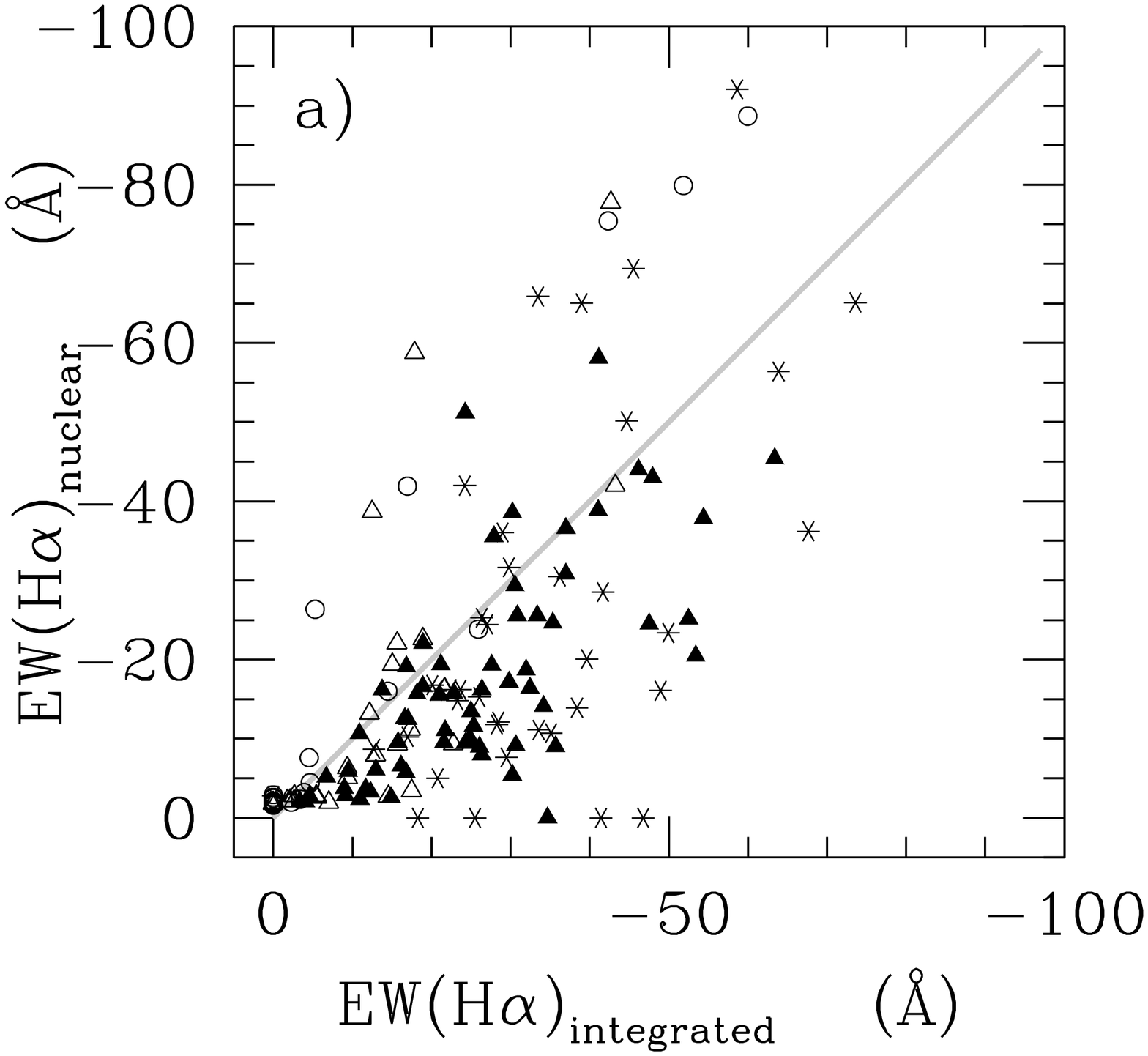,width=0.40\txw,clip=}\hfill
      \epsfig{file=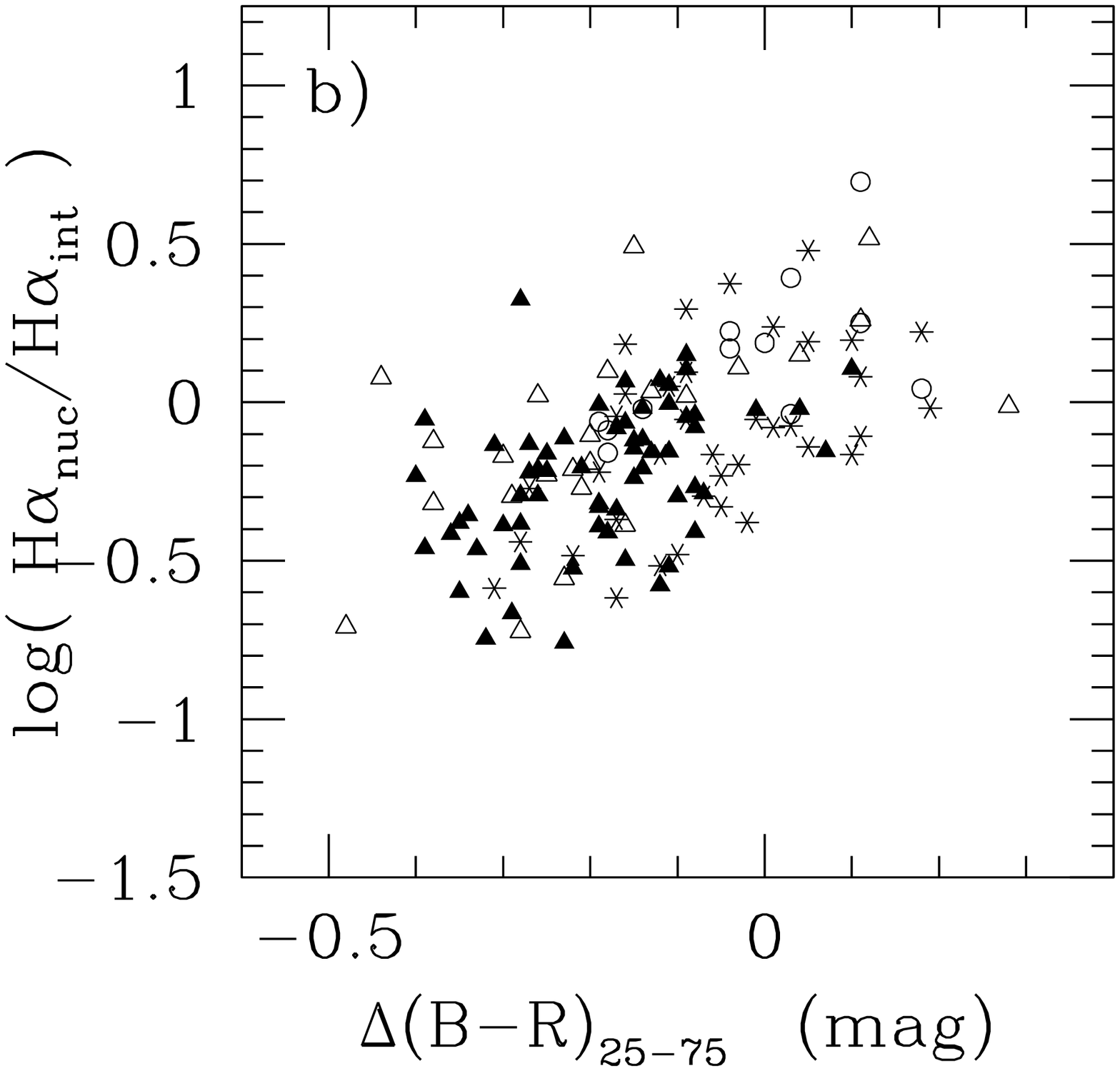,width=0.40\txw,clip=}
   }
}\par\noindent\leavevmode
\makebox[\txw]{
   \centerline{
      \epsfig{file=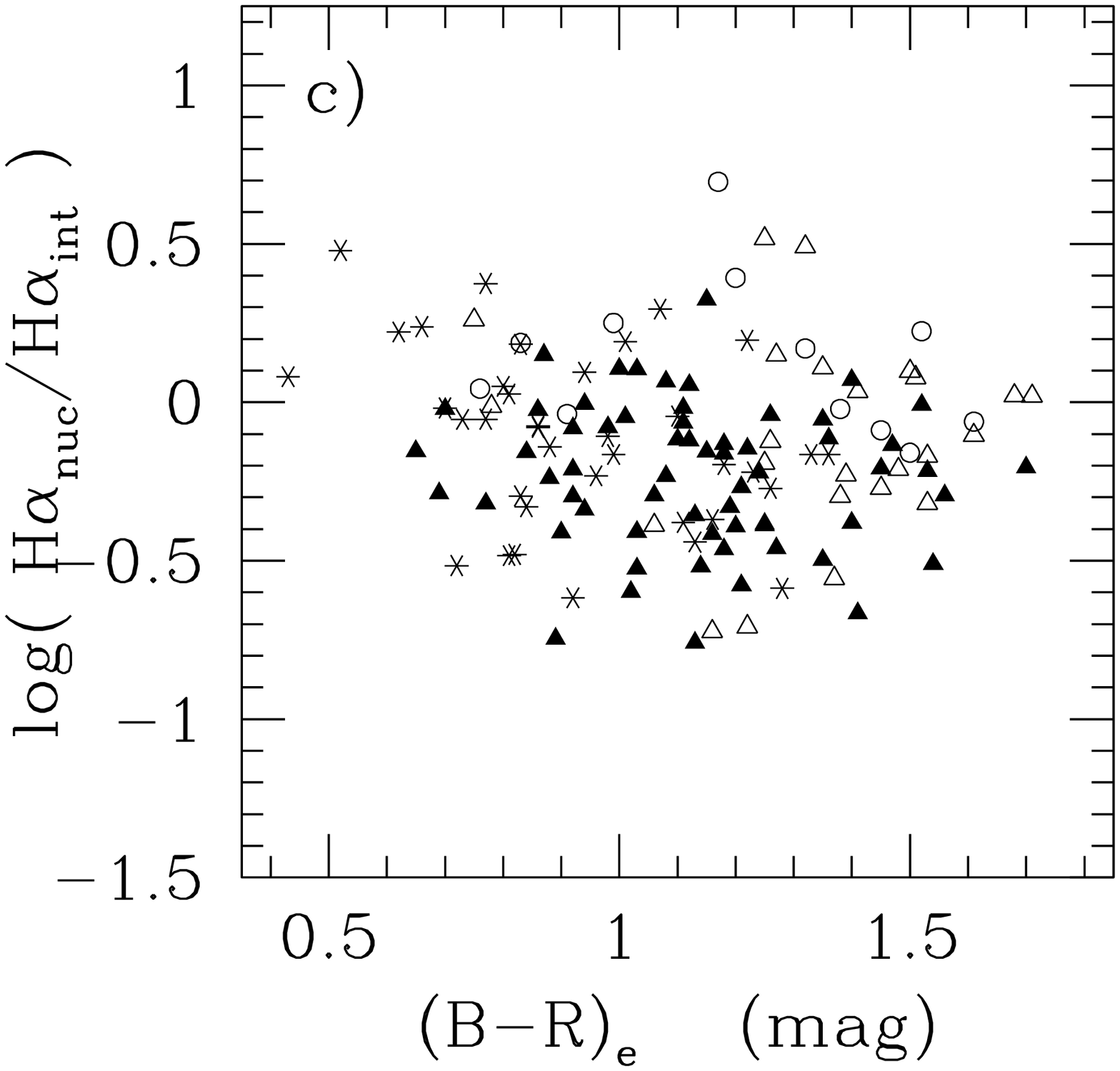,width=0.40\txw,clip=}\hfill
      \epsfig{file=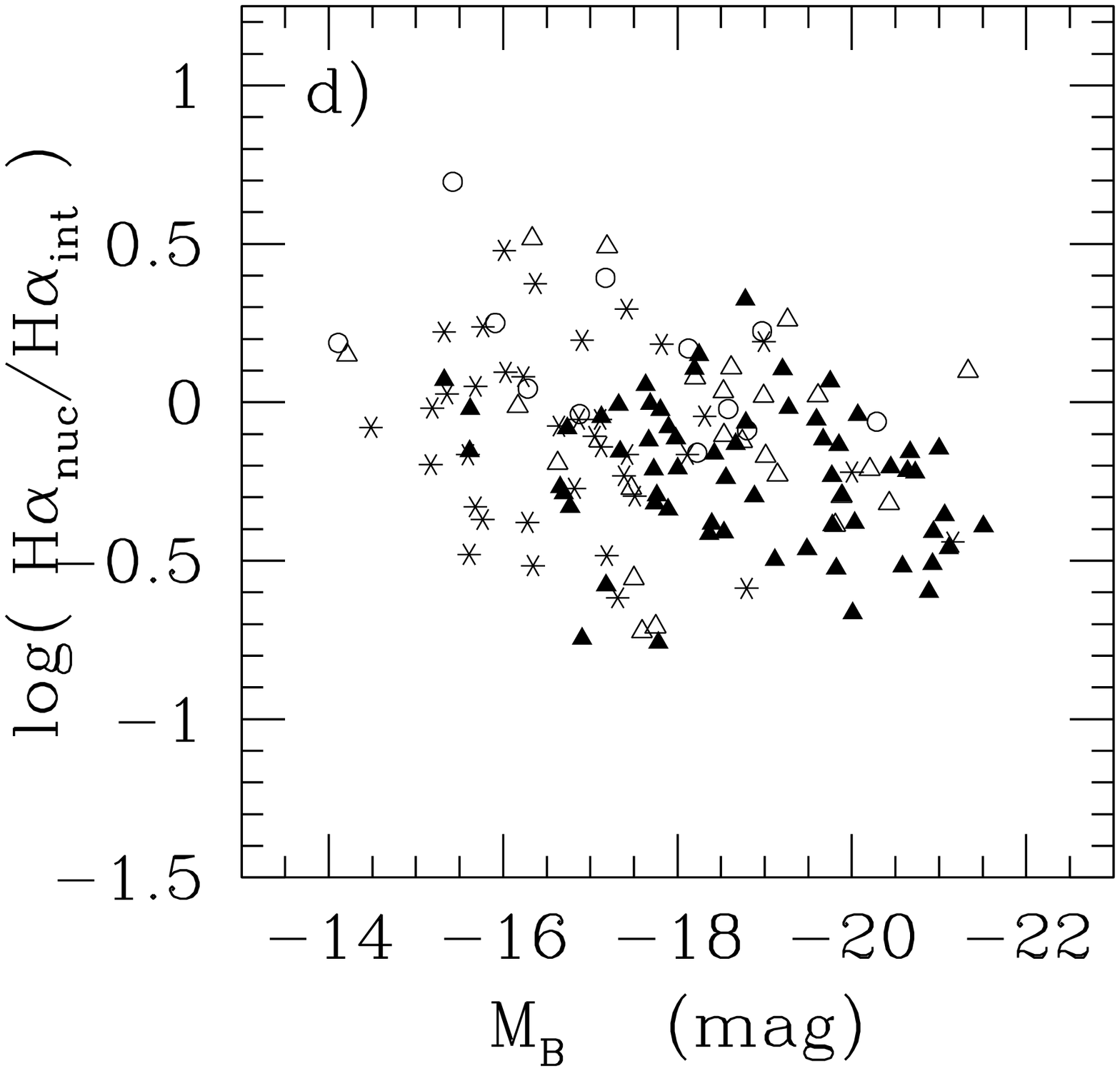,width=0.40\txw,clip=}
   }
}\par\noindent\leavevmode
\makebox[\txw]{
\centerline{
\parbox[t]{\txw}{\footnotesize {\sc Fig.~11 ---} A comparison of nuclear
and integrated emission line strengths.  a) Nuclear versus integrated
\Ha\ EW.  Note the large scatter.  The line indicates where the nuclear
EW equals the integrated EW.  Panels b--d) Ratios of nuclear and
integrated \Ha\ EW versus b) the difference in \br\ color between inner
and outer parts in a galaxy (see text).  As expected, large differences
in nuclear and integrated EW are traced by broadband colors as well.  c)
effective \br\ color.  The nuclear/integrated ratio does not depend on
galaxy color per se.  d) absolute $B$ magnitude.  Lower luminosity
systems show a tendency towards larger nuclear/integrated EW ratios, but
this trend appears to be mainly due to the earlier type galaxies
(E--Scd); the later type galaxies show no trend.  } }
}\vspace*{0.7cm}

%

\noindent $R$ photometry and with previous work (Kennicutt 1992b).  We
plan to make the data, atlas and tables available in electronic format
in due course. 

We find that the range in continuum shape and emission line
characteristics within a given morphological type class is large when
galaxies of a wide range in luminosity are included.  Our initial
results include: (1) the luminosity dependence of the \OII/\Ha\ ratio,
(2) the increase in emission line strengths and bluing of continua as
the luminosity decreases and (3) the large scatter on the expected
general correlation between nuclear and integrated \Ha\ emission line
strengths. \vfill

\null\vspace*{0.92\txw}


\section*{Acknowledgements}

This work was supported by grants from the University of Groningen, the
Leiden Kerkhoven-Bosscha Fund, the Netherlands Organisation for
Scientific Research (NWO), and by the Smithsonian Institution.\\
Many thanks to the TAC for generously allocating time for this project
and keeping faith during three long years.\\
RAJ thanks the Harvard-Smithsonian Center for Astrophysics and the
F.L.~Whipple Observatory for hospitality during numerous visits, when   
all of the observations and part of this work was done.\\
We thank the referee for helpful comments.

\newpage



\appendix
\section{Discussion of individual objects}
\label{S-App.objects}

\noindent{\bf 006 A00510$+$1225} Compact Sc galaxy with Seyfert~{\sc i}
nucleus.  This galaxy was classified as compact elliptical in ZCAT.

\noindent{\bf 031 A02464$+$1807} is a compact galaxy dominated by a
foreground star right on its center.  We succeeded neither in obtaining 
surface photometry of the underlying galaxy, nor spectroscopy for this
galaxy.

\noindent{\bf 039 NGC~2780} Nuclear starburst.

\noindent{\bf 040 A09125$+$5303} This low luminosity dwarf shows the
enhanced Balmer absorption lines and blue continuum of a young
``post-starburst'' galaxy. 

\noindent{\bf 041 NGC~2799} VV~50, Arp~283; interacting pair with
NGC~2798. Star formation is ongoing in the nuclear parts.

\noindent{\bf 048 NGC~3075} Nuclear star formation.

\noindent{\bf 074 A11017$+$3828W} Also known as Markarian 421. This is
the nearest galaxy with a BL~Lac type nucleus.

\noindent{\bf 087 A11332$+$3536} Nuclear starburst.

\noindent{\bf 088 A11336$+$5829} The positional angle listed in the UGC 
has an incorrect sign and should be --9\arcdeg (or 171\arcdeg) rather
than 9\arcdeg.

\noindent{\bf 104 A11592$+$6237} Enhanced central star formation.

\noindent{\bf 108 A12064$+$4201} Nuclear starburst.

\noindent{\bf 119 A12195$+$7535} Seyfert~{\sc i} galaxy Markarian~205,
classified as compact elliptical, seen through the spiral disk of
foreground galaxy NGC 4319.

\noindent{\bf 124 NGC~4509} Very strong emission lines, particularly in
the central parts.

\noindent{\bf 144 NGC~5338} Strong nuclear starburst.

\noindent{\bf 163 A15016$+$1037} Seyfert~{\sc i} nucleus. Also known as
Mar\-ka\-rian 841.

\noindent{\bf 169 NGC~5940} Galaxy with a weak Seyfert~{\sc i} nucleus;
also known as Markarian~9030.

\noindent{\bf 173 IC~1144} Aka.\ Markarian~491. Mild ``post-starburst''
signature. Enhanced Balmer absorption lines.

\noindent{\bf 196 NGC~7752} Strongly interacting companion of NGC 7753
with one of the tidally disrupted arms of NGC 7753 causing a starburst
on one side of NGC 7752.  Also known as AKN 585, IV Zw165 and VV~5.

\newpage

\setlength{\tabcolsep}{1.5pt}
\begin{deluxetable}{rrrrrcrrrrrcrrrrrc}
\scriptsize
\tablecolumns{18}
\tablewidth{6.8in}
\tablenum{6}
\tablecaption{Synthetic color measurements.}
\tablehead{
\colhead{ID}     & \colhead{41-50}  & \colhead{\bv}    &
\colhead{\vr}    & \colhead{\br}    & \colhead{\BRe}   &
\colhead{~~ID}   & \colhead{41-50}  & \colhead{\bv}    &
\colhead{\vr}    & \colhead{\br}    & \colhead{\BRe}   &
\colhead{~~ID}   & \colhead{41-50}  & \colhead{\bv}    &
\colhead{\vr}    & \colhead{\br}    & \colhead{\BRe}   \\
\colhead{\#}     & \colhead{}       & \colhead{synth}  &
\colhead{synth}  & \colhead{synth}  & \colhead{phot}   &
\colhead{~~\#}   & \colhead{}       & \colhead{synth}  &
\colhead{synth}  & \colhead{synth}  & \colhead{phot}   &
\colhead{~~\#}   & \colhead{}       & \colhead{synth}  &
\colhead{synth}  & \colhead{synth}  & \colhead{phot}   \\
\colhead{(1)}    & \colhead{(2)}    & \colhead{(3)}    &
\colhead{(4)}    & \colhead{(5)}    & \colhead{(6)}    &
\colhead{~~(1)}  & \colhead{(2)}    & \colhead{(3)}    &
\colhead{(4)}    & \colhead{(5)}    & \colhead{(6)}    &
\colhead{~~(1)}  & \colhead{(2)}    & \colhead{(3)}    &
\colhead{(4)}    & \colhead{(5)}    & \colhead{(6)}    }
\startdata
~~1 & 0.90 & 0.99 & 0.59 & 1.58 & 1.52 & ~~~67 & 0.63 & 0.80 & 0.51 & 1.31 & 1.40 & ~~133 & 0.34 & 0.38 & 0.30 & 0.68 & 0.77 \nl
~~2 & 0.42 & 0.57 & 0.43 & 1.00 & 0.88 & ~~~68 & 0.44 & 0.59 & 0.45 & 1.04 & 1.27 & ~~134 & 0.41 & 0.52 & 0.39 & 0.91 & 0.94 \nl
~~3 & 0.89 & 1.00 & 0.60 & 1.60 & 1.53 & ~~~69 & 0.30 & 0.45 & 0.37 & 0.82 & 0.75 & ~~135 & 0.39 & 0.56 & 0.42 & 0.99 & 1.10 \nl
~~4 & 0.42 & 0.63 & 0.47 & 1.10 & 1.06 & ~~~70 & 0.54 & 0.69 & 0.50 & 1.18 & 1.21 & ~~136 & ~--- & ~--- & ~--- & ~--- & 0.99 \nl
~~5 & 0.68 & 0.81 & 0.55 & 1.36 & 1.35 & ~~~71 & 0.40 & 0.51 & 0.38 & 0.89 & 0.92 & ~~137 & 0.68 & 0.82 & 0.51 & 1.33 & 1.38 \nl
~~6 & 0.48 & 0.52 & 0.41 & 0.93 & 0.73 & ~~~72 & 0.76 & 0.91 & 0.54 & 1.44 & 1.60 & ~~138 & 0.66 & 0.87 & 0.57 & 1.44 & 1.53 \nl
~~7 & 0.97 & 1.01 & 0.60 & 1.61 & 1.55 & ~~~73 & 0.43 & 0.46 & 0.39 & 0.86 & 0.84 & ~~139 & 0.85 & 0.99 & 0.58 & 1.57 & 1.70 \nl
~~8 & 0.77 & 0.91 & 0.60 & 1.51 & 1.53 & ~~~74 & 0.31 & 0.50 & 0.39 & 0.89 & 0.68 & ~~140 & 0.45 & 0.66 & 0.47 & 1.12 & 1.13 \nl
~~9 & 0.42 & 0.49 & 0.38 & 0.88 & 0.79 & ~~~75 & 0.73 & 0.85 & 0.51 & 1.36 & 1.41 & ~~141 & 0.35 & 0.43 & 0.37 & 0.80 & 0.83 \nl
~10 & 0.96 & 1.02 & 0.61 & 1.64 & 1.57 & ~~~76 & 0.35 & 0.49 & 0.38 & 0.87 & 0.90 & ~~142 & 0.70 & 0.87 & 0.54 & 1.41 & 1.56 \nl
~11 & 0.71 & 0.85 & 0.51 & 1.36 & 1.31 & ~~~77 & 0.57 & 0.74 & 0.52 & 1.26 & 1.26 & ~~143 & 0.45 & 0.63 & 0.45 & 1.08 & 1.12 \nl
~12 & 0.50 & 0.69 & 0.52 & 1.21 & 1.18 & ~~~78 & 0.79 & 0.95 & 0.55 & 1.50 & 1.59 & ~~144 & 0.58 & 0.73 & 0.47 & 1.20 & 1.17 \nl
~13 & 0.75 & 0.88 & 0.55 & 1.43 & 1.38 & ~~~79 & 0.26 & 0.46 & 0.37 & 0.84 & 0.84 & ~~145 & 0.63 & 0.87 & 0.60 & 1.46 & 1.45 \nl
~14 & 0.78 & 0.93 & 0.56 & 1.49 & 1.46 & ~~~80 & 0.80 & 0.89 & 0.52 & 1.41 & 1.41 & ~~146 & 0.81 & 0.92 & 0.54 & 1.46 & 1.54 \nl
~15 & 0.35 & 0.47 & 0.37 & 0.84 & 0.70 & ~~~81 & 0.61 & 0.79 & 0.54 & 1.33 & 1.40 & ~~147 & 0.85 & 0.97 & 0.56 & 1.53 & 1.59 \nl
~16 & 0.44 & 0.57 & 0.43 & 1.01 & 1.03 & ~~~82 & 0.71 & 0.89 & 0.63 & 1.52 & 1.51 & ~~148 & 0.50 & 0.65 & 0.47 & 1.12 & 1.06 \nl
~17 & 0.36 & 0.54 & 0.44 & 0.98 & 0.94 & ~~~83 & 0.35 & 0.48 & 0.37 & 0.84 & 0.76 & ~~149 & 0.88 & 0.99 & 0.57 & 1.56 & 1.52 \nl
~18 & ~--- & ~--- & ~--- & ~--- & 0.52 & ~~~84 & ~--- & ~--- & ~--- & ~--- & 0.80 & ~~150 & ~--- & ~--- & ~--- & ~--- & 1.56 \nl
~19 & 0.51 & 0.68 & 0.57 & 1.24 & 1.22 & ~~~85 & 0.58 & 0.77 & 0.51 & 1.27 & 1.36 & ~~151 & 0.62 & 0.79 & 0.53 & 1.32 & 1.41 \nl
~20 & ~--- & ~--- & ~--- & ~--- & 0.71 & ~~~86 & 0.89 & 1.04 & 0.59 & 1.63 & 1.64 & ~~152 & 0.86 & 1.00 & 0.57 & 1.58 & 1.67 \nl
~21 & 0.48 & 0.60 & 0.44 & 1.04 & 0.91 & ~~~87 & 0.57 & 0.71 & 0.48 & 1.19 & 1.20 & ~~153 & 0.43 & 0.64 & 0.49 & 1.14 & 1.14 \nl
~22 & 0.85 & 0.96 & 0.57 & 1.53 & 1.52 & ~~~88 & 0.40 & 0.51 & 0.40 & 0.91 & 0.92 & ~~154 & 0.80 & 0.93 & 0.56 & 1.49 & 1.50 \nl
~23 & 0.46 & 0.65 & 0.48 & 1.13 & 1.18 & ~~~89 & 0.36 & 0.53 & 0.41 & 0.94 & 0.92 & ~~155 & 0.32 & 0.44 & 0.34 & 0.78 & 1.16 \nl
~24 & 0.36 & 0.51 & 0.40 & 0.91 & 0.88 & ~~~90 & 0.32 & 0.52 & 0.38 & 0.90 & 1.02 & ~~156 & 0.37 & 0.52 & 0.41 & 0.93 & 0.94 \nl
~25 & 0.47 & 0.66 & 0.50 & 1.16 & 1.11 & ~~~91 & 0.52 & 0.69 & 0.48 & 1.17 & 1.19 & ~~157 & 0.79 & 0.93 & 0.55 & 1.48 & 1.53 \nl
~26 & 0.85 & 0.99 & 0.60 & 1.59 & 1.61 & ~~~92 & 0.38 & 0.53 & 0.41 & 0.94 & 0.91 & ~~158 & 0.48 & 0.62 & 0.67 & 1.29 & 1.22 \nl
~27 & 0.60 & 0.74 & 0.51 & 1.25 & 1.24 & ~~~93 & 0.50 & 0.52 & 0.40 & 0.93 & 0.83 & ~~159 & 0.39 & 0.18 & 0.22 & 0.40 & 0.43 \nl
~28 & 0.43 & 0.61 & 0.46 & 1.07 & 0.98 & ~~~94 & 0.34 & 0.44 & 0.35 & 0.79 & 0.83 & ~~160 & 0.37 & 0.45 & 0.33 & 0.78 & 0.86 \nl
~29 & 0.96 & 1.04 & 0.64 & 1.68 & 1.68 & ~~~95 & ~--- & ~--- & ~--- & ~--- & 0.94 & ~~161 & 0.52 & 0.66 & 0.46 & 1.12 & 1.16 \nl
~30 & 0.97 & 1.02 & 0.61 & 1.63 & 1.63 & ~~~96 & 0.41 & 0.58 & 0.45 & 1.03 & 0.99 & ~~162 & 0.30 & 0.37 & 0.30 & 0.67 & 0.65 \nl
~31 & ~--- & 0.93 & 0.54 & 1.47 & 1.18 & ~~~97 & ~--- & ~--- & ~--- & ~--- & 1.00 & ~~163 & 0.52 & 0.41 & 0.53 & 0.94 & 0.76 \nl
~32 & 0.43 & 0.43 & 0.40 & 0.83 & 0.84 & ~~~98 & 0.33 & 0.50 & 0.39 & 0.88 & 0.92 & ~~164 & 0.46 & 0.67 & 0.48 & 1.15 & 1.23 \nl
~33 & 0.88 & 0.98 & 0.57 & 1.55 & 1.53 & ~~~99 & ~--- & ~--- & ~--- & ~--- & 1.06 & ~~165 & 0.51 & 0.68 & 0.45 & 1.13 & 1.35 \nl
~34 & 0.68 & 0.84 & 0.56 & 1.40 & 1.48 & ~~100 & 0.46 & 0.70 & 0.50 & 1.19 & 1.20 & ~~166 & 0.33 & 0.47 & 0.38 & 0.85 & 0.86 \nl
~35 & 0.96 & 1.04 & 0.64 & 1.68 & 1.68 & ~~101 & ~--- & ~--- & ~--- & ~--- & 0.96 & ~~167 & 0.74 & 0.89 & 0.54 & 1.43 & 1.53 \nl
~36 & 0.81 & 0.91 & 0.53 & 1.44 & 1.71 & ~~102 & 0.27 & 0.37 & 0.32 & 0.69 & 0.72 & ~~168 & 0.57 & 0.75 & 0.53 & 1.28 & 1.39 \nl
~37 & 0.67 & 0.87 & 0.57 & 1.44 & 1.54 & ~~103 & 0.42 & 0.64 & 0.44 & 1.07 & 1.13 & ~~169 & 0.48 & 0.66 & 0.48 & 1.13 & 1.20 \nl
~38 & 0.34 & 0.46 & 0.36 & 0.82 & 0.86 & ~~104 & 0.30 & 0.33 & 0.31 & 0.64 & 0.77 & ~~170 & 0.49 & 0.65 & 0.46 & 1.11 & 1.25 \nl
~39 & 0.56 & 0.74 & 0.52 & 1.26 & 1.32 & ~~105 & 0.50 & 0.61 & 0.50 & 1.11 & 1.22 & ~~171 & 0.41 & 0.58 & 0.40 & 0.98 & 1.03 \nl
~40 & 0.33 & 0.49 & 0.35 & 0.83 & 1.16 & ~~106 & 0.76 & 0.88 & 0.55 & 1.43 & 1.51 & ~~172 & 0.64 & 0.78 & 0.54 & 1.32 & 1.35 \nl
~41 & 0.41 & 0.57 & 0.45 & 1.02 & 1.07 & ~~107 & 0.40 & 0.51 & 0.41 & 0.92 & 0.98 & ~~173 & 0.70 & 0.89 & 0.50 & 1.38 & 1.46 \nl
~42 & 0.75 & 0.89 & 0.56 & 1.45 & 1.52 & ~~108 & 0.55 & 0.73 & 0.50 & 1.23 & 1.25 & ~~174 & 0.52 & 0.71 & 0.46 & 1.17 & 1.27 \nl
~43 & 0.66 & 0.82 & 0.54 & 1.36 & 1.45 & ~~109 & 0.34 & 0.40 & 0.35 & 0.75 & 1.16 & ~~175 & 0.41 & 0.52 & 0.42 & 0.94 & 1.01 \nl
~44 & 0.59 & 0.70 & 0.48 & 1.18 & 1.25 & ~~110 & 0.43 & 0.53 & 0.41 & 0.95 & 0.99 & ~~176 & 0.90 & 1.02 & 0.60 & 1.62 & 1.59 \nl
~45 & 0.44 & 0.61 & 0.44 & 1.05 & 1.11 & ~~111 & ~--- & ~--- & ~--- & ~--- & 0.86 & ~~177 & 0.85 & 0.96 & 0.57 & 1.53 & 1.58 \nl
~46 & 0.49 & 0.65 & 0.56 & 1.20 & 1.26 & ~~112 & 0.40 & 0.52 & 0.41 & 0.93 & 0.87 & ~~178 & 0.88 & 0.99 & 0.56 & 1.55 & 1.55 \nl
~47 & 0.44 & 0.47 & 0.37 & 0.84 & 0.81 & ~~113 & 0.55 & 0.70 & 0.46 & 1.16 & 1.36 & ~~179 & 0.35 & 0.54 & 0.40 & 0.94 & 1.08 \nl
~48 & 0.46 & 0.62 & 0.45 & 1.07 & 1.15 & ~~114 & 0.45 & 0.64 & 0.46 & 1.10 & 1.18 & ~~180 & 0.72 & 0.90 & 0.52 & 1.43 & 1.50 \nl
~49 & 0.41 & 0.61 & 0.45 & 1.05 & 1.18 & ~~115 & ~--- & ~--- & ~--- & ~--- & 1.63 & ~~181 & 0.49 & 0.53 & 0.41 & 0.94 & 0.78 \nl
~50 & 0.34 & 0.36 & 0.32 & 0.68 & 0.79 & ~~116 & 0.33 & 0.45 & 0.36 & 0.81 & 0.80 & ~~182 & 0.93 & 1.01 & 0.59 & 1.61 & 1.60 \nl
~51 & 0.44 & 0.56 & 0.42 & 0.98 & 1.18 & ~~117 & 0.73 & 0.84 & 0.50 & 1.34 & 1.39 & ~~183 & 0.51 & 0.67 & 0.53 & 1.20 & 1.15 \nl
~52 & 0.44 & 0.54 & 0.40 & 0.94 & 1.11 & ~~118 & ~--- & ~--- & ~--- & ~--- & 0.79 & ~~184 & 0.68 & 0.85 & 0.59 & 1.43 & 1.41 \nl
~53 & 0.36 & 0.48 & 0.39 & 0.88 & 0.89 & ~~119 & 0.42 & 0.46 & 0.40 & 0.86 & 0.73 & ~~185 & 0.76 & 0.91 & 0.57 & 1.48 & 1.45 \nl
~54 & 0.76 & 0.90 & 0.51 & 1.42 & 1.56 & ~~120 & ~--- & ~--- & ~--- & ~--- & 0.70 & ~~186 & 0.41 & 0.53 & 0.36 & 0.90 & 0.86 \nl
~55 & 0.35 & 0.53 & 0.39 & 0.92 & 0.96 & ~~121 & 0.31 & 0.38 & 0.28 & 0.66 & 0.62 & ~~187 & 0.62 & 0.77 & 0.60 & 1.37 & 1.32 \nl
~56 & 0.52 & 0.68 & 0.52 & 1.20 & 1.21 & ~~122 & 0.34 & 0.42 & 0.33 & 0.75 & 0.82 & ~~188 & 1.01 & 1.07 & 0.63 & 1.69 & 1.65 \nl
~57 & 0.48 & 0.42 & 0.34 & 0.77 & 0.77 & ~~123 & 0.32 & 0.45 & 0.34 & 0.79 & 0.81 & ~~189 & 0.45 & 0.63 & 0.49 & 1.12 & 1.03 \nl
~58 & 0.71 & 0.86 & 0.58 & 1.44 & 1.52 & ~~124 & 0.38 & 0.29 & 0.28 & 0.57 & 0.52 & ~~190 & 0.52 & 0.68 & 0.53 & 1.21 & 1.25 \nl
~59 & 0.37 & 0.56 & 0.44 & 1.00 & 1.00 & ~~125 & 0.40 & 0.59 & 0.46 & 1.05 & 1.10 & ~~191 & 0.91 & 1.01 & 0.60 & 1.61 & 1.61 \nl
~60 & 0.57 & 0.74 & 0.53 & 1.27 & 1.33 & ~~126 & 0.24 & 0.38 & 0.32 & 0.70 & 0.66 & ~~192 & 0.54 & 0.73 & 0.54 & 1.27 & 1.28 \nl
~61 & 0.36 & 0.52 & 0.40 & 0.92 & 1.08 & ~~127 & 0.49 & 0.71 & 0.54 & 1.26 & 1.12 & ~~193 & 0.45 & 0.61 & 0.47 & 1.09 & 1.13 \nl
~62 & 0.32 & 0.36 & 0.30 & 0.66 & 0.70 & ~~128 & 0.86 & 0.96 & 0.56 & 1.52 & 1.49 & ~~194 & 0.91 & 1.04 & 0.62 & 1.67 & 1.61 \nl
~63 & 0.28 & 0.36 & 0.30 & 0.65 & 0.69 & ~~129 & 0.82 & 0.96 & 0.55 & 1.52 & 1.57 & ~~195 & 0.63 & 0.85 & 0.61 & 1.46 & 1.47 \nl
~64 & 0.49 & 0.70 & 0.49 & 1.19 & 1.26 & ~~130 & ~--- & ~--- & ~--- & ~--- & 1.54 & ~~196 & 0.44 & 0.56 & 0.52 & 1.08 & 1.01 \nl
~65 & 0.77 & 0.98 & 0.61 & 1.59 & 1.70 & ~~131 & 0.86 & 0.98 & 0.55 & 1.54 & 1.60 & ~~197 & 0.93 & 1.05 & 0.62 & 1.67 & 1.61 \nl
~66 & 0.62 & 0.77 & 0.50 & 1.28 & 1.37 & ~~132 & ~--- & ~--- & ~--- & ~--- & 0.93 & ~~198 & 0.40 & 0.40 & 0.33 & 0.73 & 0.73 \nl
\enddata
\end{deluxetable}
\normalsize
\clearpage

\end{document}